\documentclass[numberedappendix]{emulateapj}
\usepackage{natbib}
\usepackage{amsmath}

\newcommand{\msol}{\,M_\sun}                

\newcommand{\HST}{\emph{HST}}
\newcommand{\jkcs}{JKCS~041}
\newcommand{\Ntext}{N12}

\shorttitle{Spectroscopic Confirmation of the Rich $z=1.80$ Galaxy Cluster JKCS~041}
\shortauthors{Newman et al.}
\submitted{}

\begin{document}
\title{Spectroscopic Confirmation of the Rich $z=1.80$ Galaxy Cluster JKCS~041 using the WFC3 Grism:  Environmental Trends in the Ages and Structure of Quiescent Galaxies}
\author{Andrew B. Newman\altaffilmark{1,2}, Richard S. Ellis\altaffilmark{1}, Stefano Andreon\altaffilmark{3}, Tommaso Treu\altaffilmark{4}, Anand Raichoor\altaffilmark{3,5}, and Ginevra Trinchieri\altaffilmark{3}}
\altaffiltext{1}{Cahill Center for Astronomy and Astrophysics, California Institute of Technology, MS 249-17, Pasadena, CA 91125, USA}
\altaffiltext{2}{Current Address: The Observatories of the Carnegie Institution for Science, 813 Santa Barbara Street, Pasadena, CA 91101, USA}
\altaffiltext{3}{INAF-Osservatorio Astronomico di Brera, via Brera 28, 20121, Milano, Italy}
\altaffiltext{4}{Department of Physics, University of California, Santa Barbara, CA 93106, USA}
\altaffiltext{5}{GEPI, Observatoire de Paris, 77 av.~Denfert Rochereau, 75014 Paris, France}
\email{Email: anewman@obs.carnegiescience.edu}

\begin{abstract}
We present \emph{Hubble Space Telescope} imaging and grism spectroscopy in the field of the distant galaxy cluster \jkcs~using the Wide Field Camera 3. We confirm that \jkcs~is a rich cluster and derive a redshift $z=1.80$ via the spectroscopic identification of 19 member galaxies, of which 15 are quiescent. These are centered upon diffuse X-ray emission seen by the \emph{Chandra} observatory. As \jkcs~is the most distant known cluster with such a large, spectroscopically confirmed quiescent population, it provides a unique opportunity to study the effect of the environment on galaxy properties at early epochs. We construct high-quality composite spectra of the  quiescent cluster members that reveal prominent Balmer and metallic absorption lines. Using these, we measure the mean stellar ages in two bins of stellar mass. The quiescent cluster members' ages agree remarkably closely with that inferred by Whitaker et al.~for similarly selected samples in the field, supporting the idea that the cluster environment is more efficient at truncating star formation while not having a strong effect on the mean epoch of quenching. We find some evidence (90\% confidence) for a lower fraction of disk-like quiescent systems in \jkcs~compared to a sample of coeval field galaxies drawn from the CANDELS survey. Taking this into account, we do not detect a significant difference between the mass--radius relations of the quiescent \jkcs~members and our $z \sim 1.8$ field sample. Finally, we demonstrate how differences in the morphological mixture of quenched systems can complicate measures of the environmental dependence of size growth.
\end{abstract}


\section{Introduction}

The evolution of the structure and stellar populations of massive galaxies at high redshifts entails some of the key puzzles in galaxy evolution. The mean size of quiescent galaxies increases by a factor of about four since $z = 2.5$ \citep[e.g.,][]{Buitrago08,vanDokkum08,Toft09,Williams11,Damjanov11,Newman12}, as reflected in a progressive buildup of light in their outer envelopes \citep{vanDokkum10}, while the typical morphologies of the massive examples appear to become more spheroid-dominated \citep{vanderWel11,Chang13b,Chang13}. At the same time, star formation is being truncated in many galaxies as they transition onto the red sequence. Both the rates of structural growth and the increase in number density of quenched galaxies appear to accelerate at $z \gtrsim 1.5$ \citep[][hereafter N12]{Newman12}.

The physical mechanisms driving these changes are only partially understood. Accretion of material in low-mass, gas-poor satellites has emerged as a popular explanation for the structural changes, since this adds stars at large radii while increasing the overall mass comparatively little \citep[e.g.,][]{Naab09,Bezanson09,Hopkins09}. However, the observed (N12) and theoretical \citep{Nipoti12} rates of such minor mergers appear too low to account fully for the rate of size growth, suggesting that additional processes may be at play. The continual arrival of new galaxies onto the red sequence whose sizes may differ from those of the older population already in place further complicates the interpretation; this could lead to a type of progenitor bias whose significance is still debated observationally (e.g., N12; \citealt{Whitaker12b,Carollo13,Poggianti13}). Whether a stochastic merger history can lead to the tight scaling relations seen locally has also been questioned by some authors \citep[e.g.,][]{Nipoti09,Nair10}. 

Additional insight into the growth mechanisms arises from lookback studies that compare the rates of structural evolution as a function of environment. It is expected that the merger history of a galaxy depends on the local density or halo mass \citep{McIntosh08,Fakhouri10,Lin10,Jian12,Kampczyk13}. If size growth is primarily merger driven, it is natural to expect that it will proceed at a rate that depends on past merger activity. Internally driven growth processes such as expansion via mass loss \citep{Fan08,Fan10,Damjanov09}, on the other hand, should be less sensitive to environment. At the same time as gradual size growth proceeds, morphological transformations occur through a variety of processes that are both environmentally related (e.g., mergers, galaxy harassment, tidal interactions, and gas deprivation; see \citealt{Treu03,Moran07}, and references therein) and internally driven (e.g., secular bulge growth; \citealt{Kormendy04}). Lookback studies to $z \sim 0.5 - 1$ have been essential to determine the history of morphological change in clusters \citep[e.g.,][]{Dressler97,Postman05,Poggianti09}.

Similarly, while the cessation of star formation is clearly influenced by both environmentally related processes ---  e.g., ram-pressure stripping, gas starvation, galaxy--galaxy interactions --- as well as internal mechanisms, such as feedback from supernovae or an active galactic nucleus (AGN), the underlying physical processes and their relative importance as a function of mass and cosmic time remain uncertain. Understanding the history of star formation quenching in different environments aids in disentangling the influence of these processes. Observationally, this is constrained by the evolution of the fraction of quenched galaxies and their star formation histories in clusters, groups, and the field \citep[e.g.,][]{Finn10,Tran10,Quadri12,Muzzin12,Raichoor12b,Dressler13,Brodwin13,Bedregal13,Alberts14}.

High-redshift galaxy clusters represent excellent laboratories in which to address these questions, since they probe extreme overdensities at the epoch when quiescent galaxy growth and also the buildup of the red sequence appear most rapid. The expected decline in the number of clusters at high redshifts, coupled with the increasing difficultly of the observations necessary to locate and confirm them, has limited our knowledge of these systems. To date, only a handful of $z>1.6$ clusters hosting red galaxies are known \citep[e.g.,][see Section~\protect\ref{sec:discussion}]{Papovich10,Gobat13,Stanford12,Zeimann12,Muzzin13,Tanaka13,Galametz13}. Spectroscopic data is required not only to confirm a putative cluster and isolate its members but also to precisely constrain the stellar populations and past star formation activity. Very few red cluster galaxies have been spectroscopically studied thus far, which has been the prime limiting factor in undertaking a study of the role of the environment in their evolution.

In this paper we present imaging and grism spectroscopy for the cluster candidate \jkcs~using the Wide Field Camera 3 (WFC3) on board the \emph{Hubble Space Telescope} (\emph{HST}). \jkcs~was originally discovered as an overdensity of galaxies with similar colors \citep{Andreon09} in images from the UKIRT Infrared Deep Sky Survey \citep{Lawrence07}. It exhibits a tight red sequence coincident with diffuse X-ray emission \citep{Andreon11,Andreon11b} detected securely in a 75~ks \emph{Chandra} observation. The X-ray observations and the galaxy richness indicate a relatively high halo mass of $\log M_{200}/\msol \simeq 14.2-14.5$ \citep{Andreon13}. \jkcs~was not detected in a Sunyaev--Zel'dovich (SZ) survey conducted by \citet{Culverhouse10}, but the present upper limit on the mass is consistent with these X-ray-- and richness-based mass estimates (Section~\ref{sec:discussion}). Estimates of the redshift of \jkcs~based on different photometric techniques and data sets ranged from $z = 1.9 - 2.2$.  However, earlier attempts to confirm the reality of the cluster and to secure its spectroscopic redshift were unsuccessful.

Here we use the WFC3 grisms to show that \jkcs~is a genuine $z = 1.80$ rich cluster, confirmed via the spectroscopic confirmation of 19 member galaxies, of which 15 are quiescent. This is by far the largest number of quiescent cluster members beyond $z\simeq1.5$ with spectroscopic data, making \jkcs~a unique probe of early evolution in a dense environment. Our observations provide an ideal complement recent \emph{HST} field surveys based on similar WFC3 data, such as CANDELS \citep{Grogin11,Koekemoer11} and 3D-HST \citep{Brammer12}.

After describing our observations and methods in Sections 2 and 3, we introduce the cluster members and their basic properties in Section 4. In Section 5, we construct composite spectra of the quiescent cluster members. The stacking technique has been successfully employed in many cluster studies at lower redshifts \citep[e.g.,][]{Dressler04,Gobat08,Poggianti09,Muzzin12} to discern variations in galaxy populations and star formation histories with mass and environment. For the first time in such a distant cluster, the quality of our spectra is sufficient to measure age-sensitive stellar absorption features and derive mean stellar ages as a function of galaxy mass. Additionally, through a comparison with composite spectra assembled by \citet{Whitaker13} based on 3D-HST data, we are able to compare the stellar ages of quenched galaxies in \jkcs~and the field near the same epoch. We demonstrate that although the fraction of quiescent systems in the cluster is elevated, the mean ages of these galaxies do not differ appreciably from the field sample.

To investigate the role of the environment in structural evolution, in Section 6 we compare the shapes, sizes, and radial mass profiles of members of \jkcs~to a large sample of coeval field galaxies drawn from the CANDELS survey. By comparing the distribution of axis ratios, we find some evidence that a lower fraction of quiescent galaxies in the cluster contain a significant disk-like component. We consider the effect that variations in the morphological mixture of quenched galaxies in different environments may have on comparisons of the mass--radius relation, and conclude that there is no significant difference in the sizes of the \jkcs~members compared to the field sample, particularly when these are better matched in morphology. In Section~7 we compare to results derived in other $z > 1.6$ clusters. We discuss the physical significance of our findings in Section~8, and finally summarize them in Section~9.

Throughout we adopt a $\Lambda$CDM cosmology with $\Omega_m = 0.3$, $\Omega_{\Lambda} = 0.7$, and $H_0 = 70$~km~s${}^{-1}$~Mpc${}^{-1}$. All magnitudes are in the AB system, and stellar masses refer to a \citet{Salpeter55} initial mass function (IMF).

\section{\emph{HST} Observations and Data Reduction}

We observed \jkcs~with the infrared channel of WFC3 (GO 12927, Cycle 20, P.I.~Newman) in four visits with a common pointing center but various spacecraft orientations. One two-orbit visit was devoted to imaging in the F160W and F105W filters, and the remaining 14 orbits were divided among 3 visits comprising G141 and G102 grism observations. In addition to our new \HST~data, \jkcs~benefits from an array of earlier ground- and space-based photometry. In this section we describe our reduction of the \HST~observations and construction of a multi-wavelength catalog.

\subsection{HST Imaging}

\jkcs~was imaged through the F160W and F105W filters for approximately 4/3 and 2/3 orbits, respectively, using a four-point dither pattern identical to that adopted by the CANDELS survey \citep{Koekemoer11}. After combining these deeper exposures with the grism pre-images, described below, the total exposure times were 4.5~ks in F160W and 2.7~ks in F105W. Although the calibrated frames produced using \texttt{calwfc3} by the archive on-the-fly pipeline were mostly sufficient, we found it necessary to expand the pixel mask to include additional warm and hot pixels. The exposures were then registered and combined using \texttt{multidrizzle} with a pixel scale of $0\farcs06$.

\subsection{Photometric Catalog}
\label{sec:catalog} 

In addition to the new \emph{HST} imaging, \jkcs~has been observed in the $ugrizJHKs$ filters by the MegaCam and WIRCam instruments at the Canada--France--Hawaii Telescope (CFHT) as part of the CFHT Legacy Survey (Deep Field 1) and the WIRCam Deep Survey \citep{Bielby12}. We also made use of \emph{Spitzer} Infrared Array Camera (IRAC) observations in the $3.6\mu$m and $4.5\mu$m channels taken as part of the \emph{Spitzer} Wide-Area Infrared Extragalactic Survey (SWIRE; P.I.~Lonsdale). 

A multi-wavelength catalog was created using \texttt{SExtractor} \citep{SExtractor} with F160W as the detection band. The procedures followed those detailed in \Ntext. All images were first aligned and drizzled onto the F160W pixel scale. Colors were then measured in apertures on images of matched resolution. To account for systematic uncertainties in zeropoints and PSF matching, we added a 3\% uncertainty (10\% for IRAC) in quadrature to the random flux errors. For a few of the galaxies that we confirm to be members of \jkcs~(IDs 359, 375, 376, and 281; see Section~\ref{sec:specconfmembers}), the aperture photometry was affected by neighboring sources. In order to measure accurate colors in these cases, we used \texttt{Galfit} \citep{Peng02} to fit S\'{e}rsic profiles to all nearby sources simultaneously in each observed band.

Photometric redshifts $z_{\rm phot}$ were computed using the $z_p$ estimator provided by \texttt{EAZY} \citep{Brammer08}. Stellar population parameters were derived using a custom code for the sample of bright galaxies with strong continuum signal in the grism spectra (see Section~\ref{sec:contfit}). For fainter galaxies, we used \texttt{FAST} \citep{Kriek09b} to fit \citet[][BC03]{BC03} models with exponentially declining star formation histories, dust attenuation, and a Salpeter IMF to the photometry; details of the grid can be found in \Ntext. Finally, we use \texttt{InterRest} \citep{Taylor09} to interpolate to rest-frame colors in the \citet{Bessell90} $UBV$ and 2MASS $J$ filters.

\subsection{HST Grism Spectroscopy}
\label{sec:hstgrism}

A total of 14 orbits, split among 3 visits, was devoted to spectroscopy using the G102 and G141 grisms. The spacecraft orientations were spaced by 26~deg and 72~deg from the initial visit to facilitate the deblending of spectral traces. At the beginning of each sequence of grism exposures, a short undispersed exposure through the F160W (for G141) or F105W (for G102) filter was taken to register the grism images, which were then taken following the same dither pattern used for the imaging. The total integration time was 17.0~ks for each grism. In three exposures we noticed a rapidly increasing background in the final few reads; we successfully recovered data with the normal background level by masking the final reads and reprocessing the up-the-ramp readouts using \texttt{calwfc3}. G102 covers the wavelength range $\simeq850-1140$~nm with a dispersion of 2.4~nm per pixel, whereas G141 spans $\simeq1110-1670$~nm at 4.6~nm per pixel. The wide wavelength range provided by the combination of grisms proved essential to locating the Balmer/4000~\AA~continuum break at the redshift of \jkcs. 

The grism data were reduced using the \texttt{aXe} package \citep{Kummel09}. For each object in the catalogs described in Section~\ref{sec:catalog} and for each visit, \texttt{aXe} generates a calibrated two-dimensional (2D) spectrum and an extracted spectrum, along with estimates of the noise and the flux contamination from other objects. A vertical extraction was used, with the wavelength constant perpendicular to the grism trace. Contamination from overlapping spectra was taken into account using the Gaussian emission model, which estimates the spectrum of each object by linearly interpolating the fluxes in the $i$, $z$, F105W, $J$, and F160W filters and distributes the flux spatially according to the Gaussian shape parameters estimated by \texttt{SExtractor}. We found the extracted spectra generated by \texttt{aXe} sufficient for deriving emission line redshifts (Section~\ref{sec:emlines}); however, we made several improvements to the extraction of the brighter sources whose continuum emission we have modeled (Section~\ref{sec:contfit}). 

First, the global background subtraction performed by \texttt{aXe} often left significant residual trends, especially for the G102 grism. We improved upon this by fitting and subtracting a linear trend in wavelength to the background pixels in each 2D spectrum, omitting pixels in the extraction aperture and those with significant contaminating flux from other objects. The 2D spectra were created with larger dimensions than the \texttt{aXe} default in order to ensure they contain a significant number of blank pixels. With this improvement, the G102 and G141 spectra generally join together smoothly.

Second, \texttt{aXe} relies on a Gaussian approximation to the object light profile when it performs optimally-weighted extraction of spectra. While adequate for many objects, this is a poor representation of the extended light profiles of large spheroidal galaxies, which include many of our primary targets. Thus, for each galaxy for which we extract a continuum spectrum, we use the F160W image to measure the light profile in the cross-dispersion direction appropriate to each visit. This profile was then used to extract a one-dimensional (1D) spectrum, including error and contamination estimates, from the 2D spectrum with improved weighting. At the same time we measure the light profile of each galaxy in the dispersion direction. In grism spectroscopy this sets the line spread function (the LSF, i.e., the spectral resolution) and so is essential for the modeling we perform in Section~\ref{sec:contfit}.

\section{Redshift Measurements and Spectral Fitting}
\label{sec:z}

The WFC3 G102 and G141 grisms represent a powerful combination, particularly for faint continuum spectroscopy: they cover a wide wavelength range continuously with uniform sensitivity, reach magnitudes that remain difficult from the ground, and sample all objects in the field of view with no pre-selection of targets. In this section we describe the measurement of 98 redshifts in the field of \jkcs, which form the basis for our identification of the cluster members and the study of their properties in the remainder of the paper. The full catalog of redshift measurements is tabulated in Appendix~A. Our single WFC3 pointing covers the region within 1~arcmin, or 0.51 Mpc, of the X-ray centroid of \jkcs. This is well-matched to the virial radius $R_{500} = 0.52$~Mpc estimated by \citet{Andreon09} based on the X-ray temperature.

The galaxies included in our redshift survey consists of two distinct samples with very different selection properties: an \emph{emission line sample} of galaxies showing one or more spectral lines, and a \emph{continuum sample} of brighter galaxies for which we extract and model the continuum emission. The former is approximately limited by line flux, whereas the latter is limited by broadband flux. In Section~\ref{sec:completeness} we estimate how these selections correspond to physical galaxy properties at the cluster redshift.

\subsection{The Emission Line Sample}
\label{sec:emlines}

\begin{figure*}
\centering
\includegraphics[width=0.49\linewidth]{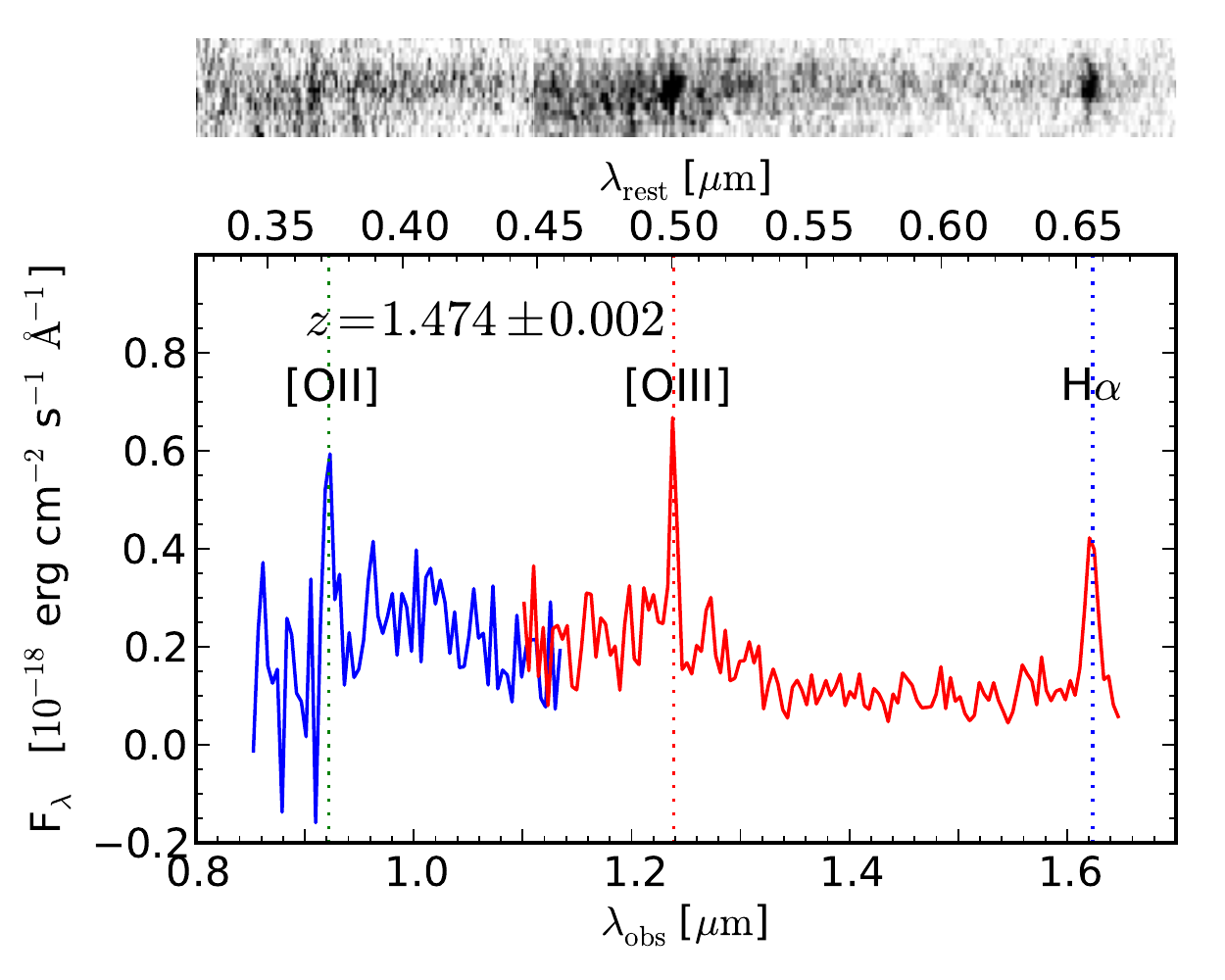}
\includegraphics[width=0.49\linewidth]{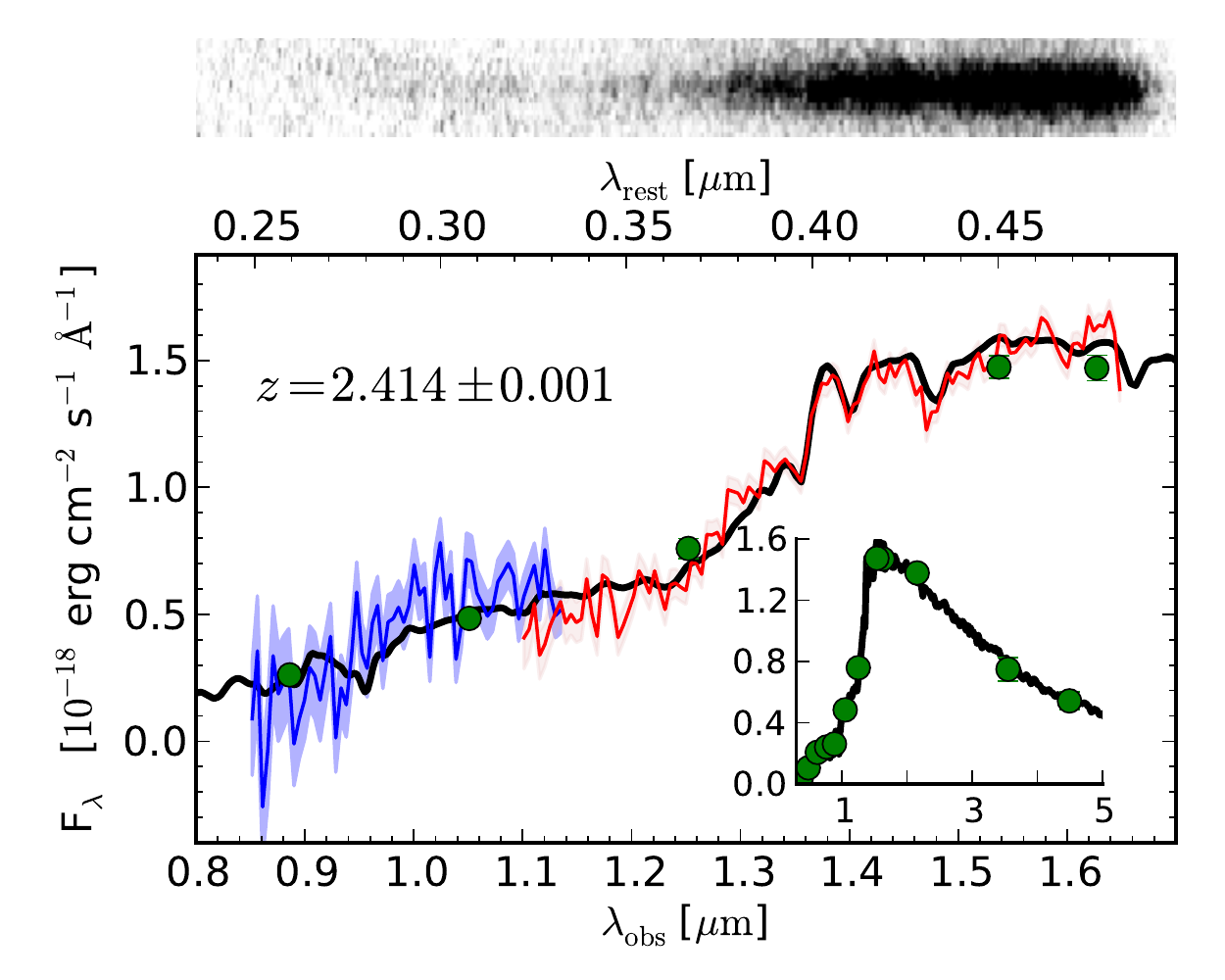}
\caption{{\bf Left:} example spectrum in which three emission lines are identified to yield an unambiguous redshift. {\bf Right:} example of a luminous ($H_{160} = 21.2$) continuum-selected galaxy at $z = 2.414$ showing a prominent continuum break and several absorption lines. Blue and red lines show the coadded G102 and G141 spectra, respectively, binned to 48~\AA~pixels with associated $1\sigma$ errors shaded. The black line shows the best-fitting model (Section~\ref{sec:contfit}), and broadband photometry is shown in green. The inset shows the full set of photometry on an expanded wavelength scale. The upper panels show the 2D spectra, displayed without applying a flux calibration.
\label{fig:zexamples}}
\end{figure*}

We searched for emission lines in the 1D and 2D spectra of all galaxies having $H_{160} < 25.5$ using the plots generated by \texttt{aXe2web}. These include contamination estimates, which are very useful for distinguishing true emission lines from overlapping zero order images of other galaxies. We additionally verified the reality of the emission lines by comparing the three independent spectra obtained for each object at the various orientations. In total we identified 63 emission line sources. An example spectrum is shown in the left panel of Figure~\ref{fig:zexamples}.

Wavelengths of emission lines were measured by fitting Gaussian profiles in \texttt{IRAF}. We averaged the wavelengths that were measured separately in each valid spectrum (i.e., each orientation at which the spectrum fell in the field of view and was not strongly contaminated). In 35 of 63 sources unambiguous redshifts were derived through the identification of multiple lines, primarily H$\alpha$, [\ion{O}{2}], and [\ion{O}{3}]. When only a single line was identified (28 sources), it was interpreted as H$\alpha$ (22 sources) or [\ion{O}{3}] (6 sources) depending on which was more consistent with the photometric redshift.

The rms redshift uncertainty was estimated internally from the scatter in independent measurements as $\sigma_z = 0.003$. For nine galaxies we can compare with redshifts measured at higher spectral resolution in the VIMOS VLT Deep Survey \citep[VVDS;][]{LeFevre13}. After excluding one outlier with $\Delta z = 0.07$, the rms scatter is $\sigma_{\Delta z} = 0.005$ with no detectable systematic bias. This is 20 times smaller than the median uncertainty in the photometric redshifts of the these galaxies.

We estimate a typical $5\sigma$ line flux limit of $5 \times 10^{-17}$ erg cm${}^{-2}$ s${}^{-1}$ in G141 data over $\lambda \approx 1.2 - 1.6 \mu{\rm m}$ and in G102 over $\lambda \approx 0.9 - 1.1 \mu{\rm m}$. By simulating artificial emission lines in the extracted spectra, we verified that we would visually identify $\sim 80\%$ of lines exceeding this flux limit. This limit applies to the spectra from each visit, which are the basis of our line search. These have 2--3 orbit depth, which is comparable to the 3D-HST \citep{Brammer12} and WISP \citep{Atek10} surveys, and these programs have estimated similar limits.

\subsection{The Continuum Sample}
\label{sec:contfit}

For all galaxies in the \emph{HST} field of view brighter than $H_{160} < 23.3$ with photometric redshifts $1.4 < z_{\rm phot} < 3$, we model the continuum emission in order to derive precise redshifts and stellar population properties. This flux limit corresponds to a typical signal-to-noise ratio of 5 per spectral pixel in the coadded spectra, suitable for continuum fitting, while the redshift range restricts the sample to galaxies for which the Balmer/4000 \AA~break is expected to fall well within the grism spectral coverage.

For each galaxy, we visually examined the spectra obtained during each of the three visits extracted using the improved weighting described in Section~\ref{sec:hstgrism}. The contamination model was subtracted from each spectrum. Heavily contaminated wavelength regions, often comprising an entire visit, were identified and masked. The spectra were then coadded using inverse variance weighting to produce a combined spectrum for each grism. The galaxy light profiles, measured for each visit along the dispersion direction (Section~\ref{sec:hstgrism}), were averaged with the same weights to estimate the LSF. The exposure times of the spectra vary significantly, since the number of visits that contribute to the stack ranges from 1 to 3. Of the 59 galaxies in the continuum sample, we were able to extract G102 and G141 spectra for 40 objects (68\%). The remaining 19 sources were either heavily contaminated by neighboring sources or dispersed off the detector.

To make optimal use of the extensive data we gathered for \jkcs, we developed a code designed to fit stellar population  models jointly to spectroscopic and photometric data with flexible models and arbitrary LSFs. \texttt{pyspecfit} is written in Python. It is Bayesian in nature and uses \texttt{MultiNest} \citep{Feroz09}, a Markov Chain Monte Carlo (MCMC) engine, to explore the parameter space and properly estimate uncertainties and degeneracies.  The details of the code are described in Appendix~B. An example fit is shown in the right panel of Figure~\ref{fig:zexamples} for a luminous red galaxy at $z = 2.414$. While our fits are based on the BC03 models, we note in passing that we experimented with using the 2007 models instead, but decided against this due to their uniformly poorer fits to the spectrophotometry. The poorer fits arise from excess light in the rest-frame near-infrared, which is consistent with other studies indicating that the contribution of the TP-AGB stars in these more recent models is overstated \citep[e.g.,][]{Kriek10,Zibetti13}.

A potential source of error in deriving redshifts from the continuum shape arises from joining spectra from the two grisms. This is of particular concern in the present sample since, as we show in Section~\ref{sec:specconf}, the 4000~\AA~break at the redshift of \jkcs~falls near the division between the grisms. We tested for errors arising from this possible confusion by reanalyzing the spectra of the 17 continuum-selected cluster members (detailed in Section~\ref{sec:specconf} below) after explicitly forcing the G102 and G141 flux levels to agree, on average, in the small wavelength range where they overlap. This process may introduce some additional noise, but it eliminates the possibility of a spurious spectral break. We found that only 2 of 17 redshifts shift by a significant amount ($> 2 \sigma$).\footnote{These are IDs 286, where there is some confusion about the location of the break (see the $P(z)$ distribution in Figure~\ref{fig:memberspectra}), and ID 375, which is likely a satellite of a nearby, luminous red cluster members whose spectrum is difficult to clearly separate.} Both galaxies are on the red sequence and are very likely cluster members.

Only five galaxies in the continuum sample show strong emission features; in these cases, we adopt the emission redshift. We slightly increased the noise estimates for the spectral data by 20\% to obtain a median $\chi^2_{\rm spec} / n_{\rm pixels} = 1.0$, while for the photometry we find a median $\chi^2_{\rm phot} / n_{\rm filters} = 1.1$. This indicates that the models provide good fits and that the noise estimates are reasonable. The median random uncertainty in $z_{\rm grism}$ is $\sigma_{z_{\rm grism}}/(1+z) = 0.0025$ for the continuum-selected galaxies, which is a factor 15 improvement over their photometric redshift errors.

\section{Spectroscopic Confirmation of \jkcs~and Identification of Member Galaxies}
\label{sec:specconf}

In this section we use our grism redshift survey to identify \jkcs~spectroscopically. Thanks to the excellent precision of the grism redshifts, which are typically $\sim 15-20$ times more precise than the photometric estimates, we will show that \jkcs~stands out as a strong overdensity of massive galaxies at $z = 1.80$ which are spatially coincident with diffuse X-ray emission, thus supporting the identification of \jkcs~as a galaxy cluster with a hot intracluster medium (ICM). We then isolate a sample of spectroscopically confirmed member galaxies and discuss its likely completeness, before turning to the color distribution and star formation activity of these cluster members.

\begin{figure}
\includegraphics[width=\linewidth]{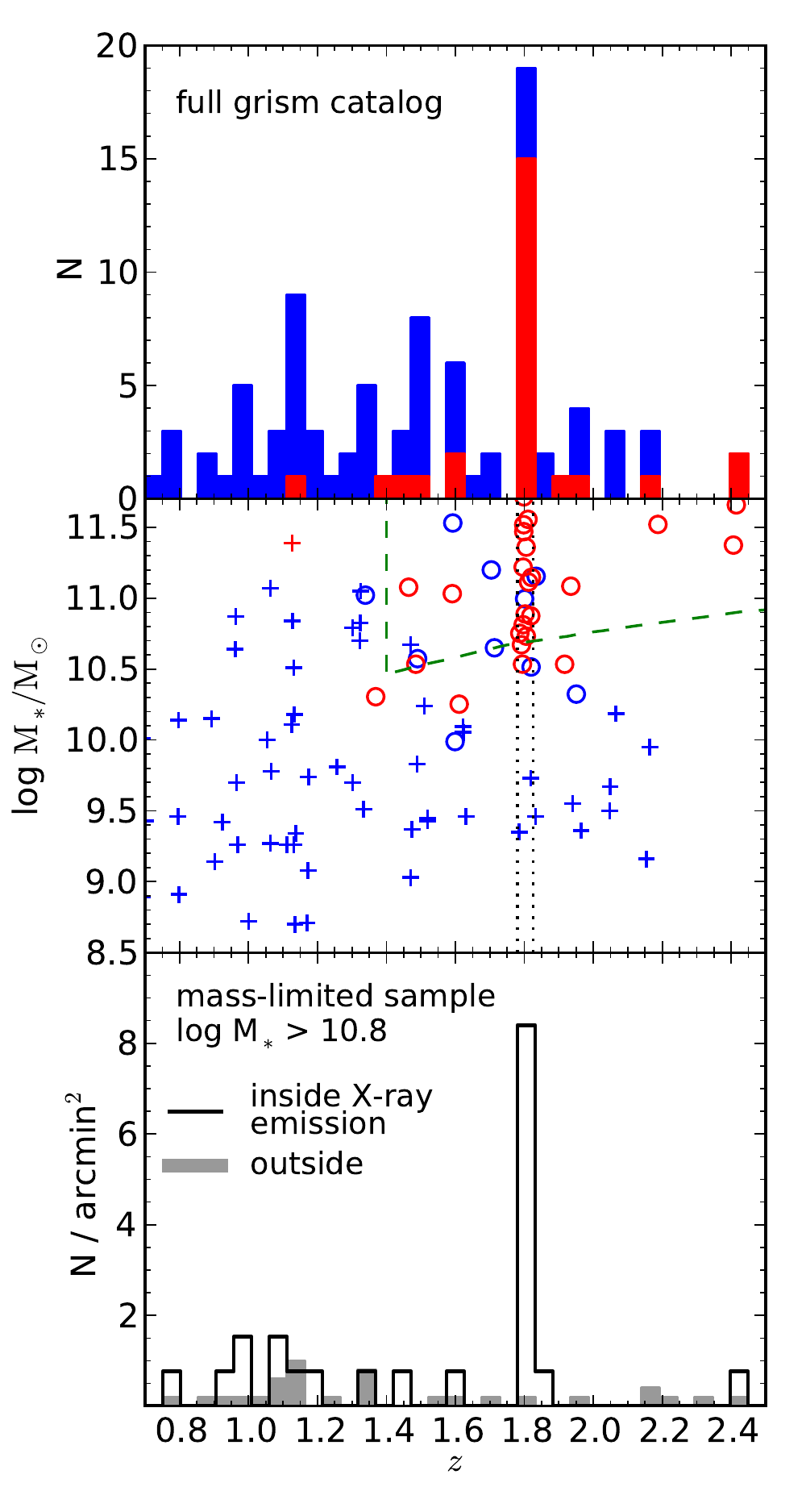}
\caption{{\bf Top:} distribution of grism redshifts at $z_{\rm grism} > 0.7$ derived from emission lines and continuum fitting. Red and blue colors refer to the $UVJ$-based quiescent and star-forming classifications, respectively (see Figure~\ref{fig:uvj}). {\bf Middle:} stellar mass and redshift distribution for the same galaxies as in the top panel. Circles and crosses denote continuum and emission line redshifts, respectively. Vertical lines encompass the 19 identified cluster members. The green dashed line approximates the mass completeness of the continuum sample ($z_{\rm phot} > 1.4$, $H_{160} < 23.3$) for a solar metallicity galaxy formed in a burst at $z_f = 5$. {\bf Bottom:} redshift distribution of a mass-limited sample of galaxies found within the WFC3 field of view, divided into those located inside and outside of the outermost contour of detected X-ray emission (Figure~\ref{fig:clusterimage}). The histograms are normalized by the area of these regions. A spectroscopic redshift is available for 83\% of sources from one of the sources described in the text; for the remainder we rely on $z_{\rm phot}$. \jkcs~is the clear excess evident at $z=1.8$.\label{fig:zdist}}
\end{figure}

\subsection{Spectroscopic Identification of \jkcs~and Alignment with X-ray Emission}

The redshift distribution of the emission line and continuum-selected samples in our grism survey is shown in the top panel of Figure~\ref{fig:zdist}. \jkcs~is the richest structure in the field, comprising 19 galaxies, and is located at $z = 1.80$. The prominence of this peak is more remarkable when one considers that many of the members are red and massive systems with $M_* > 10^{11} \msol$. Since the uncertainties in the grism redshifts are $\sigma_z \lesssim 0.01$, this shows that the $6.5\sigma$ overdensity of red galaxies discovered by \citet{Andreon09} identified a dominant structure and not a blend of several poorer ones.

\begin{figure*}
\centering
\includegraphics[width=0.9\linewidth]{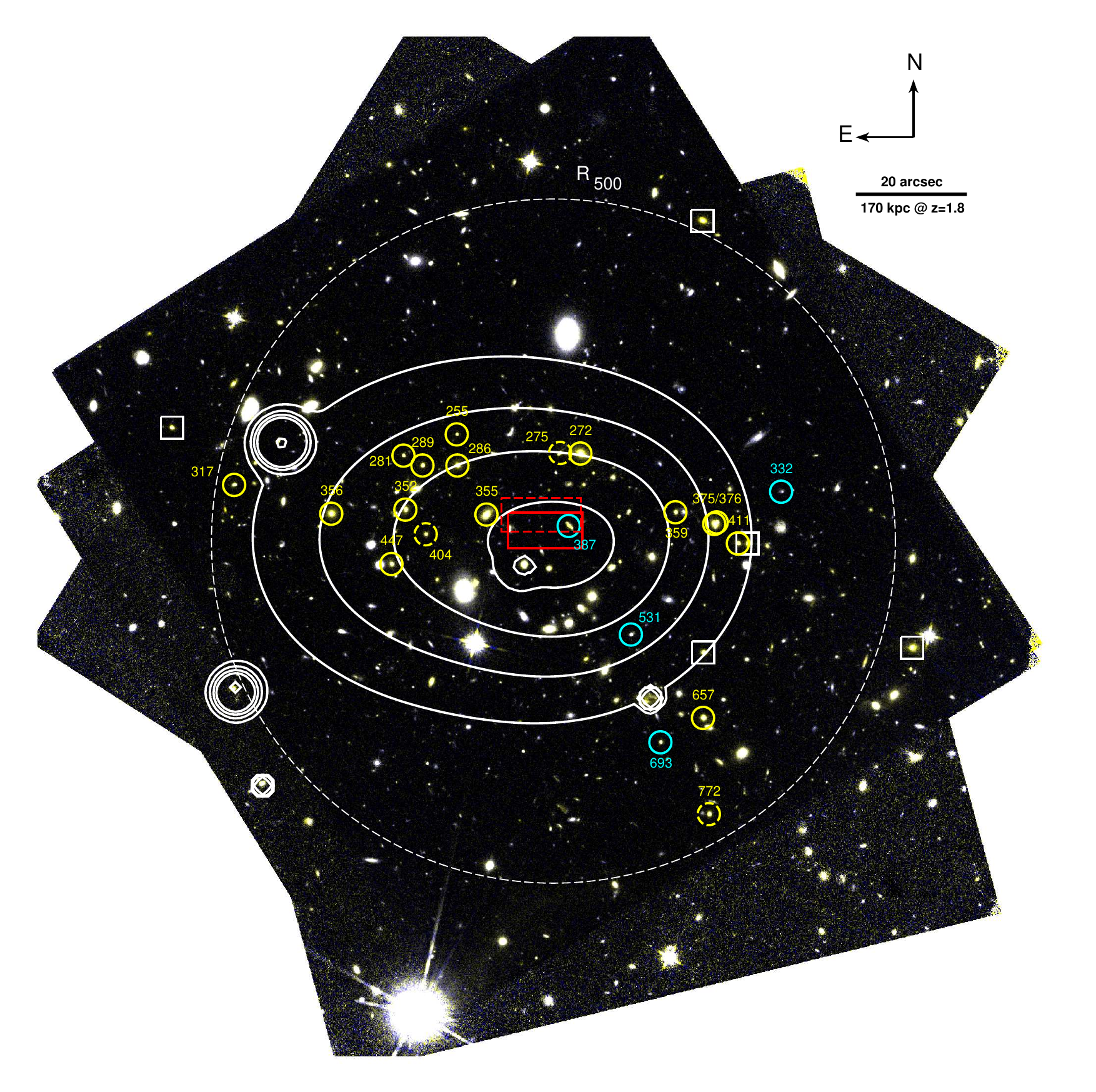}
\caption{\emph{HST}/WFC3 image of \jkcs~in the F160W and F105W filters. Confirmed cluster members are indicated by yellow (quiescent galaxies) and light blue (star-forming) circles.  The smoothed \emph{Chandra} X-ray emission \citep{Andreon09} is overlaid as contours. The centroid of the spectroscopically confirmed members and its $1\sigma$ uncertainty is shown by the red rectangle, which is well-aligned with the X-ray centroid. Similarly, the dashed rectangle shows the mass-weighted centroid of the quiescent members, including the three likely members listed in Table~1 whose positions are indicated by dashed circles (contaminated spectra preclude a spectroscopic determination for these galaxies). White squares show spectroscopically confirmed non-members that are on the cluster red sequence (Section~\ref{sec:completeness}). \label{fig:clusterimage}}
\end{figure*}

Figure~\ref{fig:clusterimage} shows the that the distribution of galaxies in the $z=1.80$ cluster is clearly centered upon the diffuse X-ray emission \citep{Andreon09}. Similar to some other high-redshift clusters \citep[e.g.,][]{Zeimann12}, \jkcs~does not have a single dominant galaxy located at the cluster center, which presumably reflects a lack of dynamical relaxation compared to lower-redshift systems. Nonetheless, the centroid of the spectroscopic cluster members is R.A.~=~02:26:44.0 $\pm$ 6 arcsec, Decl.~=~--04:41:36 $\pm$ 4 arcsec (red box in Figure~\ref{fig:clusterimage}), which coincides with the X-ray centroid determined by Andreon et al.~to within $1\sigma$. The cluster members are not distributed uniformly over the field; instead, all lie within $R_{500}$ of the X-ray center, and the majority are confined to a much smaller, elongated structure overlapping the X-ray emission. By considering a larger sample of red sequence candidate members extending to fainter magnitudes than our spectroscopic sample, \citet{Andreon13} show that the red sequence galaxies follow a smoothly declining radial profile with parameters resembling those of lower-redshift clusters.

\jkcs~is therefore a natural identification for the source of the X-ray emission. Based on our grism data, we can now assemble a highly complete redshift catalog in the zone of the X-ray emission and verify that \jkcs~is indeed the most likely origin. A stellar mass-selected sample is ideal for thus purpose, since it allows us to compare similar galaxy populations uniformly at different redshifts, and massive galaxies are better tracers of a deep gravitational potential. The green line in Figure~\ref{fig:zdist} (middle panel) shows the limiting stellar mass for our continuum flux-selected sample, estimated as described in the caption, and demonstrates that this sample is fairly complete at masses $M_* > 10^{10.8} \msol$ and redshifts $z = 1.4 - 2$. At lower redshifts, since the 4000~\AA~break lies outside our spectral coverage, we combine our grism catalog with redshifts from the VVDS \citep{LeFevre13} and the Carnegie--\emph{Spitzer}--IMACS Survey \citep{Kelson14}. This yields a spectroscopic redshift for 83\% of the mass-limited sample; for the remainder we use $z_{\rm phot}$. To assess the association of galaxies with the X-ray emission, we consider systems that are located within the outermost contour of the X-ray emission shown in Figure~\ref{fig:clusterimage}. The bottom panel of Figure~\ref{fig:zdist} shows their redshift distribution and clearly demonstrates that the $z = 1.80$ peak is dominant and concentrated within the X-ray emission.

\begin{figure}
\centering
\includegraphics[width=\linewidth]{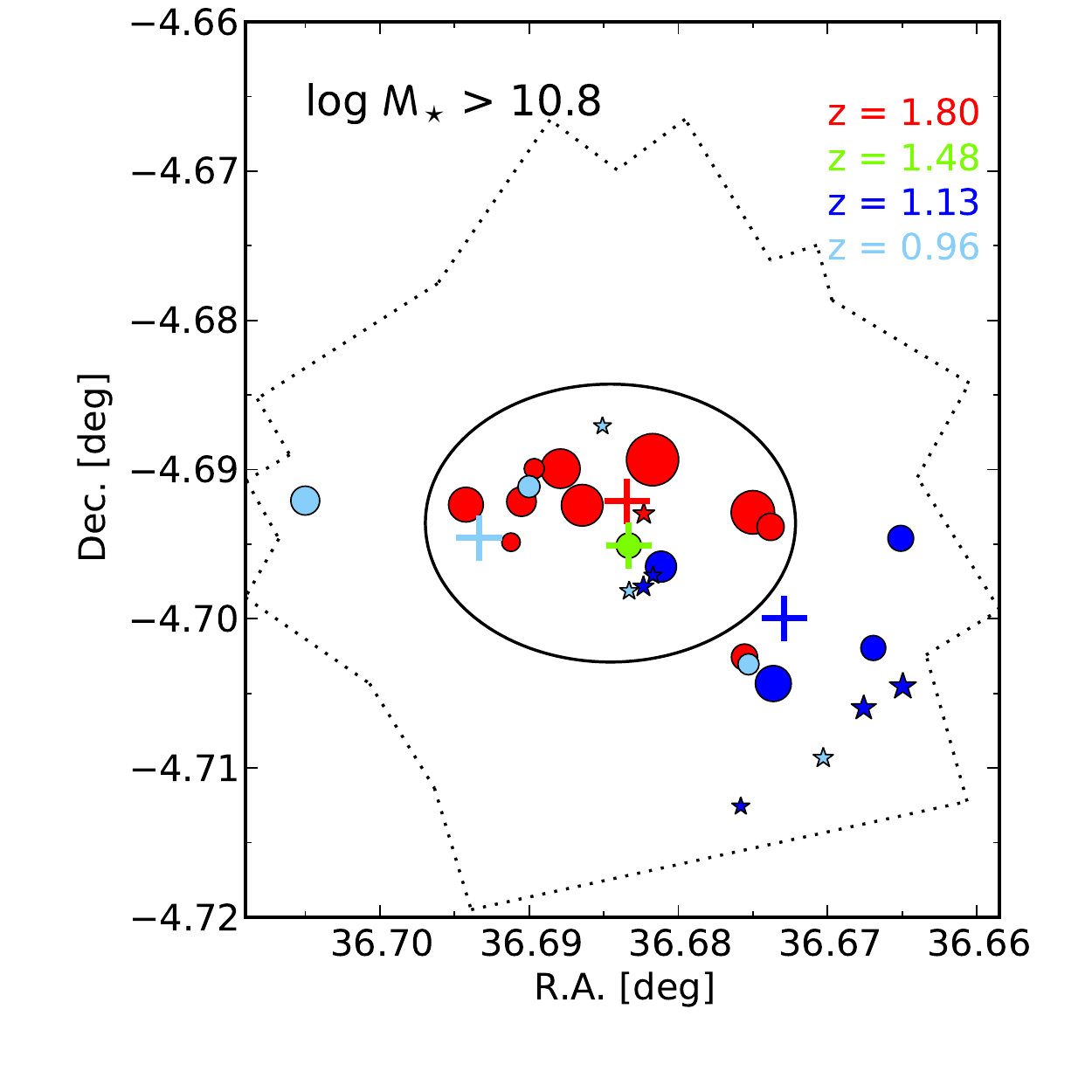}
\caption{Peaks of the redshift distribution in the field of \jkcs. The positions of galaxies with $M_* > 10^{10.8} \msol$ in four redshift peaks, as described in the text, are plotted with symbol area proportional to stellar mass. Stars and circles distinguish star-forming and quiescent galaxies, respectively, as classified by their $UVJ$ colors. The outer isophote of the X-ray emission (Figure~\ref{fig:clusterimage}) is approximated by the ellipse, and the dotted region outlines the field of the \emph{HST} imaging. The centroids of the quiescent galaxies in each peak, weighted by stellar mass, are indicated by crosses.\label{fig:groups}}
\end{figure}

Secondary peaks of the redshift distribution in the field of \jkcs~are expected and are seen in front of other high-redshift clusters \citep[e.g.,][]{Zeimann12,Mantz14}. The next strongest peaks are located at $z=0.96, 1.13$, and 1.48; each contains 1 or 2 massive galaxies within the zone of X-ray emission, compared to 11 in \jkcs~(Figure~\ref{fig:zdist}, lower panel). Figure~\ref{fig:groups} shows the positions of galaxies in these foreground structures, with crosses marking their centroids.\footnote{For galaxies with spectroscopic redshifts, we plot those within $|z_{\rm spec} - z| < 0.03$ in Figure~\ref{fig:groups}, whereas for those with only photometric redshifts, we allow $|z_{\rm phot} - z| < 0.08$.} In addition to being more sparsely populated, the $z = 0.96$ and 1.13 structures are not concentrated within the X-ray emission: the $z=0.96$ structure is very diffuse, and most galaxies in the $z=1.13$ peak lie outside of the X-ray--emitting region. The sparse $z = 1.48$ structure is better aligned with the X-ray emission than the other foreground peaks, but it seems far too poor to contribute a significant fraction of the flux. Only a single galaxy is massive enough to be included in Figure~\ref{fig:groups}. For comparison, the $z=1.62$ group discovered by \citet{Tanaka13b} in ultra-deep \emph{Chandra} data appears to be richer, yet it exhibits diffuse X-ray flux that is still $\sim 15\times$ fainter than that observed around \jkcs. 

While \citet{Bielby10} considered these foreground structures as possible sources of the X-ray emission, they were unable to locate the dominant $z = 1.80$ cluster in a ground-based optical redshift survey. With the less biased selection and dense sampling afforded by the WFC3 grisms, we have shown that \jkcs~is the most likely origin and is a genuine high-redshift cluster: it exhibits a spectroscopically confirmed population of massive, red galaxies that are concentrated within diffuse X-ray emission, and the observed X-ray properties are fairly consistent with expectations for a cluster with the observed richness of \jkcs~\citep{Andreon13}. After making a small correction to the luminosity distance, the bolometric X-ray luminosity estimated by \citet{Andreon09} is $L_X = (6.5 \pm 1.5) \times 10^{44}$~erg~s${}^{-1}$ within $R_{500}$.

\subsection{Spectroscopically Confirmed Cluster Members}
\label{sec:specconfmembers}

With the redshift of \jkcs~established, we now construct a sample of spectroscopically confirmed member galaxies that will form the basis of the remainder of the paper. The identification of cluster members is relatively unambiguous due to the high precision of the grism redshifts. We selected as cluster members those galaxies for which $>50\%$ of the integrated probability density $P(z)$ is located within $z_{\rm clus} \pm 3 \sigma_z$. Here $P(z)$ is derived from the MCMC chains for the continuum sample and is approximated as a Gaussian for the emission line sample (Section~\ref{sec:emlines}). We estimate the cluster velocity dispersion $\sigma_v = c \sigma_z / (1+z) = 800$~km~s${}^{-1}$ based on the X-ray luminosity presented by \citet{Andreon09} and the scaling relation derived by \citet{Zhang10} for nearby clusters, which is consistent with the $z \sim 1$ relation determined by \citet{Andreon08}. We began with an initial estimate of $z_{\rm clus}$ and iterate by updating $z_{\rm clus}$ with the mean redshifts of the selected members.

This procedure converged in only one iteration to yield 19 members with a mean redshift of $z_{\rm clus} = 1.803 \pm 0.003$. The selected members are precisely those in the interval $z_{\rm grism} = 1.803 \pm 0.022$, which is indicated by the vertical lines in the lower panel of Figure~\ref{fig:zdist}. We note that adopting the velocity window of $\pm 2000 (1+z_{\rm clus})$~km~s${}^{-1}$ advocated by \citet{Eisenhardt08} would remove only one galaxy from this sample. Among the several previously published estimates of the redshift of \jkcs, the {\tt EAZY} photometric redshifts with no corrections applied gave the true $z_{\rm clus}$ \citep{Raichoor12b}. Spectra, images, and $P(z)$ distributions for the 19 confirmed members are shown in Figure~\ref{fig:memberspectra}, and their coordinates and photometric properties are listed in Table~\ref{tab:memberdata}. 

\begin{figure*}
\includegraphics[width=0.49\linewidth]{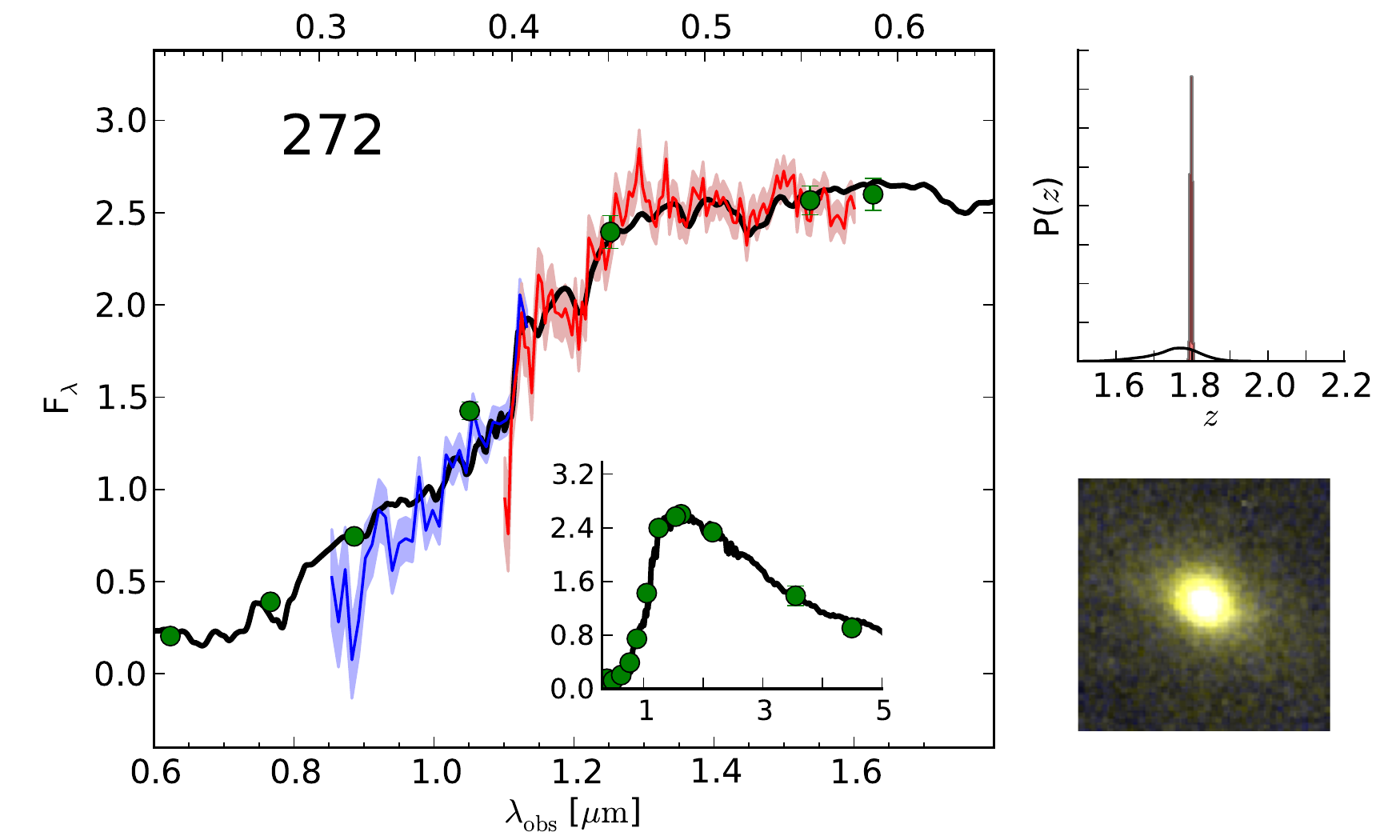} 
\includegraphics[width=0.49\linewidth]{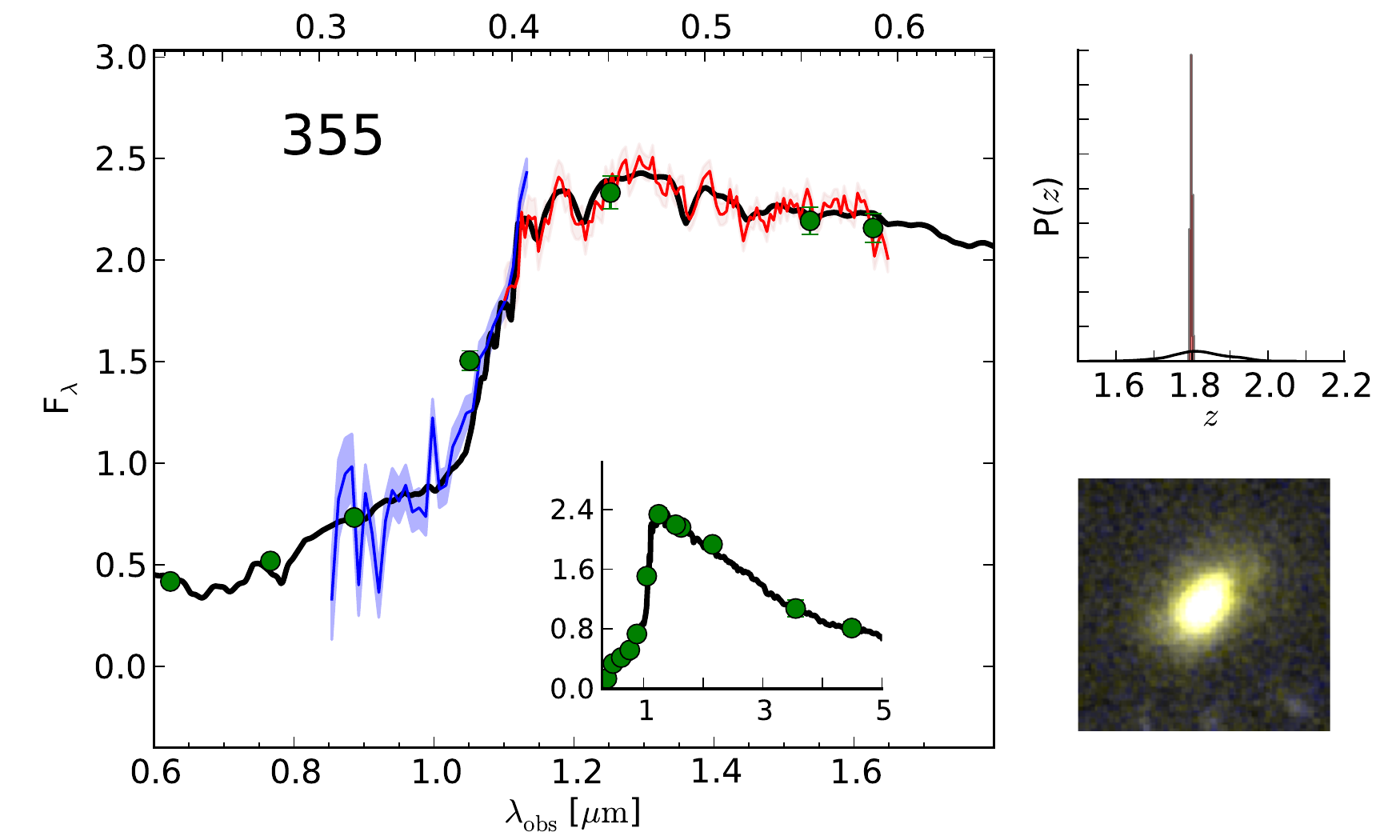} \\
\includegraphics[width=0.49\linewidth]{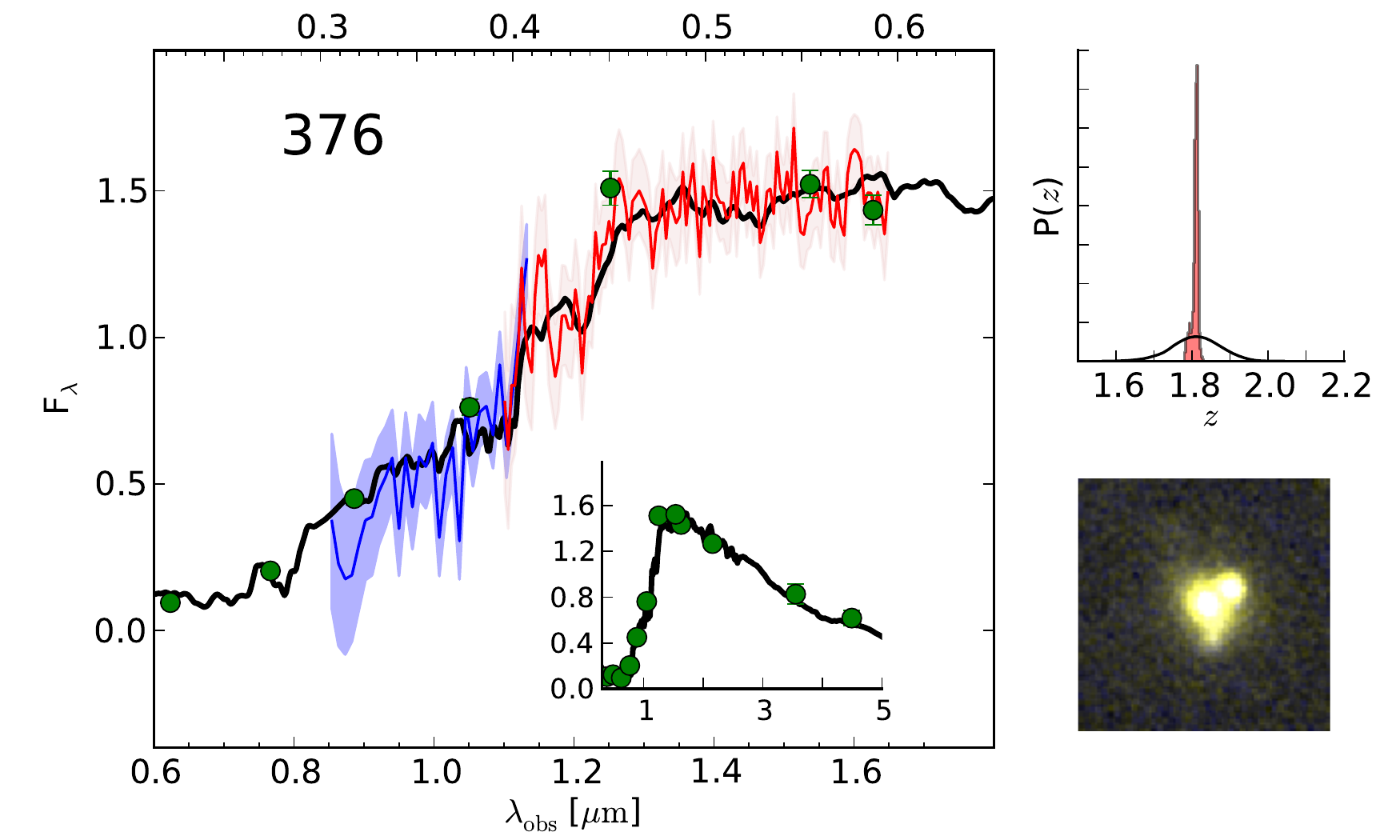} 
\includegraphics[width=0.49\linewidth]{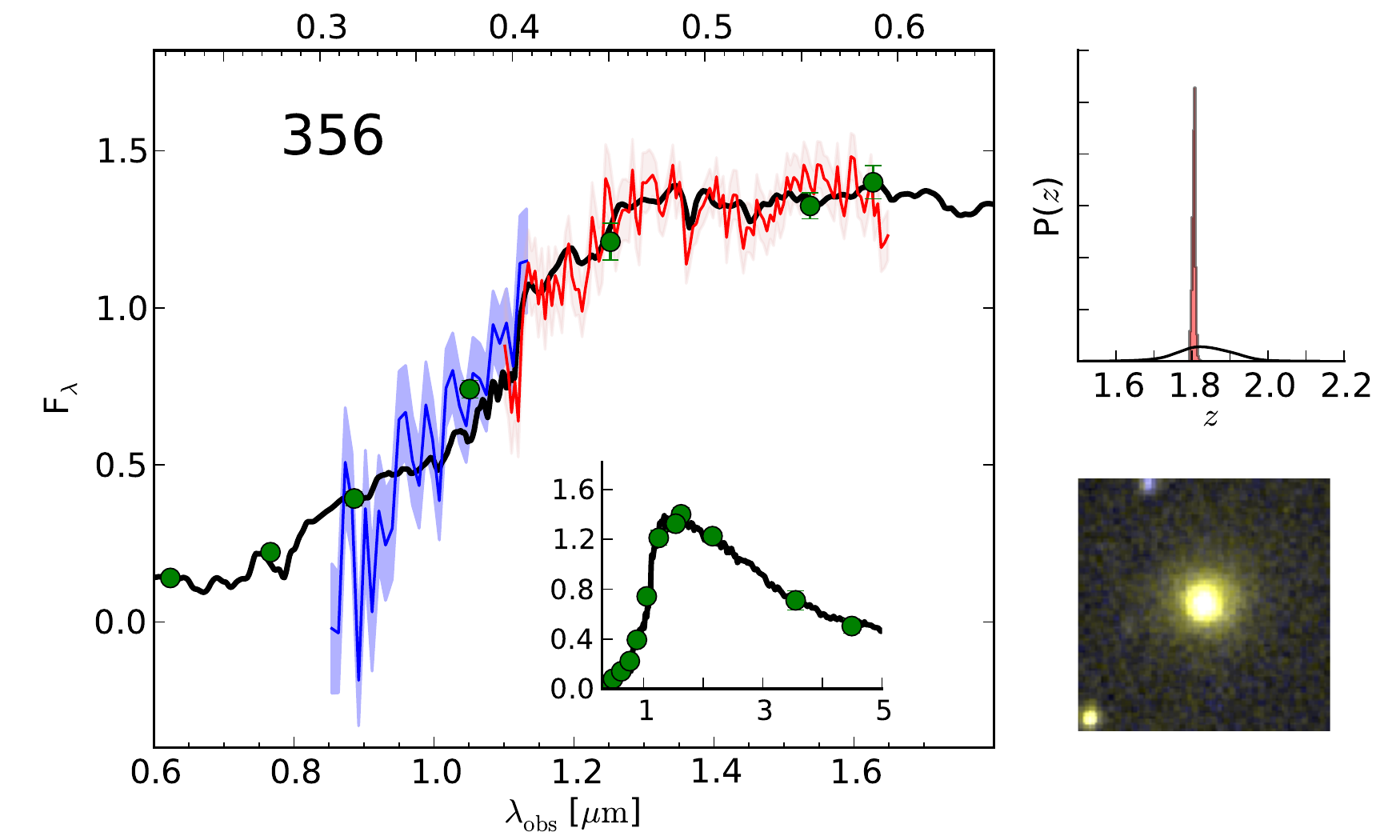} \\
\includegraphics[width=0.49\linewidth]{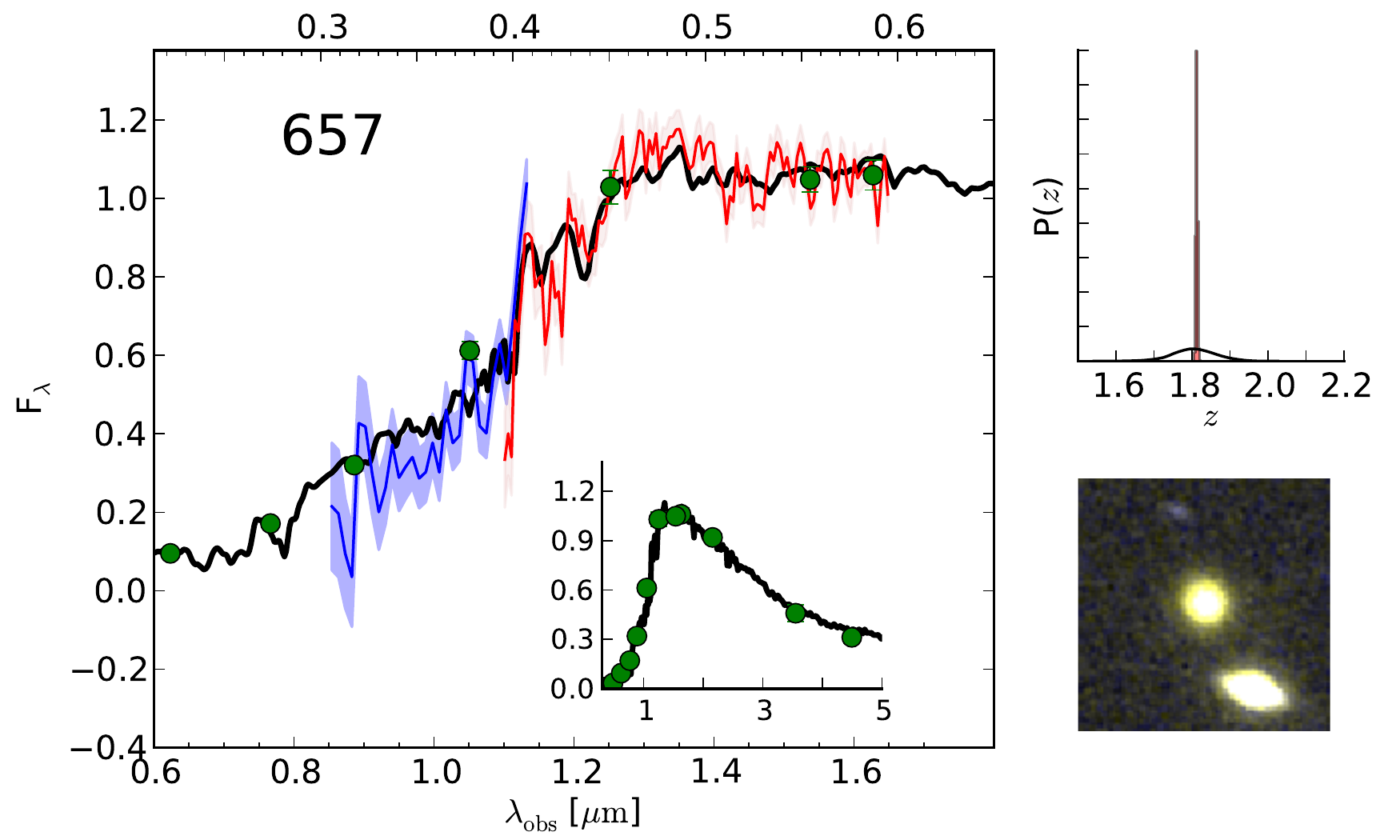} 
\includegraphics[width=0.49\linewidth]{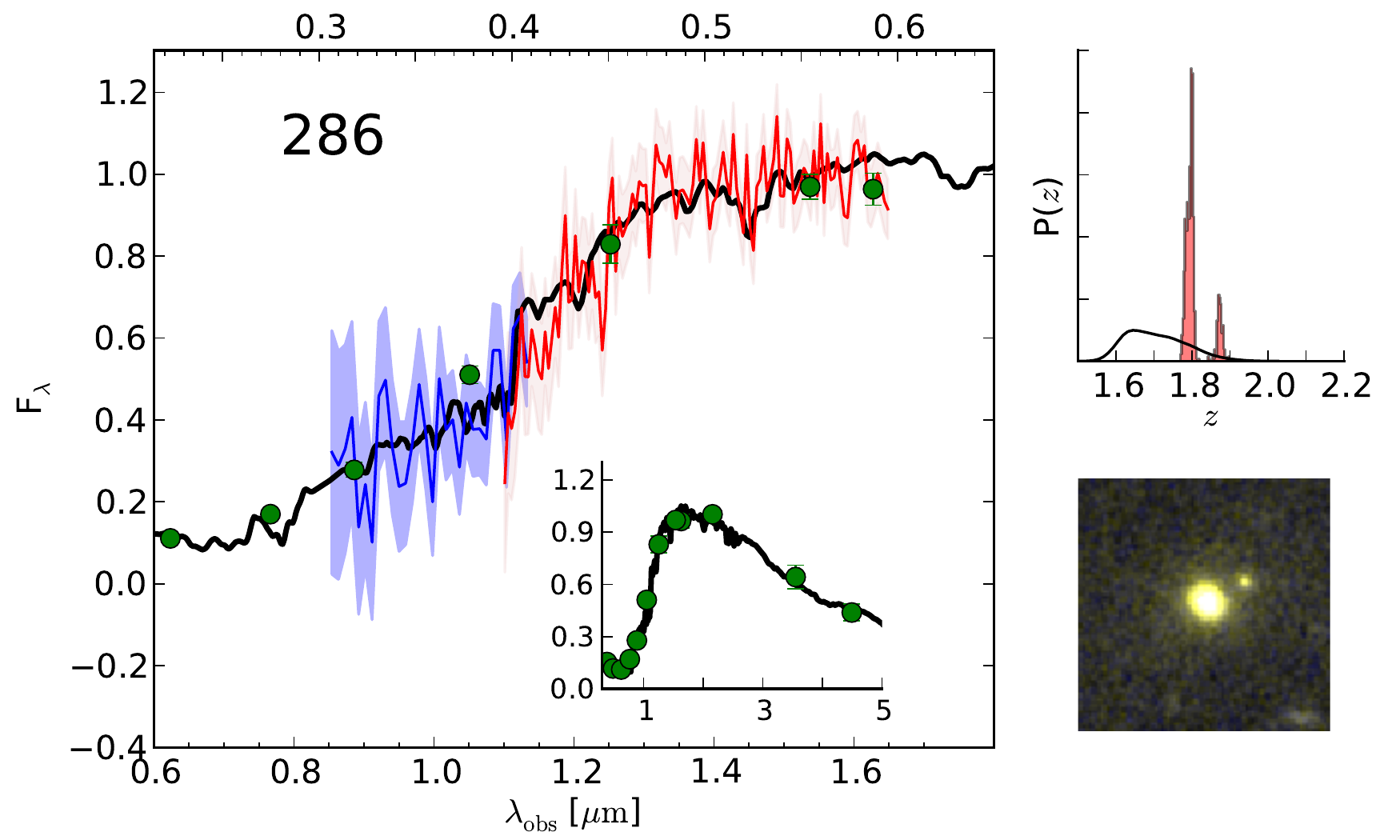} \\
\includegraphics[width=0.49\linewidth]{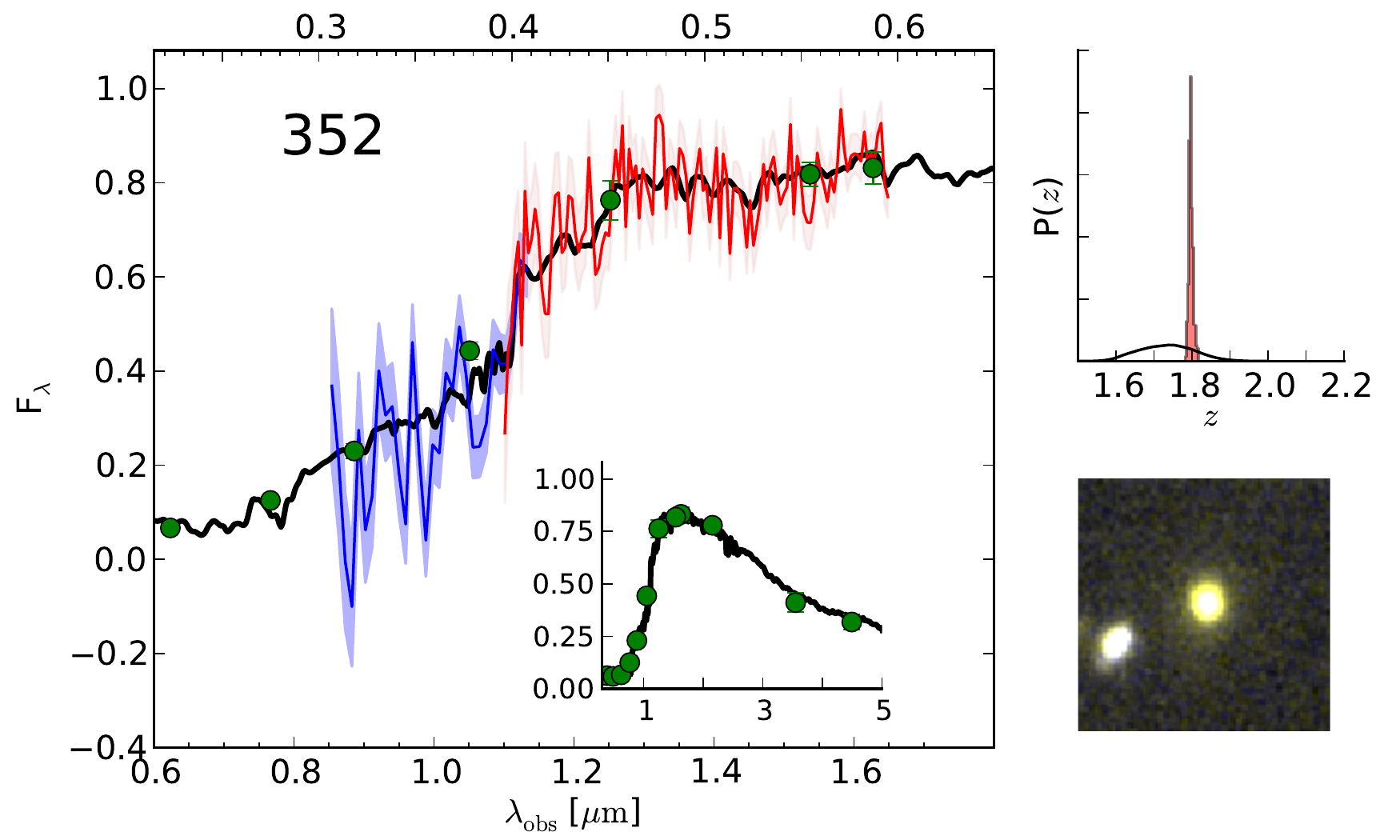} 
\includegraphics[width=0.49\linewidth]{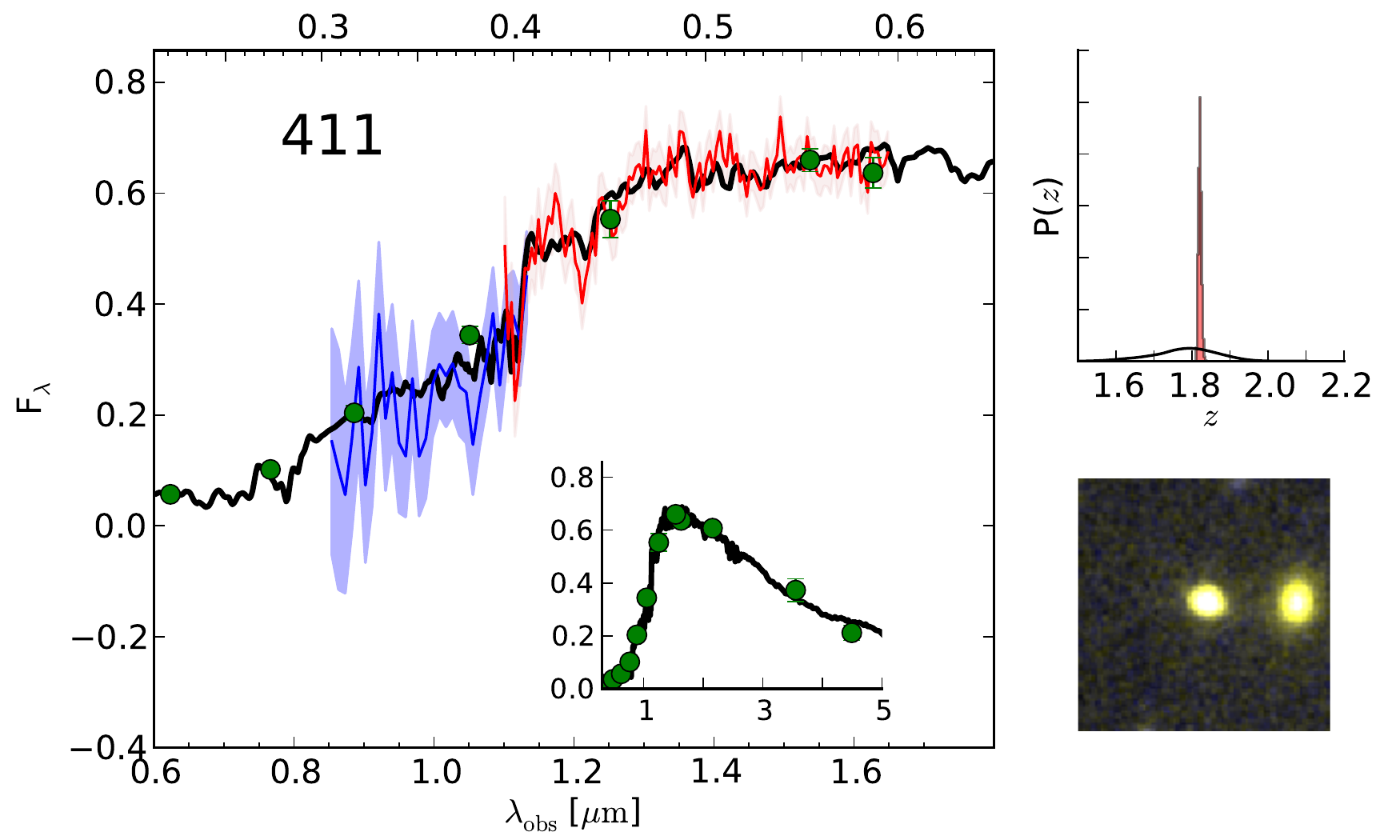}
\caption{Spectroscopically confirmed cluster members. For each object, the main panel shows the grism spectra (blue is G102, red is G141, $1\sigma$ errors are shaded) binned to 48~\AA~(red) and 96~\AA~(blue) pixels for display purposes. Photometry (green circles) and the best-fitting model (black) are overlaid. The top and bottom axes shows the rest- and observed-frame wavelength in nm, and the units of $F_{\lambda}$ are $10^{-18}$ erg cm${}^{-2}$ s${}^{-1}$ \AA${}^{-1}$. The inset shows the complete photometry on an expanded scale in the same units. Cutouts show F105W/F160W images, displayed on a linear scale, with a side length of $5\arcsec$. The $P(z)$ subpanels show the redshift probability density derived from the broadband photometry only using \texttt{EAZY} (black curves) and from our joint fits to the spectra and photometry (filled histograms). Galaxies are ordered by decreasing F160W flux. For the two galaxies in the emission line sample (IDs 332 and 531) no continuum fit is plotted.\label{fig:memberspectra}}
\end{figure*}

\addtocounter{figure}{-1}
\begin{figure*}
\includegraphics[width=0.49\linewidth]{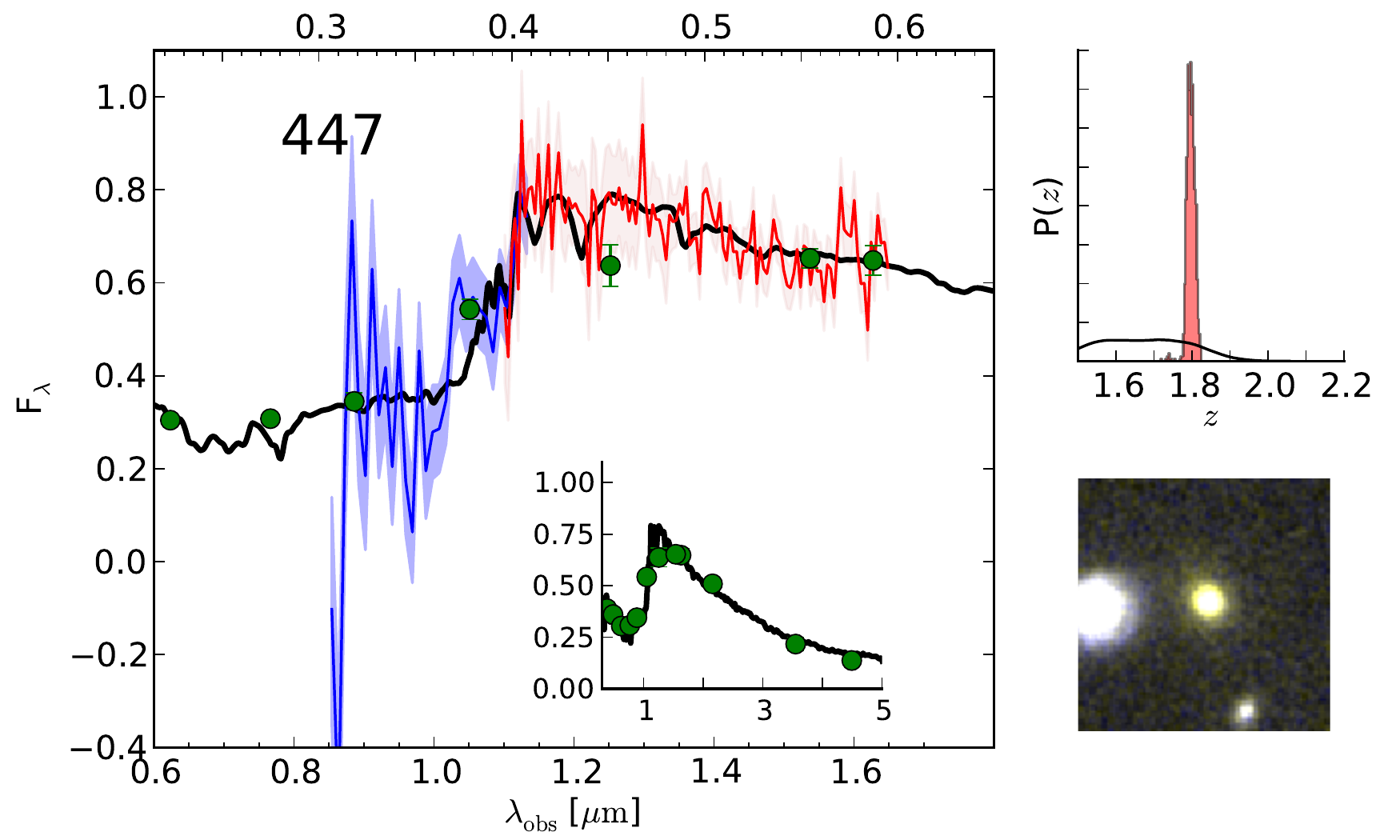} 
\includegraphics[width=0.49\linewidth]{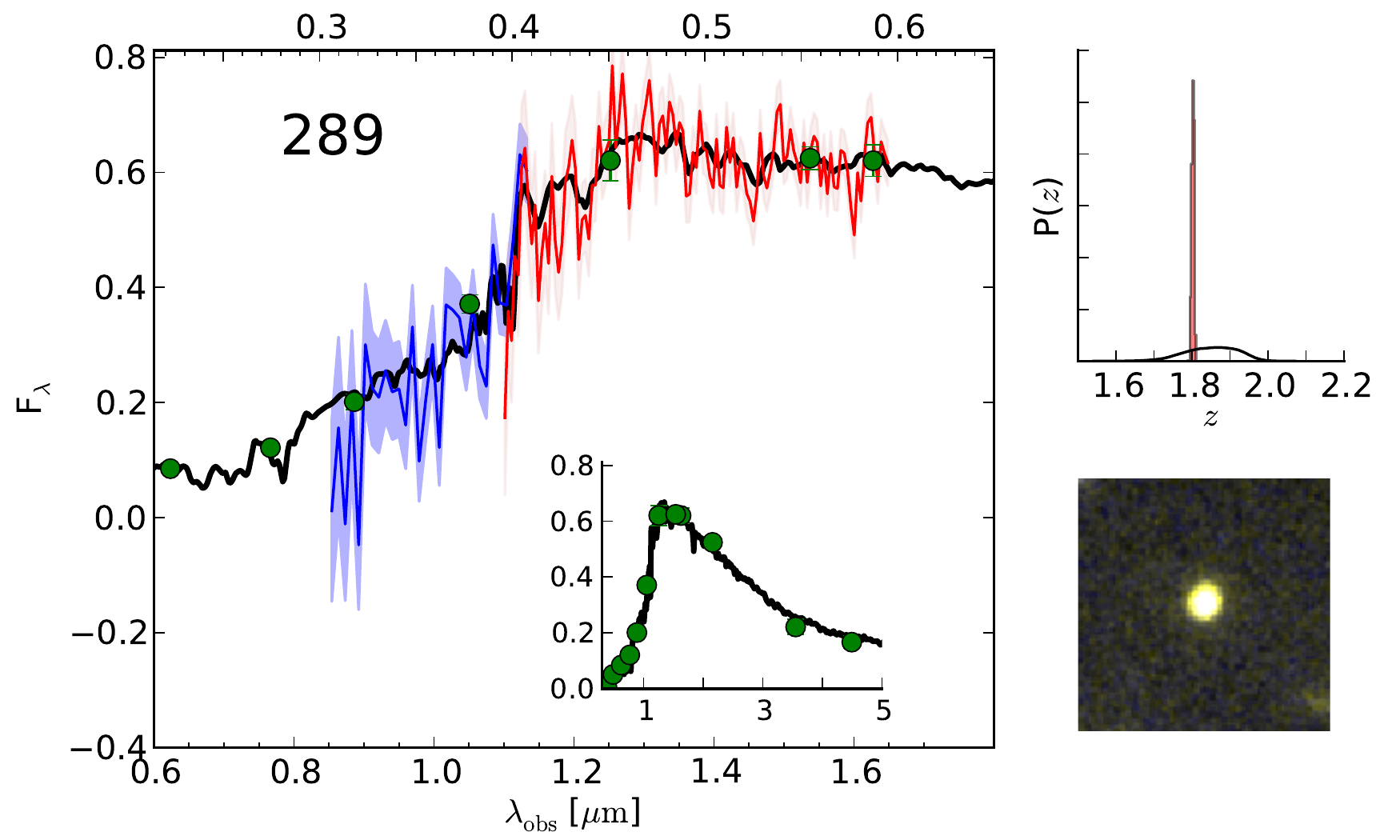} \\
\includegraphics[width=0.49\linewidth]{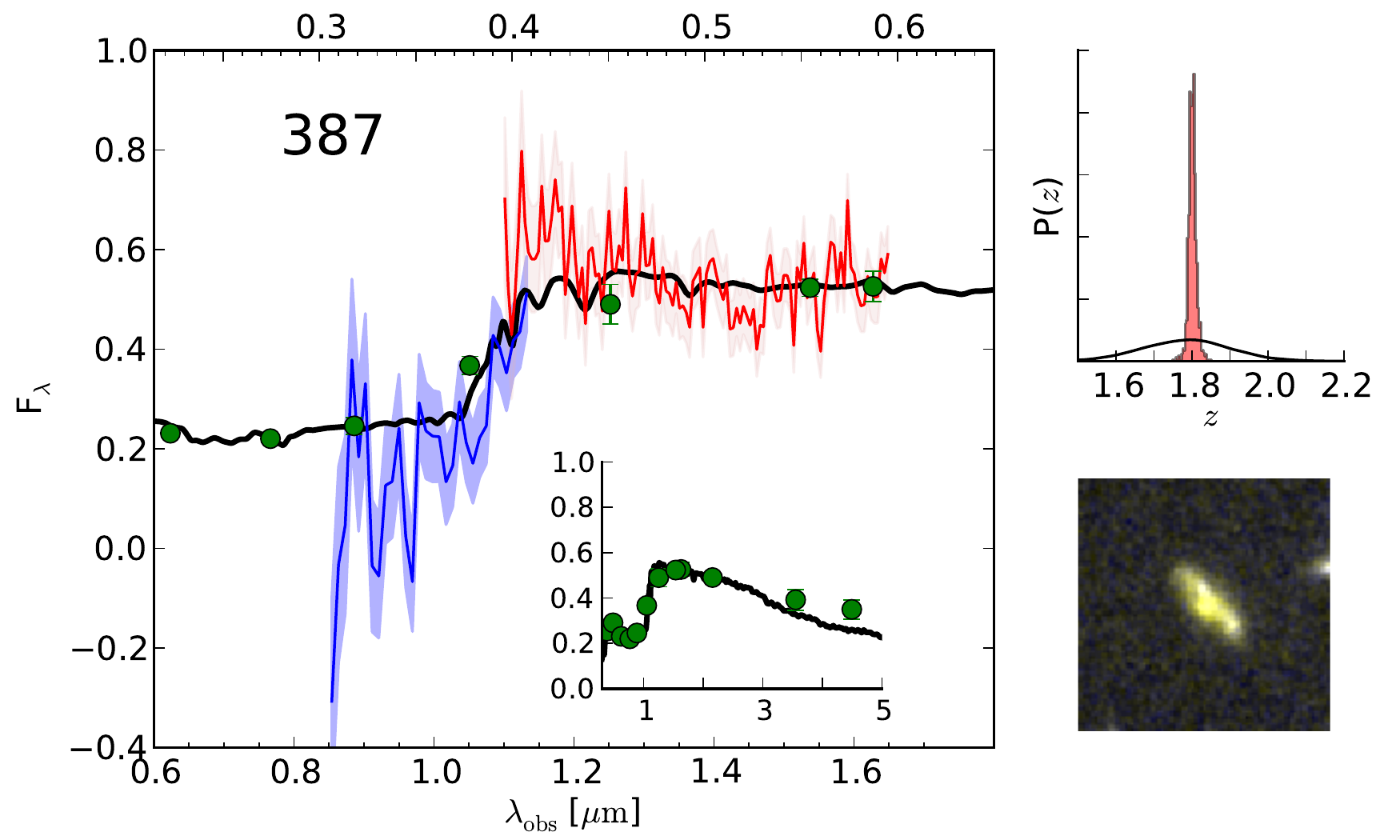} 
\includegraphics[width=0.49\linewidth]{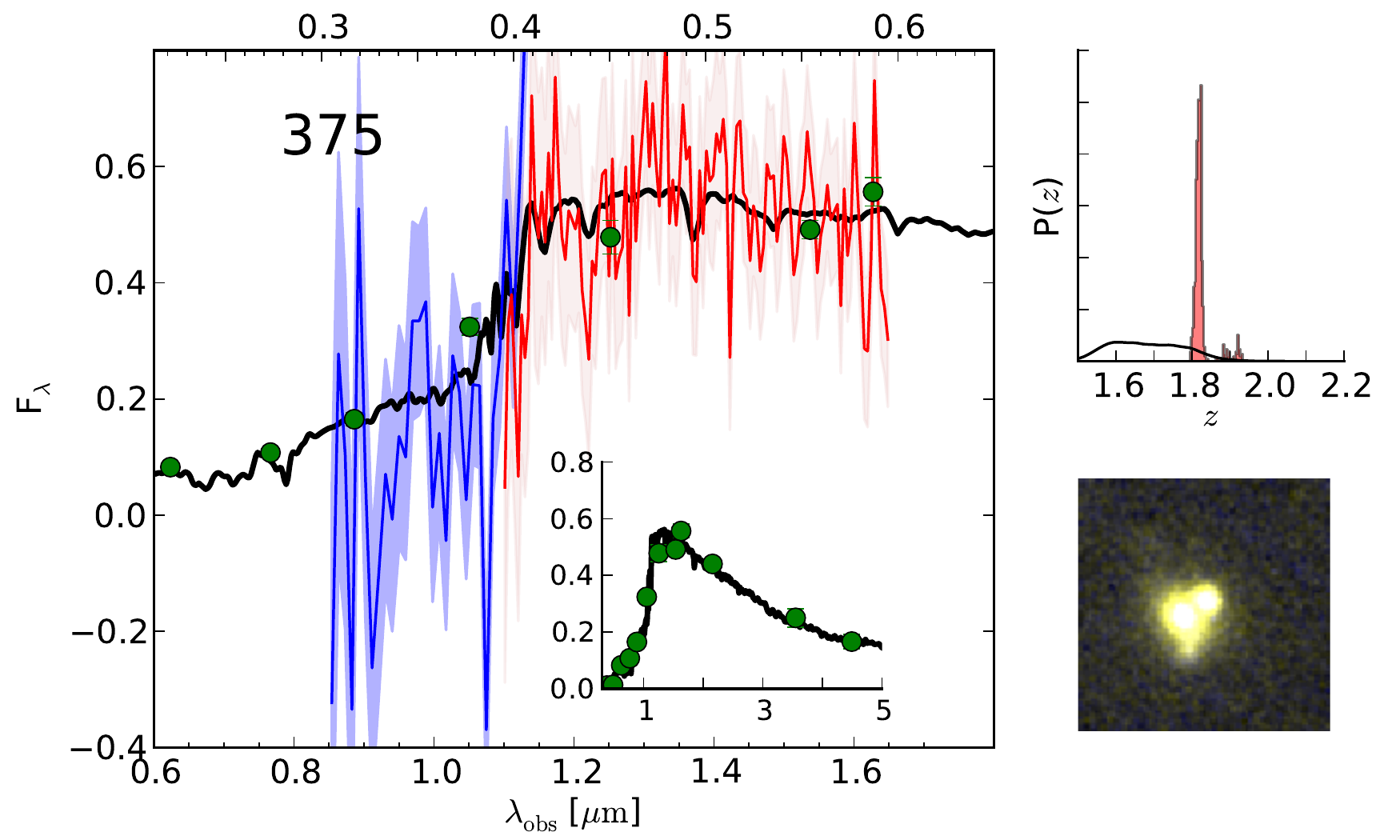} \\
\includegraphics[width=0.49\linewidth]{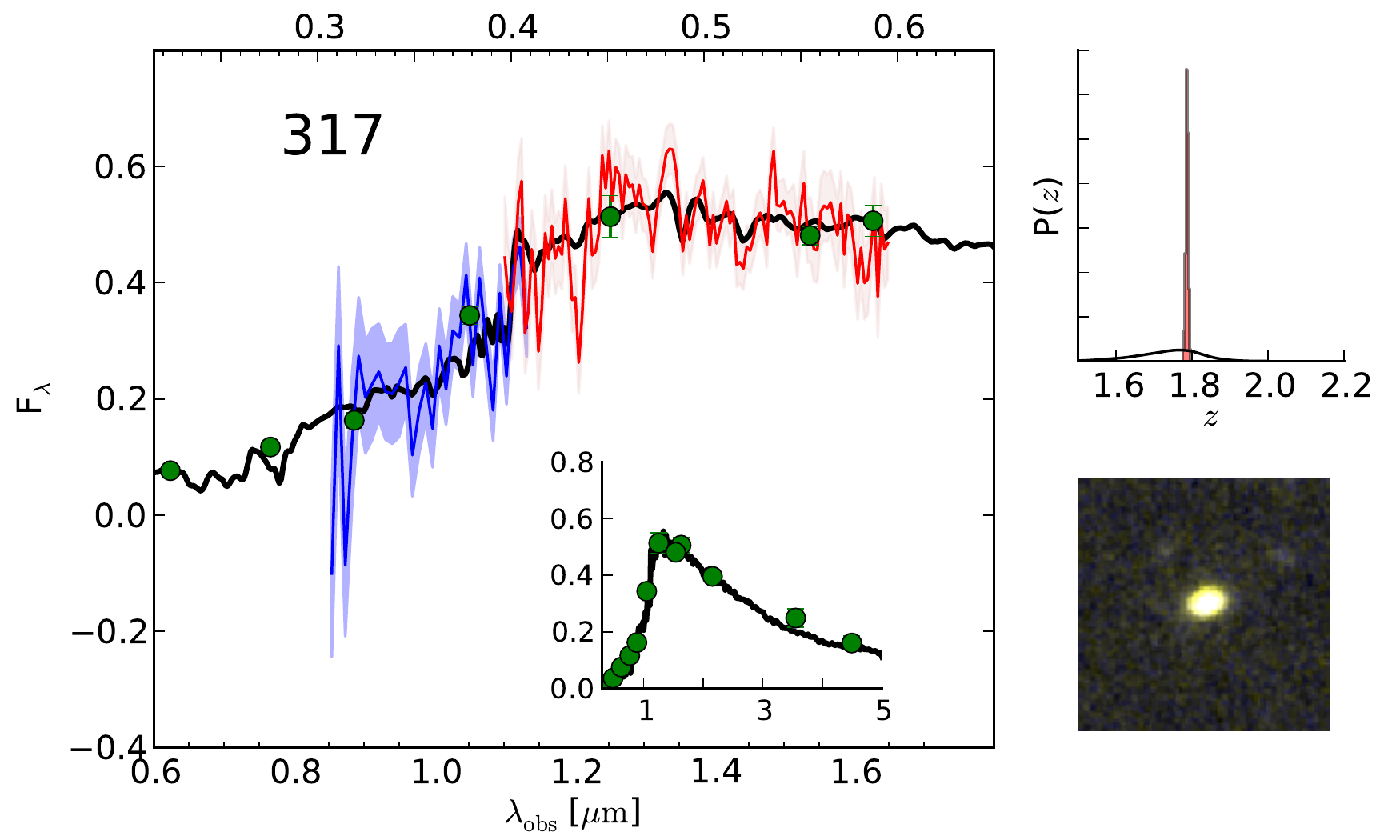} 
\includegraphics[width=0.49\linewidth]{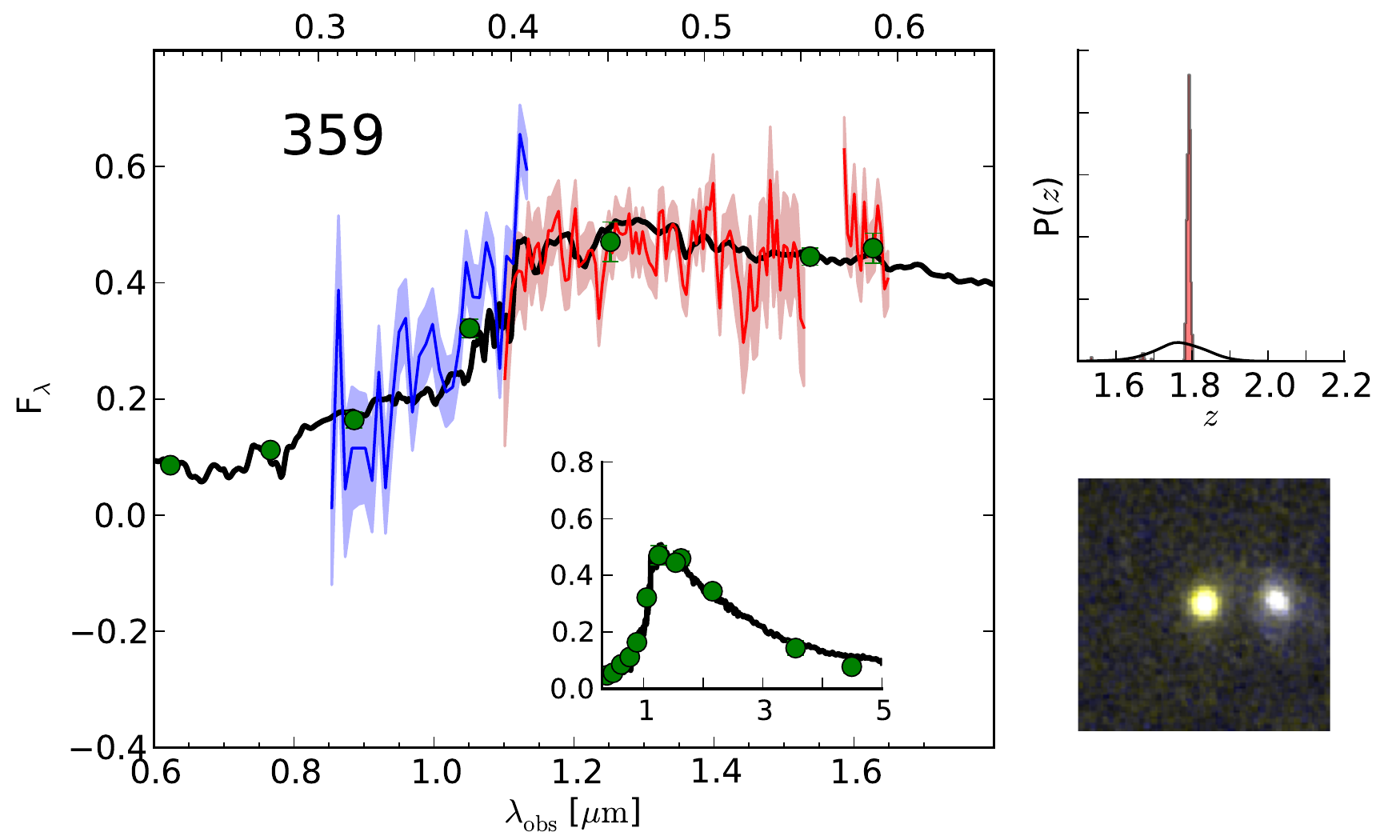} \\
\includegraphics[width=0.49\linewidth]{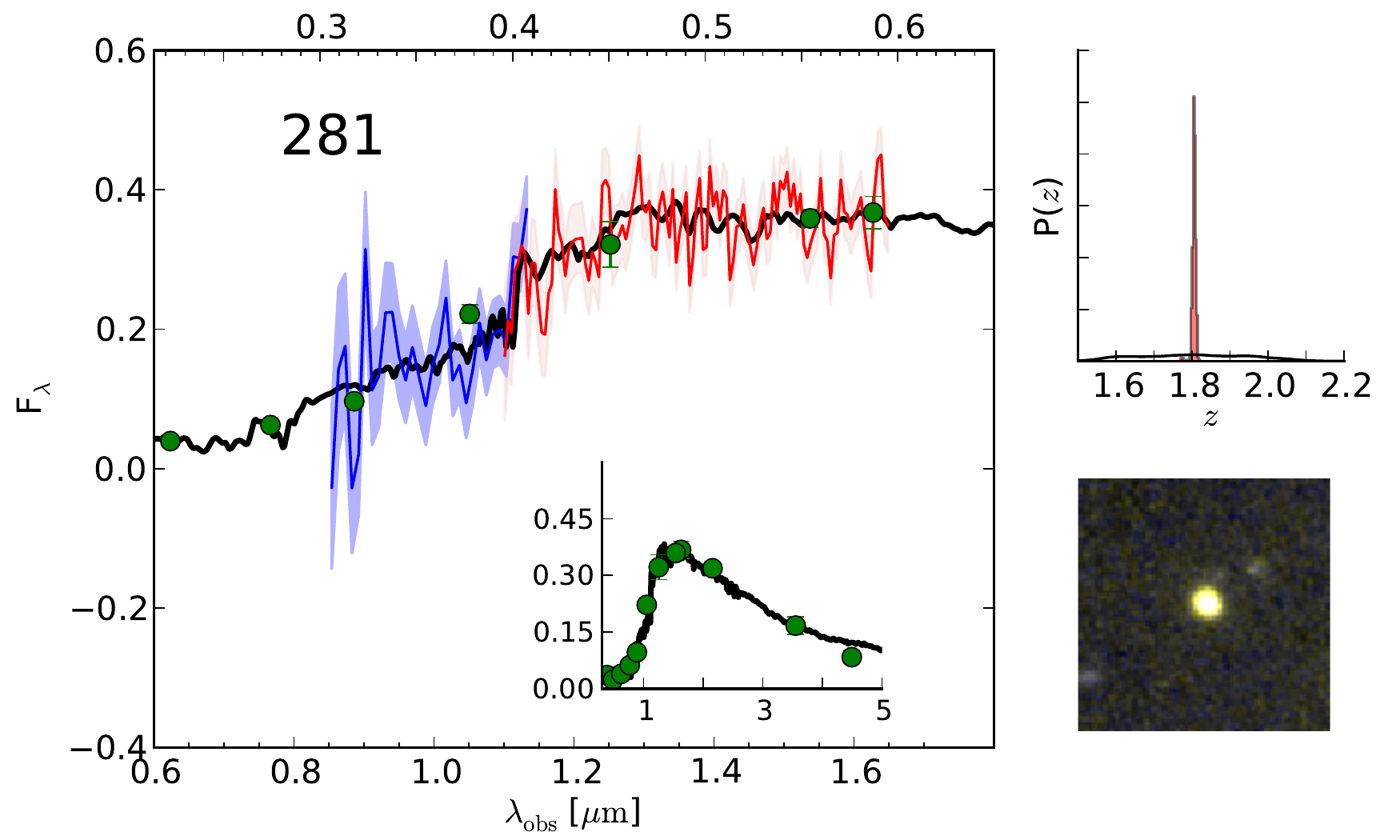} 
\includegraphics[width=0.49\linewidth]{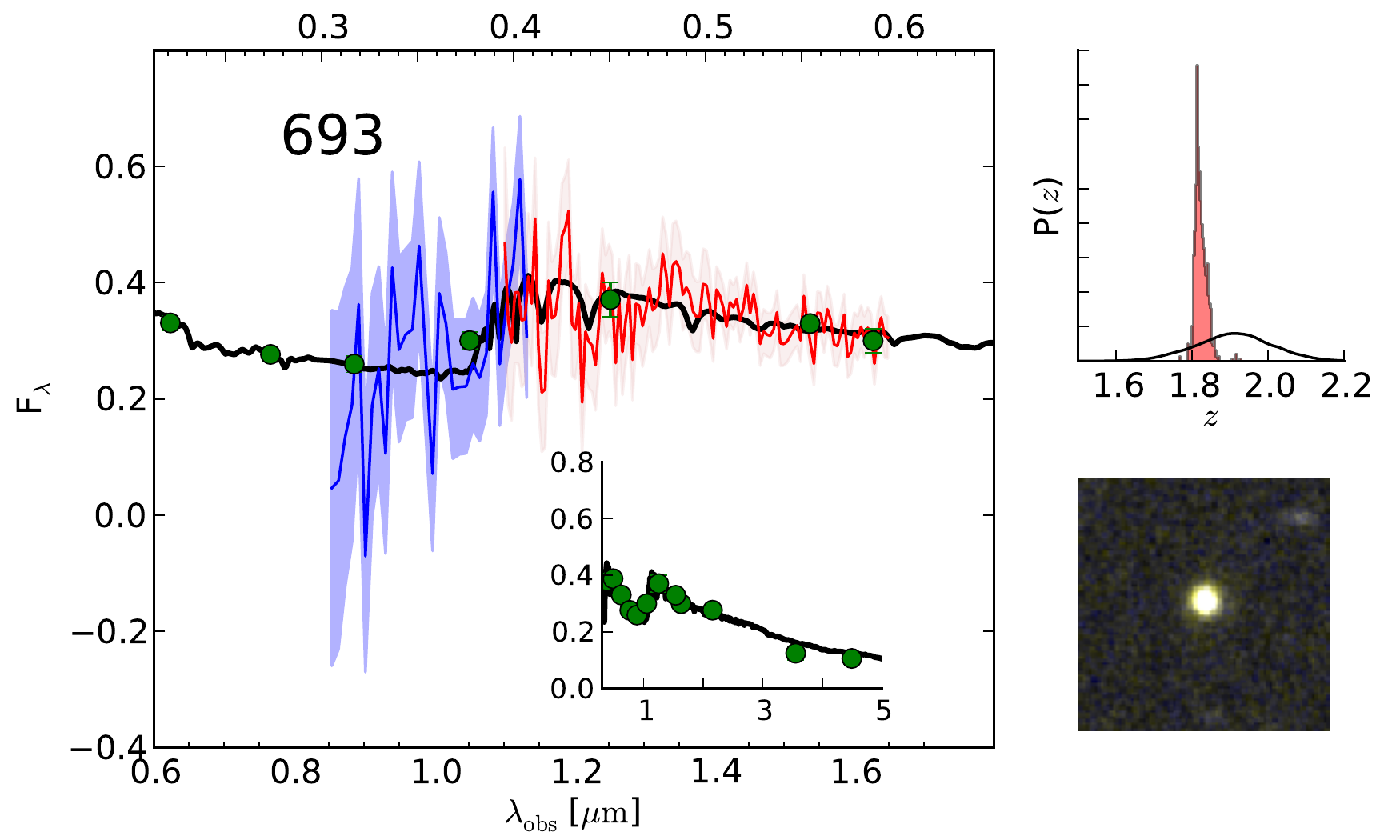}
\caption{Continued}
\end{figure*}

\addtocounter{figure}{-1}
\begin{figure*}
\includegraphics[width=0.49\linewidth]{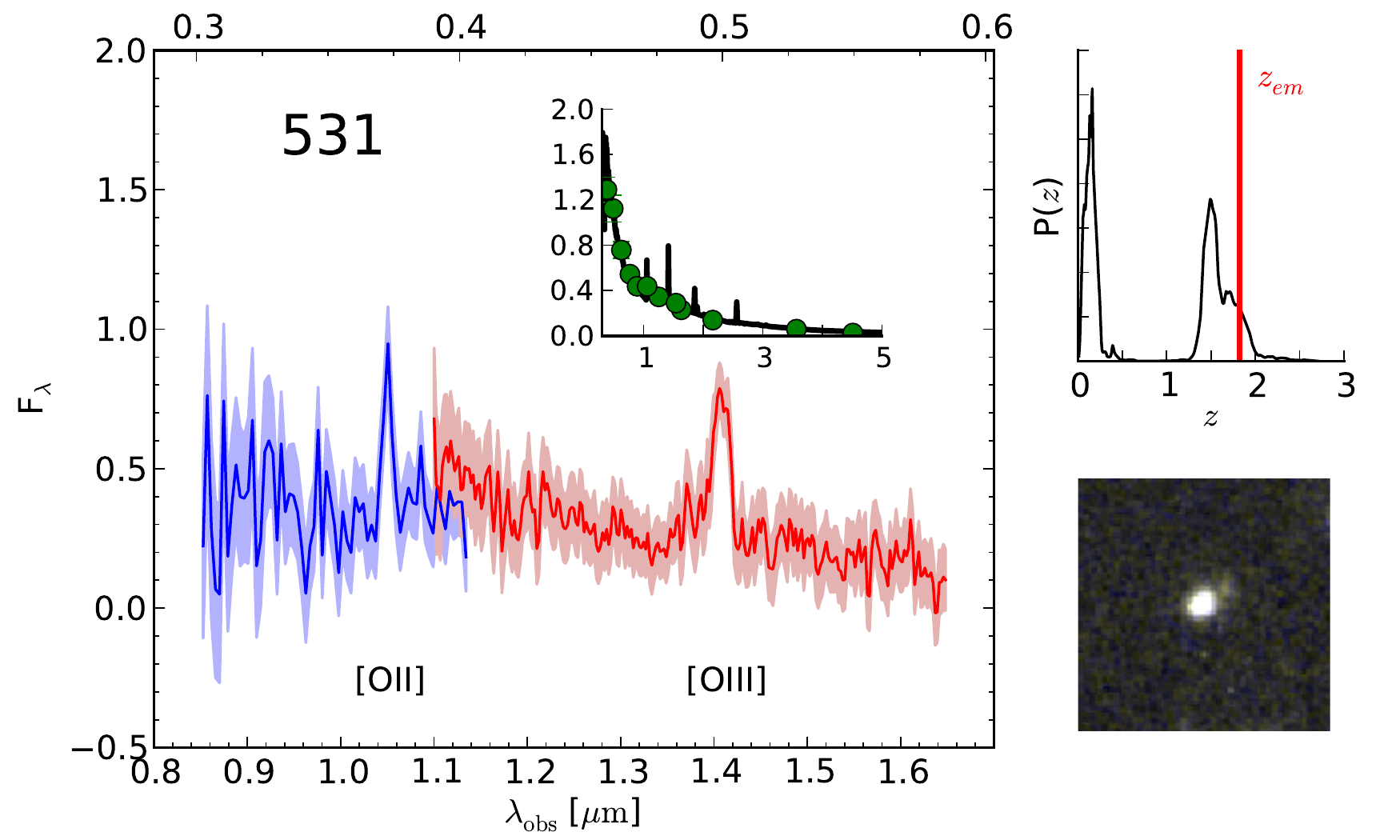} 
\includegraphics[width=0.49\linewidth]{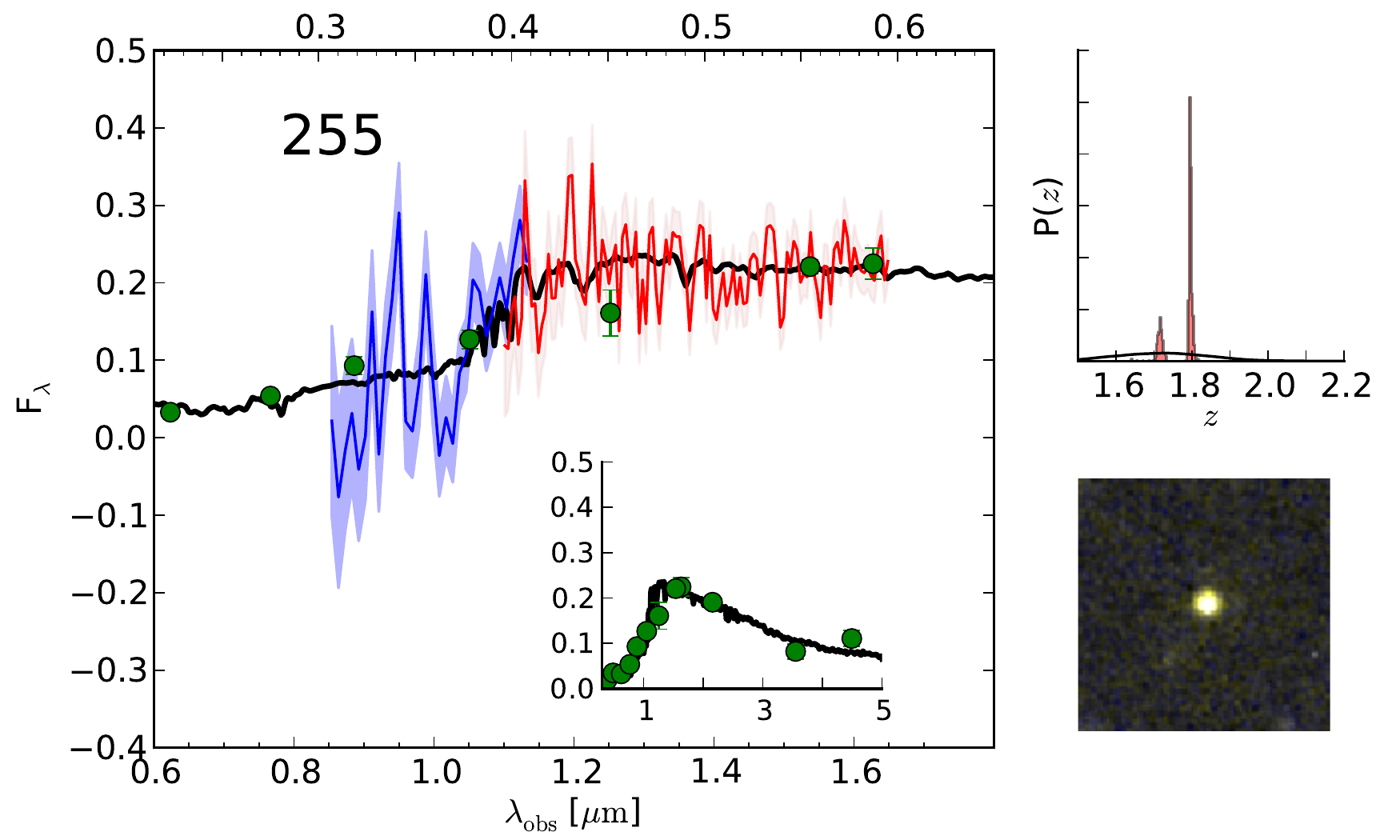} \\
\includegraphics[width=0.49\linewidth]{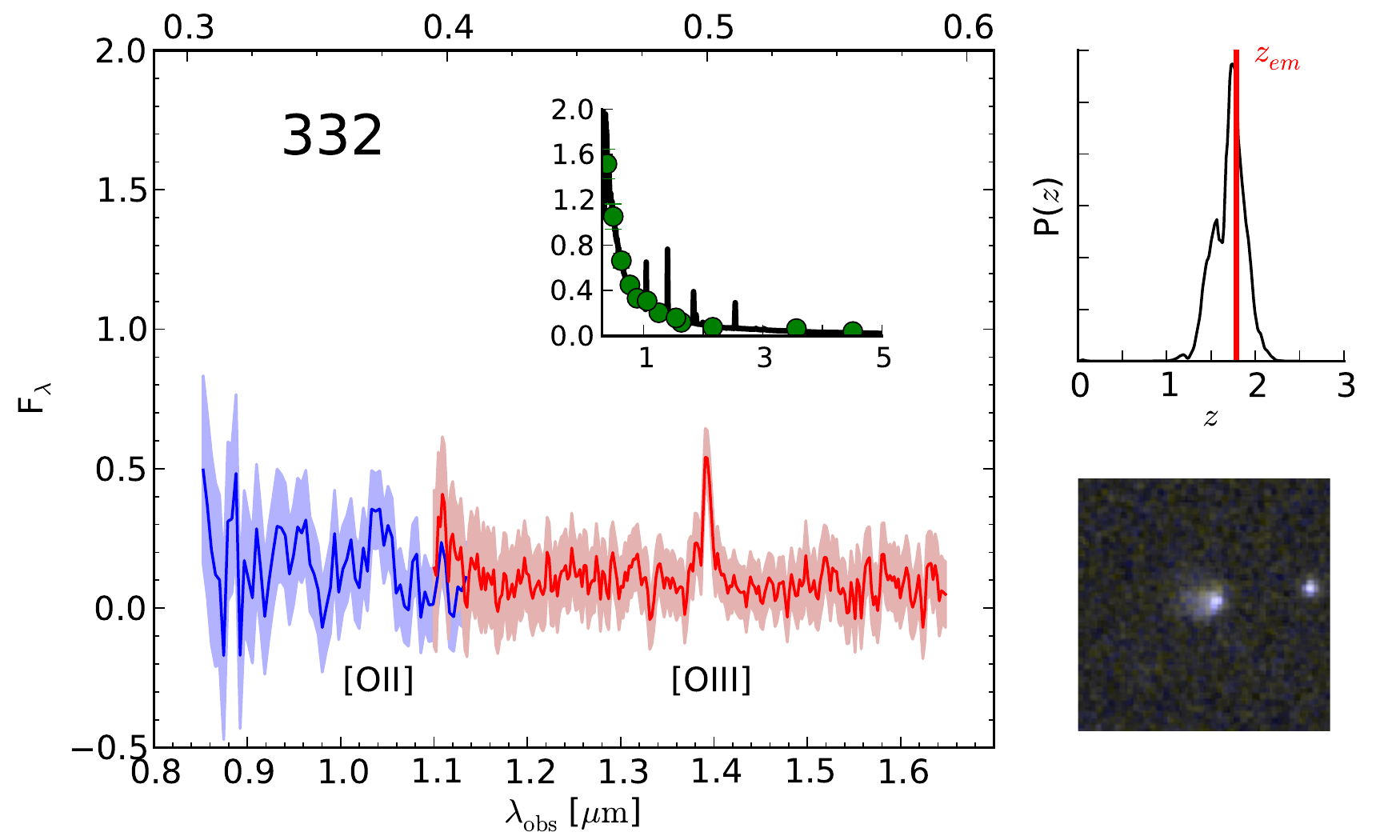}
\caption{Continued}
\end{figure*}

\begin{deluxetable*}{lllcccccccccc}
\tablewidth{\linewidth}
\tablecaption{Redshifts and Photometric Data for Spectroscopically Confirmed Cluster Members and Red Sequence Members}
\tablehead{\colhead{ID} & \colhead{R.A.} & \colhead{Dec.} & \colhead{$H_{160}$} & \colhead{$z$} & \colhead{Type} & \colhead{$\log M_*^{\rm auto}/\textrm{M}_{\odot}$} & \colhead{$z-J$} & \colhead{$(U-B)_{\rm r}$} & \colhead{$(U-V)_{\rm r}$} & \colhead{$(V-J)_{\rm r}$} & \colhead{$UVJ$} & \colhead{Quality}}
\startdata
\cutinhead{\emph{Spectroscopically confirmed cluster members}}
272 & 36.68173 & -4.68934 & 20.63 & $1.798_{-0.003}^{+0.002}$ & C & $11.71 \pm 0.03$ & $2.02 \pm 0.04$ & 1.20 & 1.84 & 1.15 & Q & A\\
355 & 36.68644 & -4.69239 & 20.80 & $1.798_{-0.002}^{+0.002}$ & C & $11.52 \pm 0.02$ & $2.01 \pm 0.03$ & 1.15 & 1.63 & 1.05 & Q & A\\
376 & 36.67501 & -4.69286 & 21.20 & $1.811_{-0.008}^{+0.004}$ & C & $11.56 \pm 0.03$ & $2.07 \pm 0.05$ & 1.34 & 1.90 & 1.14 & Q & A\\
356 & 36.69423 & -4.69235 & 21.35 & $1.805_{-0.004}^{+0.003}$ & C & $11.36 \pm 0.04$ & $1.97 \pm 0.07$ & 1.16 & 1.81 & 1.13 & Q & A\\
657 & 36.67557 & -4.70257 & 21.61 & $1.812_{-0.002}^{+0.002}$ & C & $11.11 \pm 0.02$ & $2.02 \pm 0.05$ & 1.20 & 1.77 & 0.92 & Q & A\\
286 & 36.68790 & -4.68994 & 21.69 & $1.798_{-0.013}^{+0.068}$ & C & $11.47 \pm 0.03$ & $1.94 \pm 0.08$ & 1.16 & 1.88 & 1.37 & Q & B\\
352 & 36.69051 & -4.69215 & 21.88 & $1.797_{-0.004}^{+0.006}$ & C & $11.22 \pm 0.05$ & $2.05 \pm 0.08$ & 1.23 & 1.87 & 1.08 & Q & A\\
411 & 36.67382 & -4.69384 & 22.11 & $1.821_{-0.004}^{+0.004}$ & C & $11.15 \pm 0.04$ & $1.84 \pm 0.09$ & 1.11 & 1.84 & 1.19 & Q & A\\
447 & 36.69121 & -4.69487 & 22.12 & $1.797_{-0.009}^{+0.011}$ & C & $10.81 \pm 0.03$ & $1.42 \pm 0.09$ & 0.82 & 1.34 & 0.64 & Q & A\\
289 & 36.68965 & -4.68994 & 22.17 & $1.802_{-0.004}^{+0.003}$ & C & $10.89 \pm 0.03$ & $1.97 \pm 0.08$ & 1.18 & 1.74 & 0.70 & Q & A\\
387 & 36.68231 & -4.69296 & 22.36 & $1.801_{-0.009}^{+0.009}$ & C & $11.00 \pm 0.04$ & $1.50 \pm 0.11$ & 0.94 & 1.51 & 1.49 & SF & B\\
375 & 36.67488 & -4.69278 & 22.43 & $1.819_{-0.008}^{+0.008}$ & C & $10.88 \pm 0.02$ & $1.91 \pm 0.09$ & 1.09 & 1.64 & 1.05 & Q & B\\
317 & 36.69911 & -4.69091 & 22.45 & $1.787_{-0.003}^{+0.003}$ & C & $10.75 \pm 0.04$ & $2.00 \pm 0.11$ & 1.14 & 1.61 & 1.11 & Q & A\\
359 & 36.67696 & -4.69228 & 22.54 & $1.792_{-0.005}^{+0.004}$ & C & $10.67 \pm 0.03$ & $1.90 \pm 0.11$ & 1.10 & 1.56 & 0.61 & Q & B\\
281 & 36.69061 & -4.68944 & 22.77 & $1.806_{-0.004}^{+0.004}$ & C & $10.73 \pm 0.06$ & $2.06 \pm 0.17$ & 1.12 & 1.75 & 0.98 & Q & B\\
693 & 36.67771 & -4.70379 & 22.86 & $1.820_{-0.010}^{+0.019}$ & C & $10.51 \pm 0.05$ & $1.14 \pm 0.09$ & 0.75 & 1.11 & 0.78 & SF & C\\
531 & 36.67919 & -4.69839 & 23.12 & $1.818_{-0.002}^{+0.002}$ & E & $9.73 \pm 0.06$ & $0.49 \pm 0.11$ & 0.27 & 0.46 & 0.16 & SF & A\\
255 & 36.68793 & -4.68838 & 23.30 & $1.795_{-0.075}^{+0.004}$ & C & $10.53 \pm 0.04$ & $1.35 \pm 0.24$ & 0.85 & 1.70 & 0.76 & Q & C\\
332 & 36.67165 & -4.69125 & 23.83 & $1.785_{-0.003}^{+0.003}$ & E & $9.35 \pm 0.28$ & $0.22 \pm 0.21$ & 0.11 & 0.21 & 0.82 & SF & B\\
\cutinhead{\emph{Candidate cluster members on red sequence (not spectroscopically confirmed), $H_{160} < 23.3$ and $R < R_{500}$}}
772 & 36.67527 & -4.70738 & 22.26 & $1.81_{-0.11}^{+0.08}$ & P & $10.91 \pm 0.28$ & $2.00 \pm 0.09$ & 1.20 & 1.72 & 1.00 & Q & \ldots\\
275 & 36.68274 & -4.68931 & 22.68 & $1.81_{-0.19}^{+0.12}$ & P & $10.78 \pm 0.28$ & $1.87 \pm 0.17$ & 1.00 & 1.66 & 1.02 & Q & \ldots\\
404 & 36.68949 & -4.69338 & 22.89 & $1.59_{-0.09}^{+0.17}$ & P & $10.71 \pm 0.28$ & $1.86 \pm 0.16$ & 1.22 & 1.91 & 1.33 & Q & \ldots
\enddata
\tablecomments{The ``r'' subscript denotes colors in the rest frame. C and E types indicate continuum and emission line redshifts, whereas P denotes photometric redshifts. Q and SF refer to galaxies in the quiescent and star-forming regions of the $UVJ$ color--color plane. For type C, $M_*$ is derived from fits to the full spectrophotometry (Section~\ref{sec:contfit}); for types E and P, $M_*$ is based on \texttt{FAST} fits to the photometry. Median random uncertainties in the rest-frame $U-B$, $U-V$, and $V-J$ colors are 0.07, 0.03, and 0.08 mag, respectively. $H_{160}$ is F160W magnitude in the \texttt{MAG\_AUTO} aperture, and $M_*^{\rm auto}$ is scaled here to this total flux. See Appendix~A for notes on the redshift quality flags.\label{tab:memberdata}}
\end{deluxetable*}

\begin{figure}
\includegraphics[width=\linewidth]{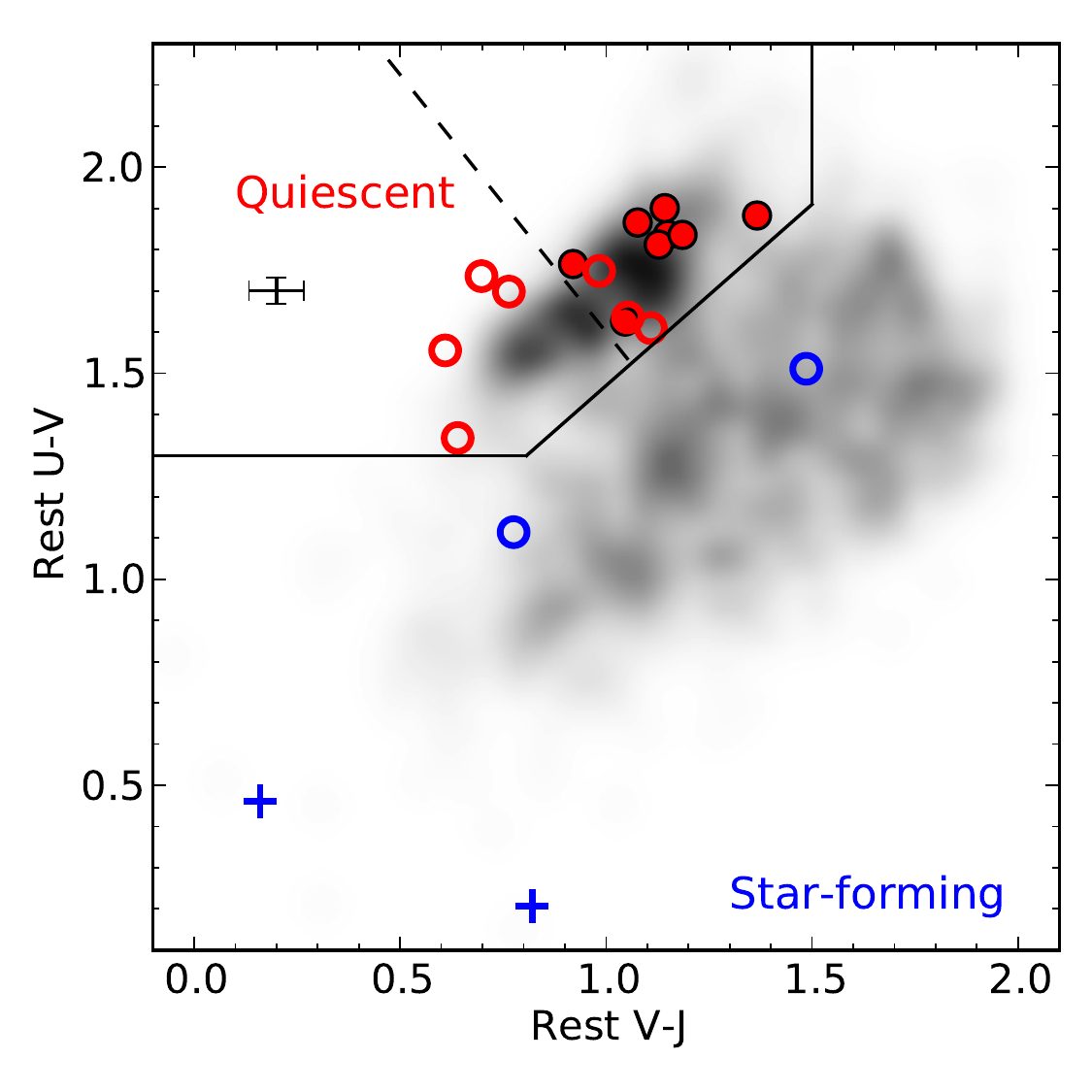}
\caption{Rest-frame colors of the spectroscopically-confirmed cluster members. Circles and crosses denote galaxies with continuum and emission line redshifts, respectively, while filled and open symbols denote massive ($M_* > 10^{11} \msol$) and less massive ($M_* < 10^{11} \msol$) systems, respectively. The grayscale shows the field distribution for galaxies drawn from the NMBS survey (see Section~\ref{sec:fq}) that have $z = 1.8 \pm 0.2$ and $M_* > 10^{10.6} \msol$. The solid line divides the quiescent and star-forming selection regions, while the dashed line shows the partition between bluer and redder quiescent galaxies used by \citet{Whitaker13}. Median color uncertainties are illustrated by the error bar. 
\label{fig:uvj}}
\end{figure}

\subsection{Completeness}
\label{sec:completeness}

Although the continuum sample is strictly flux-limited ($H_{160} < 23.3$), it forms a nearly mass-limited sample at $z = 1.80$. Based on the catalog of \Ntext~that covers a much wider area, we expect 88\% of galaxies at $z=1.8$ with $M_* > 10^{10.6} \msol$ to be brighter than $H_{160} = 23.3$. Within the WFC3 field of view surrounding \jkcs, all galaxies above this mass threshold that are photometric candidate members ($z_{\rm phot} = 1.8 \pm 0.2$) are brighter than $H_{160} = 22.8$, even though the imaging depth is $\sim 3$~mag fainter. Independently, we estimate nearly the same limiting mass using the BC03 model for a solar-metallicity galaxy formed in a burst at $z_f = 5$ (see green line in Figure~\ref{fig:zdist}, middle panel). Conversely, all confirmed cluster members in the continuum sample have $M_* > 10^{10.5} \msol$.

We thus expect the parent continuum-limited sample to be reasonably complete for stellar masses $M_* > 10^{10.6} M_{\odot}$. Additional incompleteness arises from those spectra that could not be extracted due to contamination from nearby sources. This affects 19 of the 59 galaxies in the continuum sample (Section~\ref{sec:contfit}). Three of these lie on the red sequence and are located at $R < R_{500}$. These are likely cluster members whose properties we list in Table~\ref{tab:memberdata}. Three additional bluer systems located within $R_{500}$ have $z_{\rm phot}$ consistent with \jkcs~within their 68\% confidence intervals; however, the redshift uncertainties are too large to associate them with the cluster with any confidence. None of the candidate members discussed above has a stellar mass $M_* > 10^{11} \msol$. Therefore, most likely we have spectroscopically confirmed all members with $M_* > 10^{11} \msol$ and $R < R_{500}$. At lower masses $M_* = 10^{10.6-11} \msol$, considering the three most likely photometric candidates, our estimated spectroscopic completeness is $\sim 75\%$. Given this high completeness, for the rest of the paper we focus our analysis on the spectroscopically-confirmed cluster members.

Completeness for the emission line sample is less straightforward to interpret. For this reason, we confine our quantitative analysis in Sections~\ref{sec:stellarpops} and onward to the better-defined continuum-selected sample and classify galaxies based on their colors, not on the presence of emission lines. Nonetheless, it is useful to have a rough idea of the star formation rate (SFR) corresponding to the limiting line luminosity of $3 \times 10^8 L_{\odot}$ (Section~\ref{sec:emlines}). [\ion{O}{2}] and [\ion{O}{3}] lie within our spectral coverage for \jkcs~members. For [\ion{O}{2}] emission, this limit corresponds to a SFR of $\gtrsim 30$~$M_{\odot}$~yr${}^{-1}$ according to the \citet{Kewley04} calibration with dust attenuation of $A_V = 1$. For galaxies with significantly subsolar metallicity, the [\ion{O}{2}] emission will be weaker, but [\ion{O}{3}] will be more visible. Limits will also be weaker for galaxies with higher dust content $A_V > 1$, which is expected for massive systems.

\subsection{Colors and Star Formation Properties of the Cluster Members}
\label{sec:galaxyprops}

Figure~\ref{fig:uvj} shows the distribution of the confirmed cluster members in the rest-frame $UVJ$ color--color diagram. This plane is frequently used to distinguish quiescent and star-forming systems \citep{Williams09}, and for the remainder of the paper we refer to quiescent and star-forming galaxies based on this criterion, using the specific form proposed by \citet{Whitaker11}.

Of the 19 confirmed members, 17 arise from the continuum sample, and 15 of of these fall in the quiescent region of the $UVJ$ plane. This large number of quiescent members with spectroscopic data makes \jkcs~an invaluable laboratory for studying environmental processes at high redshifts. None of the quiescent members shows unambiguous ($>3\sigma$) residual line emission above the continuum models, although there is a hint of [\ion{O}{2}] in IDs 657 and 447. Galaxy 447 is a borderline case: it falls near the edge of the quiescent selection box. It has a specific SFR of $10^{-10.2}$ Gyr${}^{-1}$ inferred from the spectrophotometric fitting, which is intermediate between the other 14 $UVJ$-quiescent members (all $<10^{-11}$ Gyr${}^{-1}$) and the star-forming members ($\sim 10^{-9}$ Gyr${}^{-1}$). Of the cluster members in the star-forming region of the $UVJ$ plane, two show emission lines (IDs 531 and 332) and have low stellar stellar masses $M_* = 10^{9.4-9.8} \msol$, while two more massive examples having $M_* = 10^{10.5-11} \msol$ were identified through continuum fitting (IDs 387 and 693). Note that we are able to secure redshifts of these bright blue galaxies even though they lack detectable emission lines.

Morphologically, virtually all of the quiescent confirmed members appear spheroid-dominated (see Figure~\ref{fig:memberspectra}). This visual impression is supported by a quantitative analysis of the galaxy shapes in Section~\ref{sec:structure}. Of the four star-forming members, two appear compact (IDs 693 and 531), ID 332 appears diffuse and irregular, and ID 387 (located near the cluster center) appears to be an inclined disk with a red bulge.

Only one spectroscopic member is detected as an X-ray point source in the 75~ks \emph{Chandra} data  \citep{Andreon09}: ID 352, a $UVJ$-quiescent galaxy with $L_{X, 0.5-2~{\rm keV}}= 6 \times 10^{42}$~erg~s${}^{-1}$. To investigate the presence of obscured star formation or AGN activity in other cluster members, particularly those classified as quiescent by their $UVJ$ colors, we measured 24 $\mu$m fluxes in the \emph{Spitzer} Multiband Imaging Photometer (MIPS) data taken for the SWIRE survey.\footnote{We used a simple $7''$ diameter aperture and applied an aperture correction factor of 2.56. The X-ray source (ID 352) has a detected close neighbor whose flux was subtracted using a PSF model.} None of the quiescent members is detected at $2\sigma$ significance ($> 0.13$~mJy), and there is no detection in a mean stack to a $2\sigma$ limit of 32~$\mu$Jy. 

The 2 more massive star-forming members (IDs 387 and 693) are detected with fluxes of $0.20 \pm 0.06$ mJy each. Based on the \citet{Wuyts08} templates, this corresponds to a total infrared luminosity of $L_{\rm IR} = (1.3 \pm 0.4) \times 10^{12} L_{\odot}$ for each source and SFRs of $140 \pm 44$ $M_{\odot}$ yr${}^{-1}$ for a \citet{Chabrier03} IMF \citep{Bell05}. These are typical for star-forming galaxies in this mass and redshift range \citep[e.g.,][]{Reddy06}. Thus, among the galaxies in our continuum-selected sample, we see a one-to-one correspondence between those which lie in the quiescent region of the $UVJ$ plane and those which lack detectable 24$\mu$m emission, albeit in fairly shallow MIPS imaging. \citet{Papovich12} also found a good correspondence between these diagnostics in a $z=1.62$ proto-cluster using deeper MIPS data, and \citet{Fumagalli13} recently showed that $UVJ$-quiescent galaxies at high redshift generally lack mid-infrared emission to very deep limits. We conclude that the $UVJ$ diagram provides reasonable classifications of cluster and field galaxies and is suitable for making differential comparisons, as we do in Sections 5 and thereafter.

\subsection{The Red Sequence}
\label{sec:candidates}

\begin{figure}
\centering
\includegraphics[width=\linewidth]{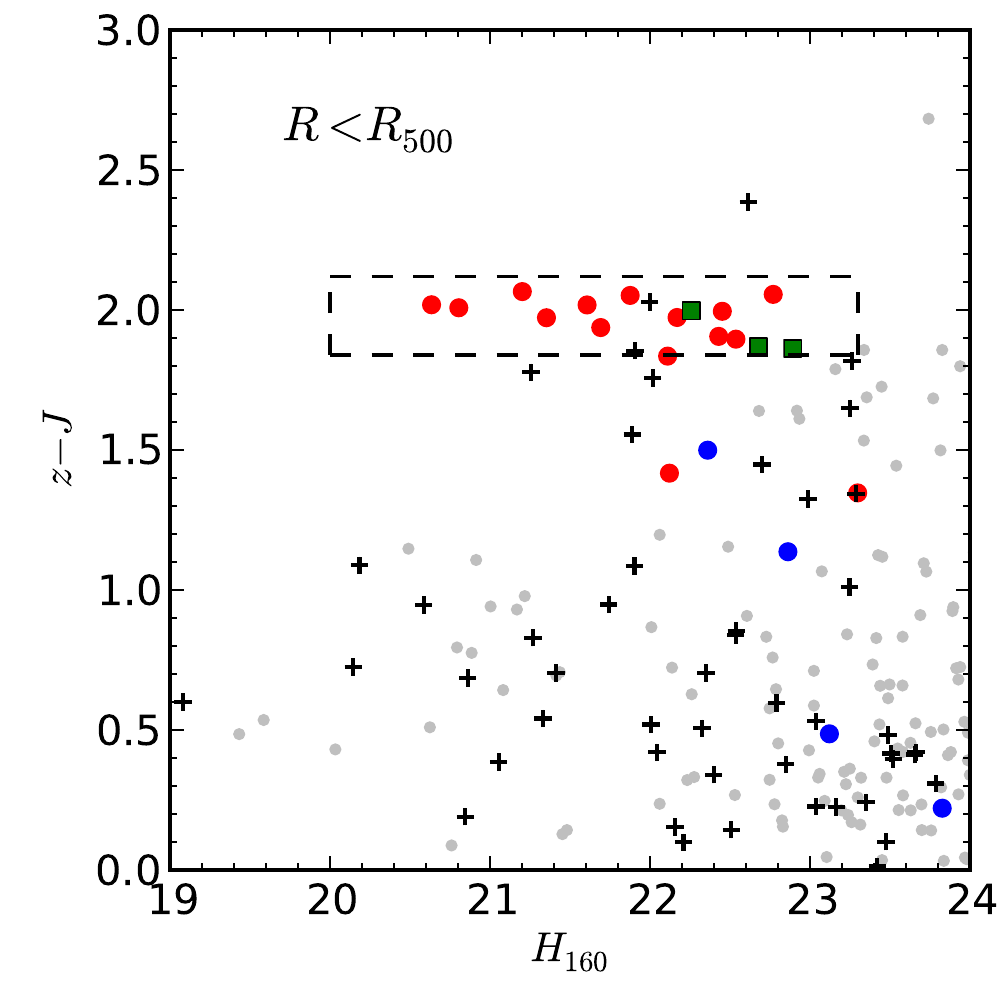}
\caption{Red sequence of \jkcs. \emph{Red circles:} spectroscopically-confirmed quiescent cluster members. \emph{Blue circles}: Confirmed star-forming members. \emph{Black crosses:} confirmed non-members. \emph{Green squares:} candidate cluster members on the red sequence (dashed region) that lack a grism redshift due to contamination of their spectra. \emph{Gray circles:} remaining galaxies with no grism redshift. Only galaxies within $R_{500}$ of the cluster center are plotted; this includes all confirmed members.\label{fig:redseq}}
\end{figure}

In the absence of spectroscopic data, members of high-redshift clusters are frequently identified based on the red sequence. With our grism observations we can assess the purity and completeness of this method. Figure~\ref{fig:redseq} shows the color--magnitude diagram for galaxies with $R < R_{500}$, where $R$ is the distance from the X-ray centroid. 

\jkcs~shows a clear red sequence with a mean observed color $\langle z-J \rangle = 1.98 \pm 0.02$ and a measured scatter of $\sigma_{z-J} = 0.07$. This is comparable to the rms measurement error of $\delta_{z-J} = 0.09$, indicating that the intrinsic scatter is low \citep{Andreon11b}. We define red sequence galaxies as those within $\pm 2\sigma$ of the mean color to a limiting magnitude of $H_{160} < 23.3$ (dashed in Figure~\ref{fig:redseq}). 

The majority of the spectroscopically confirmed members in the continuum sample (13 of 17) are on the red sequence. However, in addition to the two star-forming members, two galaxies that are classified as quiescent according to their $UVJ$ colors are bluer than the $z-J$ red sequence (IDs 255 and 447). These are likely systems where star formation has been most recently truncated. Naturally, some galaxies located on the $z-J$ red sequence will not be associated with the cluster. Using the grism redshifts, we identified five interlopers over the full field of view, which are indicated by boxes in Figure~\ref{fig:clusterimage}. Only two of these are located at $R < R_{500}$. Thus, a red sequence selection yields a fairly pure and complete sample (13 of 15, or 87\%) of quiescent members within $R < R_{500}$, as anticipated from the high overdensity of red sequence galaxies compared to the field \citep{Andreon11}. At larger radii contamination is more severe.

\section{Stellar Populations of Quiescent Galaxies: \jkcs~Compared to the Field}
\label{sec:stellarpops}

Having identified a well-defined set of cluster members based on grism spectroscopy, we now turn to the effect of the cluster environment on their stellar populations. We first consider the fraction of quenched systems in \jkcs~relative to coeval field galaxies of matched stellar mass. Additional insight can then be gained from the ages of the quiescent cluster members. We construct composite spectra that reveal age-sensitive stellar absorption lines at high signal-to-noise for the first time in such a distant cluster. Using these, we investigate the mean stellar age both as a function of mass within the cluster, and relative to similar field galaxies whose composite spectrum was constructed by \citet{Whitaker13} using 3D-HST grism data.
The 17 spectroscopically confirmed cluster members in the continuum-selected sample ($H_{160} < 23.3$) , which is approximately mass-limited ($M_* \gtrsim 10^{10.6} \msol$, Section~\ref{sec:completeness}) and confined to $R < R_{500} \approx 500$~kpc, form the basis for the following comparisons.

\subsection{The Quiescent Fraction}
\label{sec:fq}

\begin{figure}
\centering
\includegraphics[width=\linewidth]{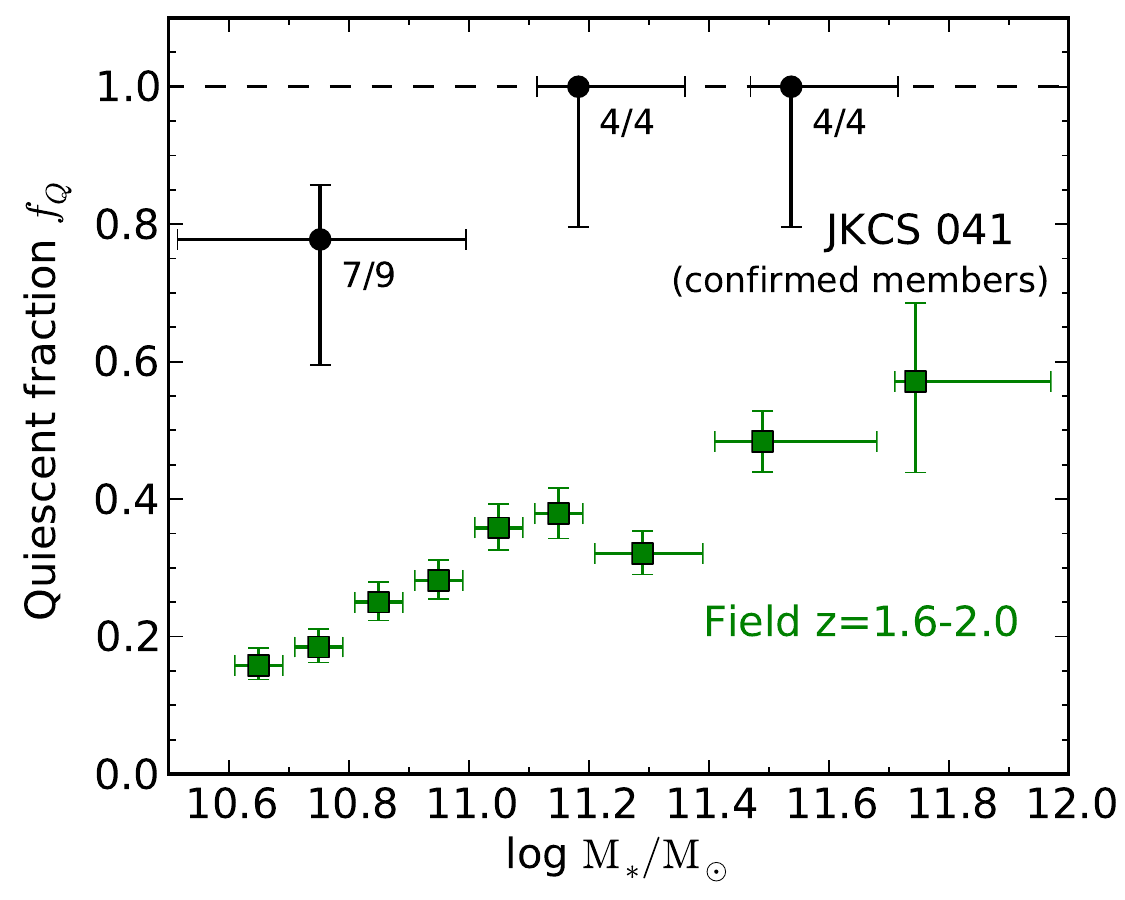}
\caption{Fraction of galaxies classified as quiescent by their $UVJ$ colors in several stellar mass bins. Spectroscopic members of \jkcs~(black) are compared to coeval field galaxies drawn from the NMBS survey (green). Horizontal error bars show the range of masses in each bin, with points placed at the median mass, while vertical $1\sigma$ errors are based on binomial statistics.\label{fig:fq}}
\end{figure}

Figure~\ref{fig:fq} compares the fraction $f_Q$ of galaxies in \jkcs~with quiescent $UVJ$ colors to that of field galaxies in the same range of stellar mass and redshift. The comparison sample is drawn from the NEWFIRM Medium Band Survey catalogs in the AEGIS and COSMOS fields \citep{Whitaker11}, selected from $z_{\rm phot} = 1.8 \pm 0.2$ and converted to a Salpeter IMF. Although this ``field'' sample includes galaxies that inhabit a range of environments, a differential comparison is still informative because \jkcs~is a strong overdensity.\footnote{For example, 9 members having $M_* > 10^{11} \msol$ lie within 1~arcmin of the cluster center, whereas only 1.8 are expected from the mean surface density in the field.}

Clearly, the cluster environment has had a powerful role in determining the number of quenched systems: 88\% (15 of 17) of the cluster members in the continuum sample are quiescent, whereas this fraction is less than half in the field. Roughly half of the quiescent cluster members were thus quenched by environmentally-related processes. 
Recalling that our spectroscopic sample may be missing some cluster members with masses $M_* = 10^{10.6-11} \msol$ due to contamination of their spectra, we have tested the effects of adding in the six unconfirmed candidate members described in Section~\ref{sec:candidates}. This would move $f_Q$ in the lowest-mass bin only with the plotted $1\sigma$ uncertainty, resulting in a fraction that would still be elevated above the field. Using a photometric redshift selection and a statistical background subtraction, \citet{Raichoor12b} also estimated a high quiescent fraction $f_Q \gtrsim 85\%$ ($1\sigma$ limit) among massive galaxies ($M_* \gtrsim 10^{11} \msol$) in the core of \jkcs~($R < 0.5 R_{200}$), consistent with our spectroscopic sample.

\subsection{Composite Spectra of Quiescent Cluster Members}
\label{sec:stackspec}

Having determined that the efficiency of quenching in \jkcs~is high, we now consider the ages of its quiescent members by constructing composite spectra of these galaxies. 
Stacking increases the signal-to-noise ratio and averages over residual contamination or background subtraction errors that may affect individual spectra. Rather than stacking the flux-calibrated spectra and photometry, we average continuum-normalized spectra covering $\simeq 4000-5900$~\AA~redward of continuum break. This technique has several advantages. First, we are able to measure the age-sensitive Balmer (H$\beta$,$\gamma$,$\delta$) and Mg~\emph{b} absorption lines; since these are narrowband features, they are more robust against errors in the continuum shape and uncertainties in dust attenuation. Second, we avoid the rest-frame near-infrared where model uncertainties related to the TP-AGB phase can influence the derived ages around 1~Gyr. Third, we are able to make a homogeneous comparison to coeval, quiescent galaxies in the field, whose composite continuum-normalized spectrum was measured by \citet{Whitaker13} using 3D-HST survey data.

In order to investigate mass-dependent trends, we split the sample of 15 confirmed quiescent members into a higher-mass subsample consisting of 8 galaxies with $M_* > 10^{11} \msol$, and a lower mass subsample whose 7 members span $M_* = 10^{10.5-11} \msol$. The continuum of each spectrum was first determined by fitting a third order polynomial to the models shown in Figure~\ref{fig:memberspectra}, excluding the strong absorption lines. Each spectrum was then divided by the continuum, shifted to the mean redshift of the cluster, and interpolated onto a grid with 48~\AA~pixels (17~\AA~in the rest frame), which is close to the native dispersion. The spectra were then combined by averaging each spectral pixel, excluding the highest and lowest measures. Uncertainties were estimated by bootstrapping. The LSFs of the galaxies entering the stack (Section~\ref{sec:hstgrism}) were also averaged to construct a mean LSF.

We fit the stacked spectra to simple stellar population (single-burst) models using \texttt{pyspecfit}, taking the redshift, age, and metallicity as free parameters. Although the actual star-formation histories are possibly more complex, using the burst models enables us to make a direct comparison with other work, particular that of \citet[][Section \ref{sec:ages}]{Whitaker13}. The model spectra were continuum-normalized using the same method that was applied to the data. A broad, log-uniform prior was placed on the age. We allow the metallicity to vary to quantify the degeneracy with age. Since these galaxies are expected to evolve into the cores of present-day massive ellipticals \citep[e.g.,][]{Bezanson09,Hopkins09}, which are metal-enriched to ${\rm [Z/H]} \approx 0.1 - 0.3$ \citep[e.g.,][]{Thomas10, Conroy14}, we place a broad uniform prior on [Z/H] over the range 0--0.3.

The top left panel of Figure~\ref{fig:stacks} shows the spectrum of the more massive ($M_* > 10^{11} \msol$) quiescent members of \jkcs. The quality of the spectrum is remarkably high, with a signal-to-noise ratio of 55 per pixel, and it clearly shows several absorption lines as indicated in the figure. The model (black curve) fits the data well with an age of $1.45^{+0.24}_{-0.18}$~Gyr, marginalized over metallicity, which corresponds to a formation redshift $z_f = 3.0^{+0.4}_{-0.2}$. 

The lower left panel displays the mean spectrum of the lower-mass ($M_* = 10^{10.5-11} \msol$) quiescent members. Although the spectrum is necessarily noisier, with a signal-to-noise ratio of 22, it is clearly different from that of the higher-mass galaxies. The clearest difference is the enhanced strength of the Balmer absorption lines: H$\beta$, H$\gamma$, H$\delta$ are all markedly deeper in the lower mass sample. We derive a younger luminosity-weighted mean age of $0.90^{+0.19}_{-0.10}$ Gyr, corresponding to a formation redshift $z_f = 2.4^{+0.2}_{-0.1}$. The Mg~\emph{b} absorption in this spectrum is too deep to be matched even by a maximally old, metal-rich model; this may be due to residual non-Gaussian noise in the stack. In any case, masking Mg~\emph{b} shifts our age inference by only $\sim 1\sigma$ to $0.79 \pm 0.19$~Gyr (dashed lines in Figure~\ref{fig:stacks}).

The quiescent galaxies in \jkcs~thus have a range of ages that follow the well-known mass-dependent trends seen in the field, in which lower-mass early type galaxies typically have younger luminosity-weighted ages \citep[e.g.,][]{Treu05L,Thomas10}. Although the absolute ages depend somewhat on metallicity, the right panel of Figure~\ref{fig:stacks} shows that the age difference of $0.52 \pm 0.26$~Gyr between the two subsamples is more robust, provided that they have broadly similar metallicity. We indeed expect the mean metallicities of our mass-selected subsamples to differ by $\lesssim 0.1$~dex, based on abundance studies at low redshift.\footnote{Given the ratio of the median stellar masses entering our two bins, we estimate a velocity dispersion ratio of $\Delta \log \sigma \approx 0.2$, which corresponds to abundance variations of $\Delta {\rm [Fe/H]} \approx 0.03$ and $\Delta {\rm [Mg/Fe]} \approx 0.08$ in $z \sim 0$ ellipticals \citep{Conroy14}.}

Two additional pieces of data support this conclusion. First, the ages of the individual galaxies as measured from fits to their grism spectra and photometry (Section~\ref{sec:contfit}) show the same trend: the median age is 1.6~Gyr and 0.96~Gyr for the high- and low-mass subsamples, respectively, which is consistent with the ages derived from their mean continuum-normalized spectra. Second, the lower-mass galaxies have bluer colors, as shown in Figure~\ref{fig:uvj}. We can predict the mean color differences between the mass-selected subsamples that should arise purely from the difference in ages inferred from their absorption lines. The predicted $\Delta \langle U-V \rangle = 0.14 \pm 0.08$ and $\Delta \langle V-J \rangle = 0.26 \pm 0.12$ are consistent with the measured values of $\Delta \langle U-V \rangle = 0.20$ and $\Delta \langle V-J \rangle = 0.29$. Thus, the color trend can be explained by a mass-dependent trend in age, rather than metallicity or dust content.

These results should be interpreted with the usual understanding the ages are luminosity-weighted and so skew toward the most recent star formation episode. Our focus is robustly constraining the mean age as a function of mass, and some cluster members at given mass may of course be older or younger. (For example, the spectrum of  ID 355, shown in Figure~\ref{fig:memberspectra}, is clearly younger than that of the first-rank cluster member.) The tightness of the red sequence led \citet{Andreon11b} to infer that the spread in ages at a fixed mass is quite small. Their analysis, however, is sensitive to the assumed cluster redshift, which we have now revised to $z = 1.80$. For further details and a revised estimate of the age scatter based on our spectroscopic data, we refer to \citet{Andreon13}.

\subsection{Age and Line Emission in Quiescent Galaxies as a Function of Environment}
\label{sec:ages}

\begin{figure*}
\centering
\includegraphics[width=0.63\linewidth]{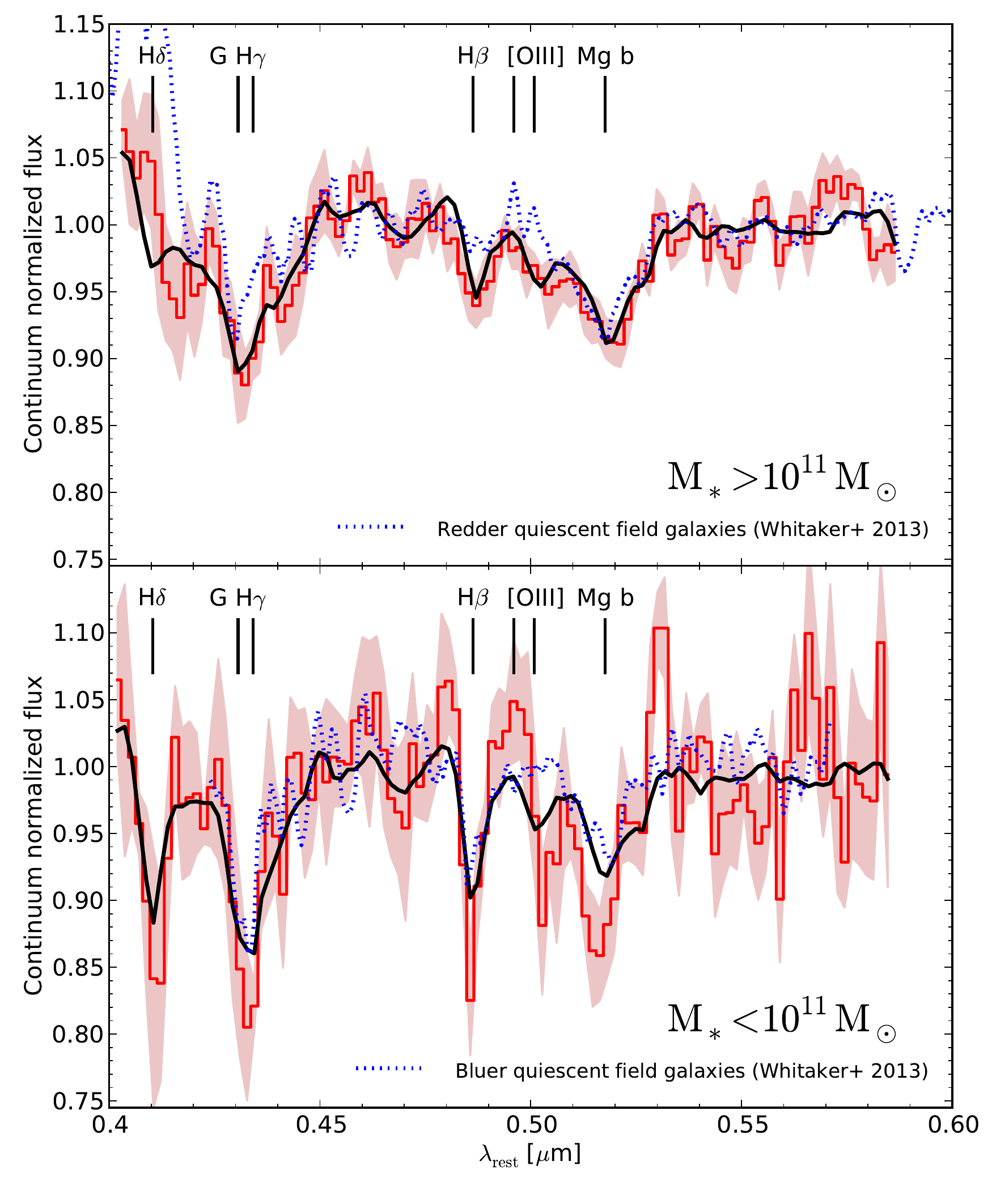}
\includegraphics[width=0.36\linewidth]{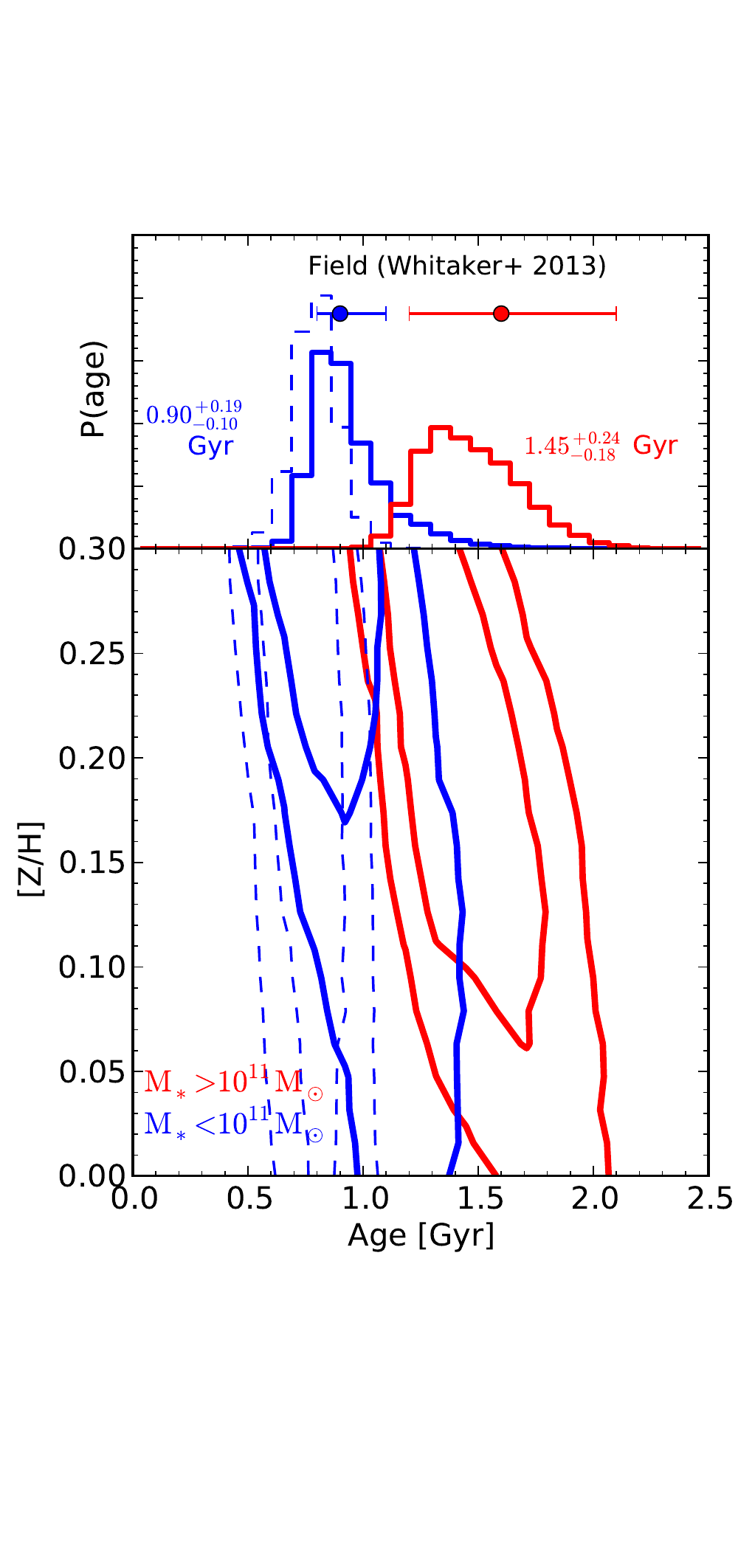}
\caption{{\bf Left:} composite spectra of confirmed quiescent members of \jkcs~in two bins of stellar mass. Red curves show the data and $1\sigma$ uncertainties, and black lines show the model fit. Dashed blue curves show composite spectra of quiescent field galaxies from \citet{Whitaker13}: the upper and lower panels show their stacks of redder and bluer quiescent galaxies, respectively. {\bf Right:} constraints on the simple stellar population model derived for the two mass-selected subsamples. Contours show $1\sigma$ and $2\sigma$ constraints; dashed contours show results for the lower-mass subsample when Mg~\emph{b} is masked. The upper panel shows the marginalized posterior distribution for the age and compares to field constraints derived by \citet{Whitaker13} for their bluer and redder quiescent galaxy subsamples ($1\sigma$ error bars).\label{fig:stacks}}
\end{figure*}

\citet{Whitaker13} recently constructed composite spectra of 171 quiescent field galaxy observed in the 3D-HST grism survey. This presents an interesting opportunity to compare quenched field and clusters galaxies at the same early epoch. The Whitaker et al.~data are very well suited for this comparison. In addition to being observed with the same instrument, they selected quiescent galaxies using the same $UVJ$ color selection, and their limiting magnitude of $H_{140} < 22.8$ (measured in the F140W filter) is similar to our limit of $H_{160} < 23.3$. Their median stellar mass $10^{11.08} \msol$, converted to a Salpeter IMF, matches the $10^{11.11} \msol$ of our sample. The main difference is that the Whitaker et al.~stacks combine field galaxies spanning a wide range in redshift, $z = 1.4 - 2.2$,  whereas the members of \jkcs~are obviously at a single redshift. Nonetheless, the median redshift of the galaxies in their stacks is $\langle z \rangle \simeq 1.6 - 1.7$, close to \jkcs. 


Rather than subdividing their sample by stellar mass, Whitaker et al.~split the quiescent selection region of the $UVJ$ plane into two regions indicated by the dashed line in Figure~\ref{fig:uvj}. Among the quiescent \jkcs~members, such a color division is very similar to a division in stellar mass: the eight quiescent members with $M_* > 10^{11} \msol$ would all fall in the redder subsample of Whitaker et al., and the seven less massive members fall in or near their bluer region. The mean color difference between the galaxies in their blue and red subsamples ($\Delta \langle U-V \rangle = 0.2$, $\Delta \langle V-J \rangle = 0.3$) is consistent with that described above for our mass-selected subsamples.

With this in mind, in the top left panel of Figure~\ref{fig:stacks} we compare our composite spectrum of massive \jkcs~members to the composite field spectrum of redder quiescent galaxies investigated by Whitaker et al. First, we note that the Mg~\emph{b} lines are nearly identical. Correspondingly, Whitaker et al.~derived an age of $1.6^{+0.5}_{-0.4}$~Gyr for their redder field sample, consistent with our measurement (see right panel). Interestingly, the field stacks show faint line emission in [\ion{O}{3}] $\lambda\lambda 4959, 5007$ and in filling of H$\beta$, whereas the spectrum of the \jkcs~members clearly lacks this emission and instead follows the stellar population model closely.\footnote{We note that the \citet{Whitaker13} stacks are median spectra and so should be relatively immune from strong line emission in a small fraction of the field sample.}

In the lower left panel of Figure~\ref{fig:stacks} we compare our composite spectrum of lower-mass \jkcs~members to the composite spectrum of bluer quiescent field galaxies. The strong Balmer lines seen in the cluster members are also evident in the field. Whitaker et al.~derived a reduced age of $0.9^{+0.2}_{-0.1}$ Gyr, again consistent with our measurement for the lower-mass ($M_* = 10^{10.5-11} \msol$) quiescent cluster members. Whitaker et al.~infer [\ion{O}{3}] emission in their bluer subsample as well, although the signal there is more ambiguous. Our stack of lower-mass members shows no clear evidence of emission, but the lower signal-to-noise ratio makes this distinction marginal.

Comparing the ages derived in our two stacks to the Whitaker et al.~measurements in the upper right panel of Figure~\ref{fig:stacks}, we find that the cluster and field samples span a very similar range. Quantitatively, the differences in luminosity-weighted mean stellar ages are $\Delta t = {\rm age}_{\rm JKCS041} - {\rm age}_{\rm field} = -0.2 \pm 0.5$~Gyr and $0.0^{+0.3}_{-0.1}$~Gyr for the more-massive/redder and less-massive/bluer subsamples, respectively. These results are marginalized over a range of metallicity, whereas Whitaker et al.~instead fixed the metallicity to solar abundance in their analysis. If we do the same, these age differences shift to $\Delta t = 0.2 \pm 0.5$~Gyr and $0.3^{+0.3}_{-0.2}$~Gyr, respectively. In this solar metallicity case, however, the age of the lower-mass cluster members is strongly influenced by the Mg~\emph{b} region, where we noted that the fit is poor. Masking Mg~\emph{b} and relying on Balmer line indictors yields $\Delta t = 0.0^{+0.2}_{-0.1}$~Gyr for the lower-mass subsample.

In each of these comparisons, we do not detect a difference between the field and cluster mean ages at the $\sim 1\sigma$ level, or about 0.5~Gyr and 0.3~Gyr for the more- and less-massive subsamples, respectively. Because the median redshift of the galaxies entering the Whitaker et al.~stacks is slightly lower than that of \jkcs, comparing ages is not precisely the same as comparing formation times. However, the difference in median lookback time is $\sim0.3$~Gyr for the massive/redder subsample and only 0.1~Gyr for less-massive/bluer examples; both are less than the statistical uncertainties. We also note that the mean ages derived above will not include any galaxies that were very recently truncated and are in transition to the quiescent region of the $UVJ$ plane.

In summary, the mean luminosity-weighted ages of the quiescent members of \jkcs~varies with mass, with lower-mass galaxies having younger ages. The cluster members span a remarkably similar range of ages to that seen in quiescent field galaxies near the same redshift. Intriguingly, however, the line emission seen in quiescent field samples is absent in \jkcs, at least among its more massive members where the high quality of the spectrum permits a comparison. We discuss the physical significance of these findings in Section~\ref{sec:discussion}.

\section{Structure of Quiescent Galaxies: \jkcs~Compared to the Field}
\label{sec:structure}

\begin{figure*}
\centering
\includegraphics[width=\linewidth]{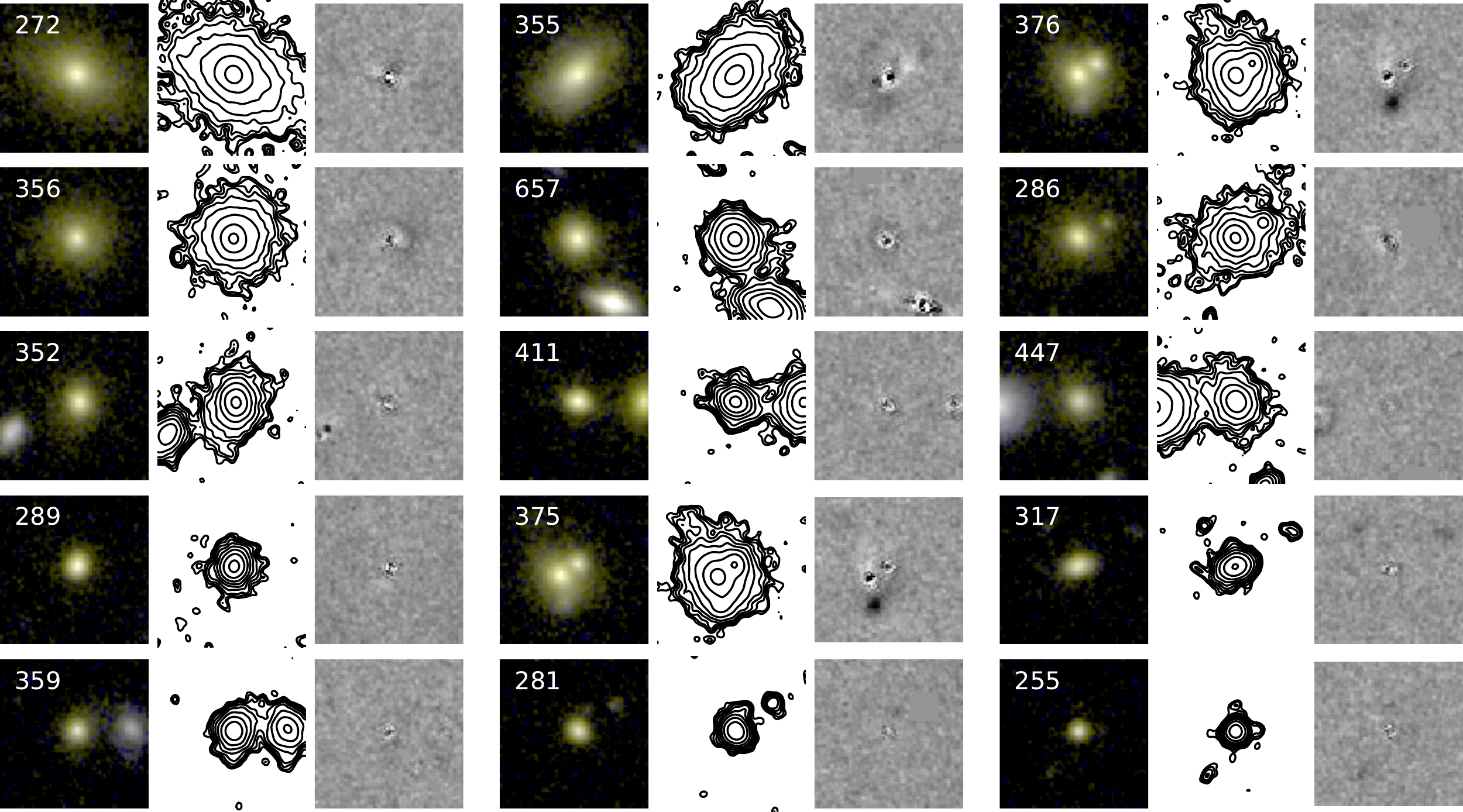}
\caption{F160W/F105W images (left panels) of the 15 confirmed quiescent members of \jkcs~ordered by F160W flux, displayed with a logarithmic scaling. Center panels show logarithmically spaced F160W isophotes. Right panels show residuals of the S\'{e}rsic fits to each F160W image, scaled linearly over $\pm 23$~mag~arcsec${}^{-2}$. Pixels masked in the fits are set to zero. The cutout side length is $4\arcsec \approx 34$~kpc.\label{fig:galfits}}
\end{figure*}

To gain insight into the role of the environment in the rapid structural evolution of quiescent galaxies at $z \sim 2$, we now compare the structural properties of the members of \jkcs~to their field counterparts. In addition to our \emph{HST} imaging of the cluster, this comparison requires a large field sample. Furthermore, in order to minimize systematic differences, the structural measurements should be conducted following the same procedures in the cluster and field. The CANDELS data provide an excellent basis for such a comparison, since the survey has imaged a large area using \emph{HST}/WFC3 to a depth similar to our F160W observations. Here we assemble a sample of 225 galaxies spanning $z = 1.8 \pm 0.3$ drawn from the CANDELS fields. Using this large sample, we are able to make a precise and homogeneous comparison between galaxy structure in \jkcs~and the field.

\subsection{Structural measurements and field sample}
\label{sec:sizemethod}

We used \texttt{Galfit} to fit 2D S\'{e}rsic profiles to the F160W images of all spectroscopically confirmed quiescent cluster members (Figure~\ref{fig:galfits}). The detailed procedures for PSF construction and masking or simultaneous fitting of nearby galaxies follow those described in N12. The only procedural difference is that we estimate the sky in a larger rectangular annulus around the object, with a width of 80 pixels, and mask objects more aggressively when the sky level is estimated. The derived structural parameters are listed in Table~\ref{tab:sersic}. Throughout this section, we refer sizes using the semi-major axis $a = R_e^{\rm maj}$ of the ellipse enclosing half of the light, and \emph{not} a ``circularized'' effective radius $\sqrt{ab}$ that is also frequently quoted in the literature. We prefer $R_e^{\rm maj}$ because it is independent of inclination for oblate objects, which form one focus of our analysis, whereas the circularized radius is very sensitive to viewing angle for flattened systems. For the lowest-mass confirmed quiescent member (ID 255), we were unable to secure a reliable size measurement, since this galaxy is essentially unresolved. Based on our simulations, its size is likely $R_h \lesssim 1~{\rm pixel} \approx 0.5$~kpc. Our comparison to the field is limited to galaxies having $M_* > 10^{10.7} \msol$, so this low-mass galaxy does not enter our analysis.

In this section we refer to stellar masses $M_*^{\rm tot}$ that are scaled to the total flux in the S\'{e}rsic profile fit. This is preferable when constructing the mass--radius relation, since the size and luminosity are derived  consistently from the same light profile. For the largest galaxies, we note that $M_*^{\rm tot}$ can exceed the {\tt MAG\_AUTO}-scaled masses $M_*^{\rm AUTO}$ (Table~1) by up to 0.25 dex.

\begin{deluxetable}{cccccc}
\tablewidth{\linewidth}
\tablecaption{S\'{e}rsic Fits to Confirmed Quiescent Cluster Members} 
\tablehead{\colhead{ID} & \colhead{$R_e^{\rm maj}$ (kpc)} & \colhead{$q$} & \colhead{$n$} & \colhead{$H_{160}^{\rm tot}$} & \colhead{$\log M_*^{\rm tot}/M_{\odot}$}}
\startdata
272 & 14.7 & 0.71 & 6.8 & 20.03 & 11.96 \\
376 & 5.00 & 0.70 & 6.5 & 20.84 & 11.70 \\
286 & 5.27 & 0.83 & 8.0 & 21.17 & 11.68 \\
356 & 10.6 & 0.97 & 7.7 & 20.71 & 11.62 \\
355 & 4.72 & 0.56 & 2.7 & 20.65 & 11.58 \\
352 & 2.45 & 0.74 & 5.2 & 21.56 & 11.34 \\
411 & 0.85 & 0.57 & 4.1 & 21.93 & 11.22 \\
657 & 1.56 & 0.91 & 3.2 & 21.45 & 11.18 \\
289 & 0.83 & 0.65 & 3.8 & 21.97 & 10.97 \\
447 & 3.13 & 0.81 & 3.3 & 21.95 & 10.88 \\
317 & 1.43 & 0.47 & 1.9 & 22.30 & 10.81 \\
281 & 0.89 & 0.75 & 3.0 & 22.61 & 10.80 \\
375 & 0.62 & 0.95 & 3.4 & 22.64 & 10.79 \\
359 & 1.47 & 0.86 & 6.9 & 22.29 & 10.77 \\
255 & \multicolumn{5}{l}{(unresolved --- see text)}
\enddata
\tablecomments{Stellar masses in the final column are scaled to the total S\'{e}rsic magnitude and so differ from the \texttt{MAG\_AUTO}-scaled masses in Table~\ref{tab:memberdata}. See Section~\ref{sec:sizemethod} for estimates of uncertainties.\label{tab:sersic}}
\end{deluxetable}

Our field comparison sample is drawn from four of the CANDELS survey fields. We have augmented the UDS and GOODS-S catalogs in \Ntext~by adding data in COSMOS and GOODS-N, where we make use of the NMBS and MOIRCS Deep Survey \citep{Kajisawa11} photometry. In each field, photometric redshifts, stellar masses, and rest-frame colors were estimated using the same procedures described in Section~\ref{sec:catalog}, based throughout on the BC03 models and a Salpeter IMF. S\'{e}rsic profiles were fit to the CANDELS F160W images using the same methods applied to \jkcs. Our field comparison sample consists of 225 galaxies with $M_* > 10^{10.7} \msol$ in the redshift interval $z = 1.8 \pm 0.3$ that are classified as quiescent according to their $UVJ$ colors. Galaxies within 1 Mpc of the known $z=1.62$ cluster at the edge of the UDS field (\citealt{Papovich10,Tanaka10}; see Section~\ref{sec:discussion}) were removed. For 17 galaxies in the field sample (7.5\%) and 1 of the cluster members, the S\'{e}rsic index reached the maximum value $n = 8$ allowed in our fits. Since the radii derived in such cases are often unreliable (see N12, \citealt{Raichoor12}), we indicate these galaxies separately in our plots and omit the $n=8$ field galaxies when fitting the mass--radius relation.

To validate our fitting method, we inserted hundreds of simulated galaxies with S\'{e}rsic profiles into the UDS and \jkcs~images with a distribution of parameters similar to that in our sample. We found that $n$, $R_h$, and the total flux are recovered with negligible biases, i.e., less than a few percent. The typical $1\sigma$ uncertainties in $R_h$ are $\sigma_{R_h} = 10\%$ for the majority of systems having $R_h < 0\farcs5$, increasing to 17\% for larger galaxies. In about 7\% of cases, $R_h$ differs from the true value by more than factor of 1.5. The S\'{e}rsic index $n$ is recovered with errors of $\sigma_n = 0.4$ when $n < 5$, increasing to $\sigma_n = 0.9$ for more extended profiles having $n = 5-7$. Total fluxes are recovered with a scatter of $\sigma_{\rm mag} \simeq 0.1$~mag. These estimates can be applied to the measurements in Table~\ref{tab:sersic}.

\begin{figure*}
\centering
\includegraphics[width=0.8\linewidth]{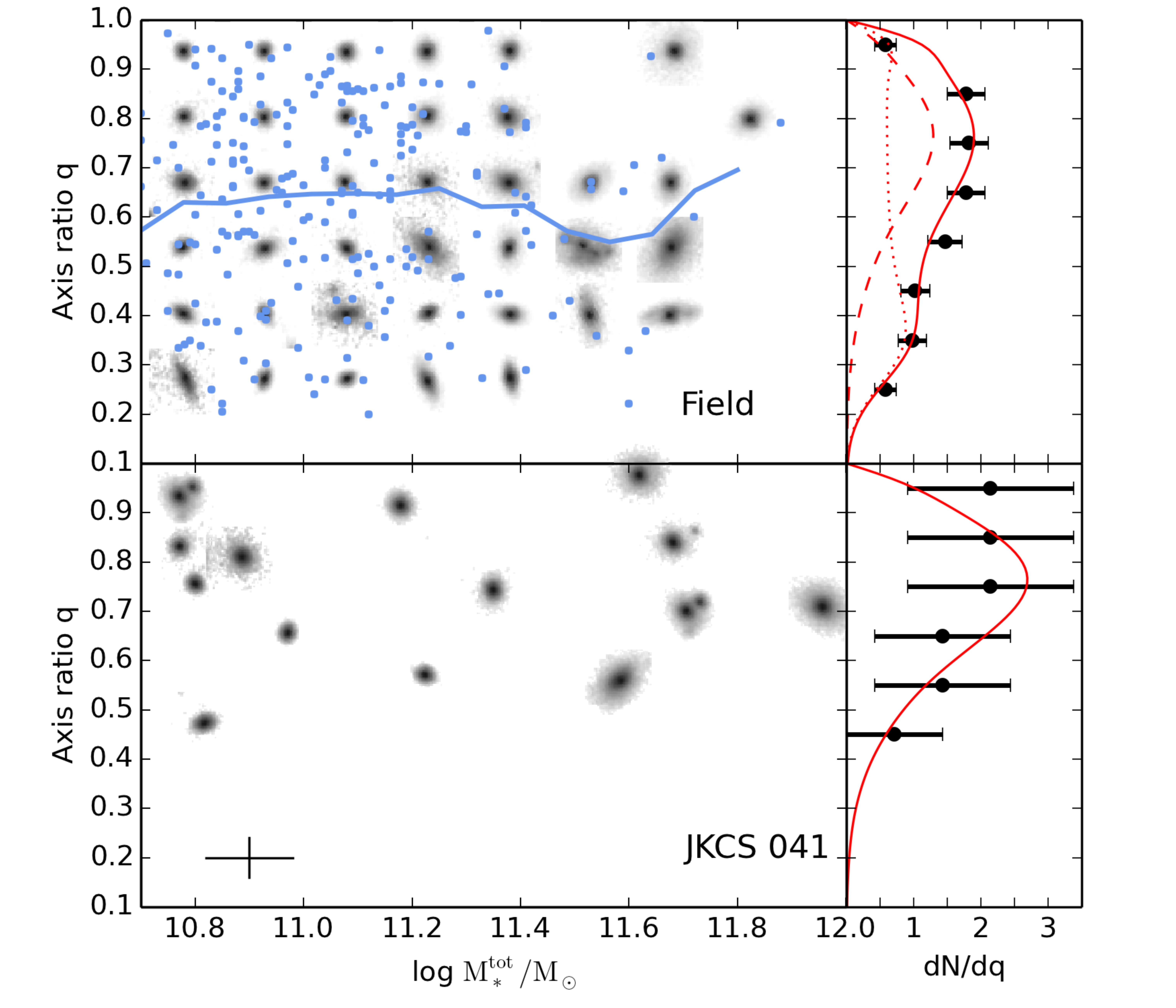}
\caption{Projected axis ratios $q$ as a function of stellar mass for the quiescent galaxies in our field sample (top panel) and in \jkcs~(bottom). In the top panel, a grid of randomly selected cutouts having the corresponding $M_*$ and $q$ is shown, with the blue points denoting the actual parameters of the field galaxies and the blue line indicating the running mean. A representative error bar in shown in the lower panel, which includes only random uncertainties in $M_*$. Histograms in the right panels show the $q$ distributions with Poisson error bars. Red curves show the best-fitting two-component model described in the text: dotted and dashed curves denote the disk-like, oblate population and the spheroid population, respectively, while solid curves show their sum. The \jkcs~members are best fit by a pure spheroid population, whereas about half of the field sample belongs to the oblate population in this model (see Figure~\ref{fig:fobl}).\label{fig:qplot}}
\end{figure*}

\begin{figure}
\centering
\includegraphics[width=\linewidth]{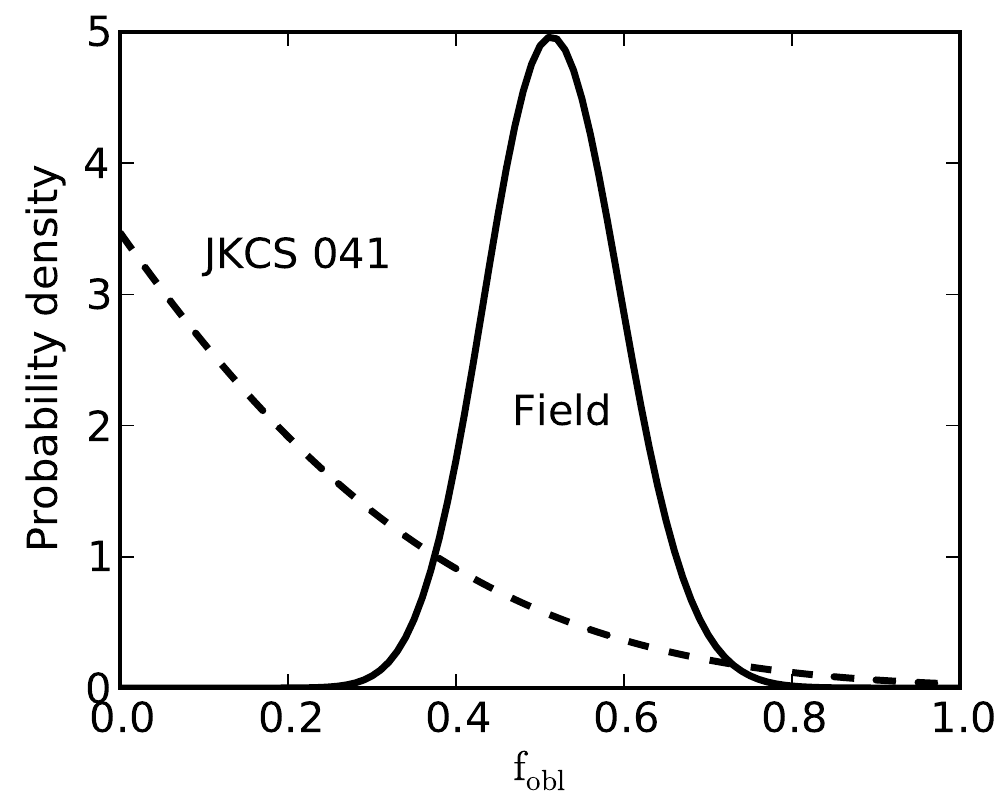} \hfill
\caption{Posterior probability density for the fraction $f_{\rm obl}$ of quiescent galaxies that belong to the disk-like, oblate population, based on the model proposed by \citet{Chang13}.\label{fig:fobl}}
\end{figure}

\subsection{Shapes of Quiescent \jkcs~Members versus the Field}
\label{sec:shapes}

We begin our structural comparison of quiescent field and cluster galaxies by considering their shapes. Figure~\ref{fig:qplot} compares the projected axis ratios $q = b/a$ of the two samples. The top panel shows that the field sample spans a wide range of shapes that extends to highly flattened systems with low $q$. This suggests that many quiescent field galaxies at $z \sim 1.8$ harbor a significant disk component. A visual inspection of images of the systems having $q \lesssim 0.5$ supports this conclusion. Other authors have noted evidence of significant disk-like structures in quiescent galaxies at $z > 1$, even at the highest stellar masses, based on both their projected axis ratio distribution \citep{vanderWel11,Weinzirl11,Buitrago13,Chang13,Chang13b} and on results from two-component bulge/disk decompositions \citep{Stockton08,McGrath08,Bruce12,Papovich12}

Turning to the \jkcs~members in the lower panel of Figure~\ref{fig:qplot}, there appear to be fewer flattened galaxies: only one, for example, has $q < 0.5$. Quantitatively, the difference in mean projected axis ratios is $\langle q_{\rm JKCS} \rangle - \langle q_{\rm field} \rangle = 0.11 \pm 0.04$, and we derive a $p$-value of 0.03 from a permutation test that indicates this difference is moderately significant.\footnote{The $p$-value is the fraction of random permutations of the field and cluster identifications for which $\langle q_{\rm JKCS} \rangle - \langle q_{\rm field} \rangle$ exceeds that which is observed in absolute value (i.e., a two-sided test).} This suggests a probable difference in the underlying morphological composition of the cluster and field galaxies.

More physical insight can be gained from the $q$ distribution using a model for the distribution of intrinsic galaxy shapes. \citet{Chang13} have shown that the $q$ distribution of quiescent galaxies can be understood as arising from a two-component population viewed at random angles. One component consists of mildly triaxial galaxies that are nearly spherical, and the other consists of a highly flattened, oblate population. In the following, we refer to these as the spheroid and disk-like components, respectively, although it should be kept in mind that the quiescent disk-like galaxies are likely composite objects containing significant bulges \citep{Bruce12} and may be related to the lenticular population at lower redshift; we note that these passive disk-like galaxies appear to span a range of S\'{e}rsic indices $n \approx 1 - 5$. This decomposition of the $q$ distribution is not unique, but it is motivated by more detailed photometric and kinematic classifications at lower redshift and serves as a useful starting point for understanding the $z > 1$ population. Chang et al.~showed that the fraction $f_{\rm obl}$ of quiescent systems belonging to the disk-like population appears to be roughly independent of mass over the range of masses and redshifts relevant for the present paper. In support of this, we see no trend in $\langle q \rangle$ with mass in Figure~\ref{fig:qplot}.

Overlaid on the histograms in Figure~\ref{fig:qplot} are fits based on this two population model.\footnote{We use the distribution of intrinsic axis ratios within the oblate and triaxial populations from the first entry in Table 3 of \citet{Chang13}.} Figure~\ref{fig:fobl} shows the inferred fraction $f_{\rm obl}$ of disk-like galaxies. We find that about half ($f_{\rm obl} = 0.52 \pm 0.08$) of the $z \sim 1.8$ field sample belongs to the disk-like population, consistent with Chang et al., whereas in \jkcs~the $q$ distribution is best fit with a pure spheroid population ($f_{\rm obl} = 0$), with $f_{\rm obl} < 0.28$ at 68\% confidence. Comparing the two samples, we find that $f_{\rm obl}$ is lower in the cluster at 90\% confidence. 

\subsection{Sizes and Radial Profiles of Quiescent \jkcs~Members versus the Field}
\label{sec:sizes}

\begin{figure*}
\centering
\includegraphics[width=0.7\linewidth]{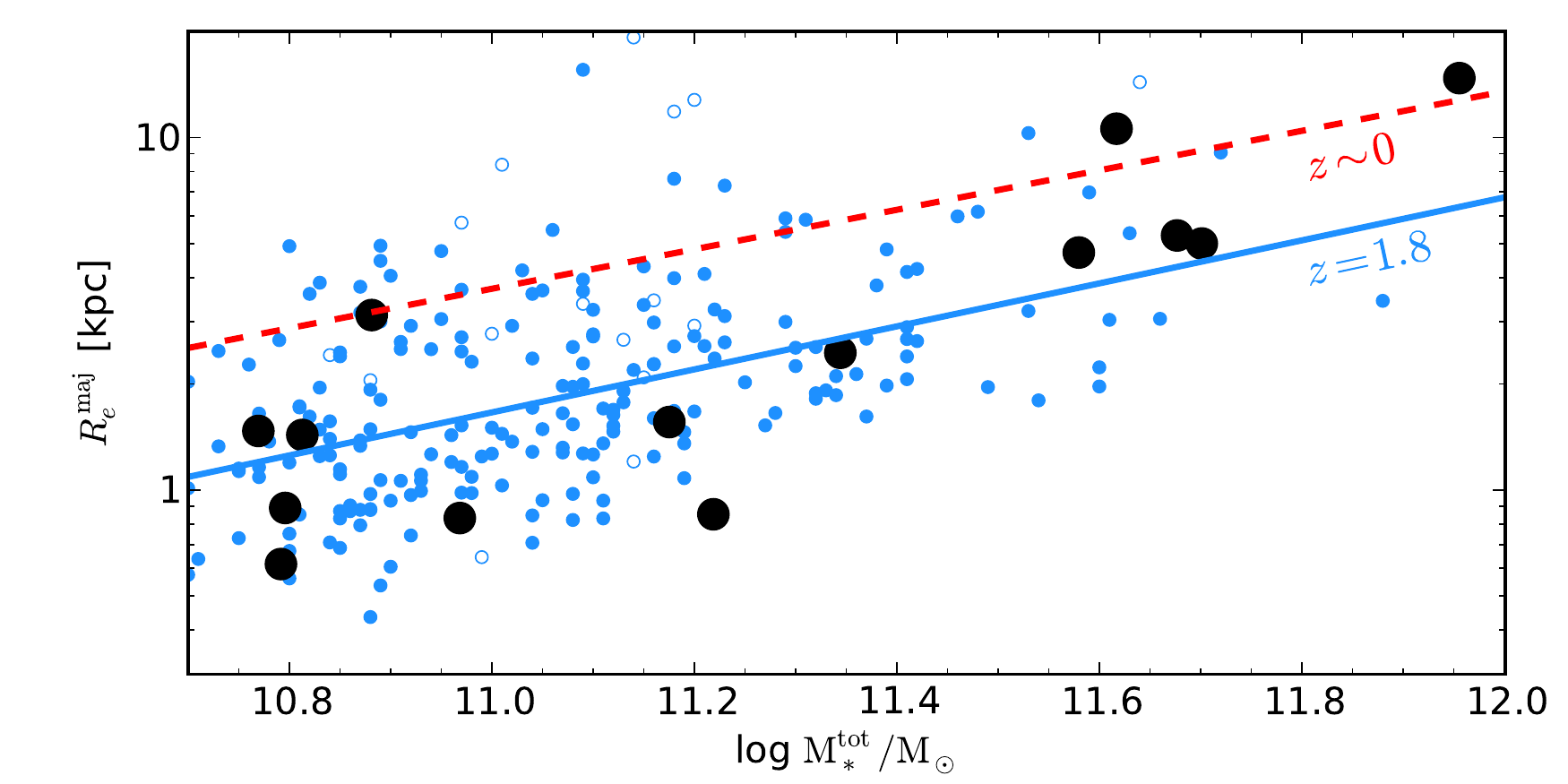} \\
\includegraphics[width=0.7\linewidth]{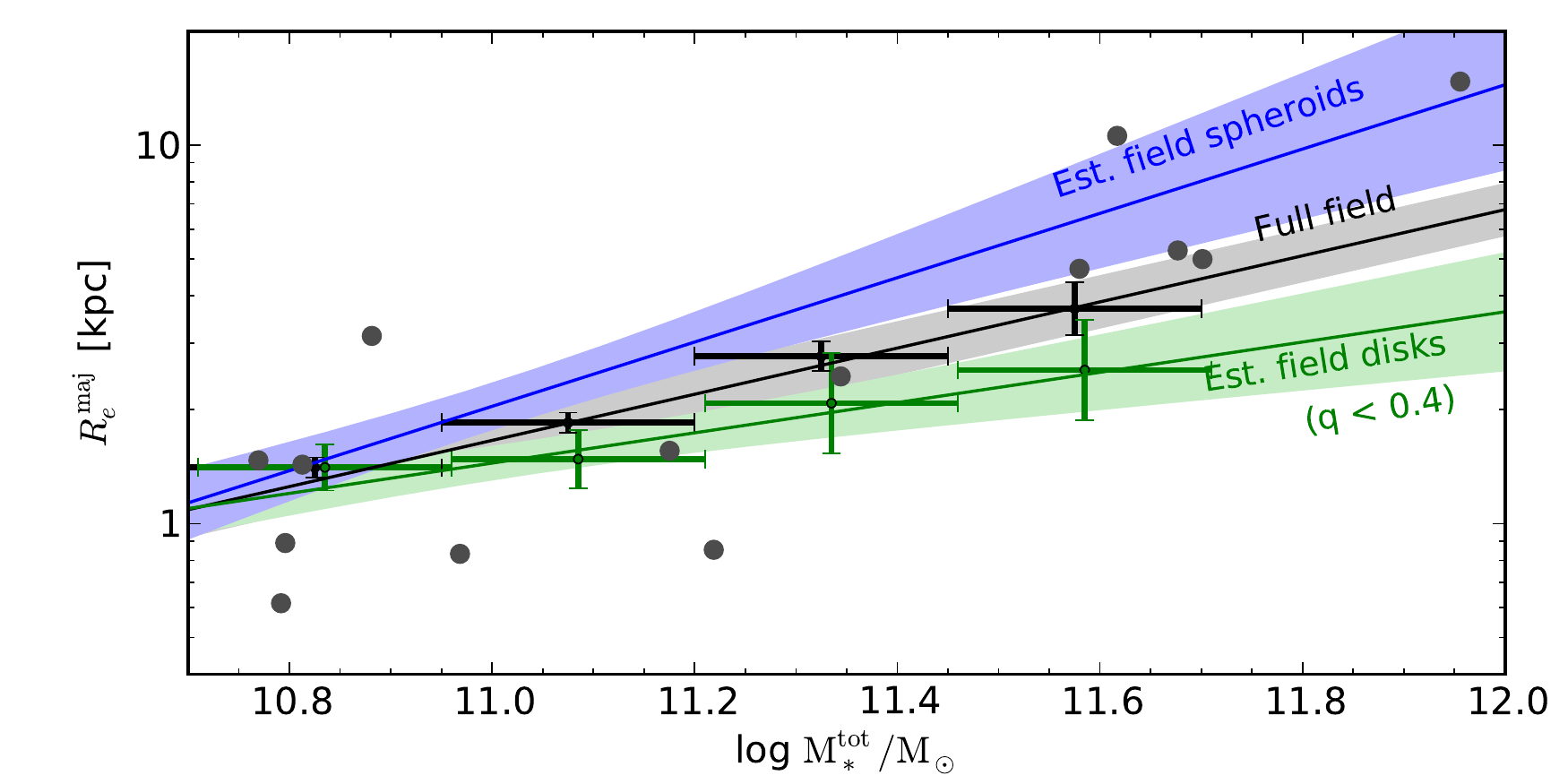}
\caption{{\bf Top:} stellar mass--$R_e^{\rm maj}$ relation for quiescent galaxies in \jkcs~(black symbols) and in our $z = 1.8 \pm 0.3$ field sample (blue). The solid line shows the field relation at $z = 1.8$ (Equation~\ref{eqn:fit_full_field}), and the dashed red line shows the $z \sim 0$ relation for early type galaxies from \citet{Shen03}, where we have converted their circularized radii to $R_e^{\rm maj}$ estimates by assuming a mean axis ratio of $\langle q \rangle \approx 0.75$ \citep[e.g.,][]{Padilla08}. Open symbols denote field galaxies best fit with $n = 8$, whose sizes may be unreliable. {\bf Bottom:} stellar mass--$R_e^{\rm maj}$ relation for our color-selected sample of quiescent field galaxies (black symbols with error bars) is compared to that defined by the subset of flattened galaxies with $q < 0.4$ (green) and to our inferred relation for the spheroid population (blue). Bands indicate $1\sigma$ uncertainties, and gray circles show the \jkcs~members as in the top panel.\label{fig:simpleMSR}}
\end{figure*}

The stellar mass--radius relations for the quiescent field galaxies and the quiescent \jkcs~members are shown in Figure~\ref{fig:simpleMSR}. As a first step toward comparing the two, we fit a linear relation with Gaussian scatter $\mathcal{N}(\sigma)$ to the field sample:
\begin{align}
\log R_e^{\rm maj} / {\rm kpc}  &= \alpha + \beta \log {\rm M}_*^{\rm tot} / 10^{11} {\rm M}_{\odot} \nonumber\\
 & - 0.26(z-1.8) + \mathcal{N}(\sigma),\label{eqn:fit_full_field}
\end{align}
where $\beta = 0.61 \pm 0.07$, $\alpha = 0.22 \pm 0.02$, and $\sigma = 0.23 \pm 0.01$. Here we have taken into account the mild redshift evolution $\partial \log R / \partial z = -0.26$ expected within field sample based on the results by N12. This fit is shown by the blue line. Comparing the \jkcs~members to the mean field relation, there is no evidence for a systematic difference between the two: $\langle \Delta \log R_e^{\rm maj} \rangle = 0.01 \pm 0.09$.\footnote{Throughout, the uncertainty in the mean $\langle \Delta \log R_e^{\rm maj} \rangle$ is estimated as $\sqrt{\sigma_{\rm clus}^2 + \sigma_{\rm field}^2}$. Here the uncertainty $\sigma_{\rm clus} = 0.23 / \sqrt{N_{\rm clus}}$ in the mean cluster galaxy offset is based on the scatter seen in the field relation (Equation~\ref{eqn:fit_full_field}), and the uncertainty $\sigma_{\rm field}$ in the mean field relation is derived from the fit parameters.} There is a hint, however, of a mass-dependent trend: the five most massive galaxies are all displaced above the mean field relation, by an average $\langle \Delta \log R_e^{\rm maj} \rangle = 0.21 \pm 0.12$. 

Since the axis ratio distribution suggests that the morphological mix of quiescent galaxies may be different in \jkcs~and the field (Section~\ref{sec:shapes}), it is important to consider what effect this may have on a comparison of sizes. If the morphological compositions indeed differ, then a simple comparison of radii --- such as that performed above --- will conflate the sizes of spheroids and disks, rather than isolating the effect of the environment on galaxies of comparable morphologies. While nearly edge-on disk-like galaxies are easily identified, it is not easy to locate the same systems viewed at lower inclination. A division in S\'{e}rsic index is not very effective, since flatted ($q \lesssim 0.4$) quiescent galaxies are seen in the field over a wide range of $n \approx 1-5$. Therefore, rather than attempting to morphologically classify the individual galaxies in the distant field sample, we proceed from the model of the underlying shape distribution discussed in Section~\ref{sec:shapes} and follow its implications for the mass--radius relation.

The lower panel of Figure~\ref{fig:simpleMSR} demonstrates that the flattened quiescent field galaxies having $q < 0.4$ (green symbols) appear to follow a different mass--radius relation: they have smaller $R_e^{\rm maj}$ than the bulk field sample (black symbols), and increasingly so at higher masses.\footnote{For a single population of triaxial objects, the smallest $q$ is seen when longest and shortest axes are in the plane of the sky, and the projected $R_e^{\rm maj}$ is maximal. The fact that small-$q$ galaxies have \emph{smaller} $R_e^{\rm maj}$ thus supports the notion that they are a distinct population with a different size distribution. We also emphasize that our discussion is confined to \emph{quiescent} galaxies, and star-forming disks are well known to have larger sizes (e.g., \citealt{Williams10}, N12, and references therein). \citet{Chang13} present evidence that highly-inclined galaxies with quiescent $UVJ$ colors are transparent and are not preferentially affected by obscured star formation (excluding the small fraction of MIPS sources does not alter the $q$ distribution).} We expect the $q < 0.4$ galaxies to be a fairly pure ($f_{\rm obl} = 0.89$, according to the decomposition in Section~\ref{sec:shapes}) but incomplete sample of the disk-like population. Since $R_e^{\rm maj}$ is independent of inclination for transparent, oblate objects, those galaxies in the disk-like population that are viewed more nearly face-on, i.e., with higher $q$, should follow the same mass--radius relation. Assuming that a fraction $f_{\rm obl} = 0.52 \pm 0.08$ of quiescent field galaxies --- of all inclinations --- belong to this disk-like population, it is then straightforward to estimate the mass--radius relation for the spheroids. Specifically, at each mass we consider the mean $\langle \log R_e^{\rm maj} \rangle$ as a weighted average: $f_{\rm obl} \langle \log R_{e,{\rm obl}}^{\rm maj} \rangle + (1 - f_{\rm obl}) \langle \log R_{e,{\rm sph}}^{\rm maj} \rangle$. 

The blue band in Figure~\ref{fig:simpleMSR} shows the resulting constraint on the relation for quiescent field spheroids. If the cluster galaxies are indeed dominated by spheroids, as suggested by their axis ratio distribution, it is clear that any difference between the field and cluster relations at high masses is much reduced. Quantitatively, 
the sizes of the five most massive cluster members do not differ systematically ($\langle \Delta \log R_e^{\rm maj} \rangle = -0.06 \pm 0.19$) from the field spheroid relation, although the uncertainties are necessarily increased, and when considering the full range of masses, the cluster members are slightly smaller but still consistent with the field spheroids ($\langle \Delta \log R_e^{\rm maj} \rangle = -0.14 \pm 0.10$). We regard our morphological separation of the mass--radius relation of quiescent galaxies as a first approximation, since it relies on a very simple model for the underlying distribution of shapes (Section~\ref{sec:shapes}; \citealt{Chang13}) and its apparent invariance with mass at $z \sim 2$. More data is needed to test this model and its implication that the fraction of massive, quiescent galaxies with significant disk components increases with redshift. However, it is clear that a difference in the morphological mixtures of the field and cluster samples could significantly affect comparisons of their mass--radius relations.

\begin{figure}
\centering
\includegraphics[width=\linewidth]{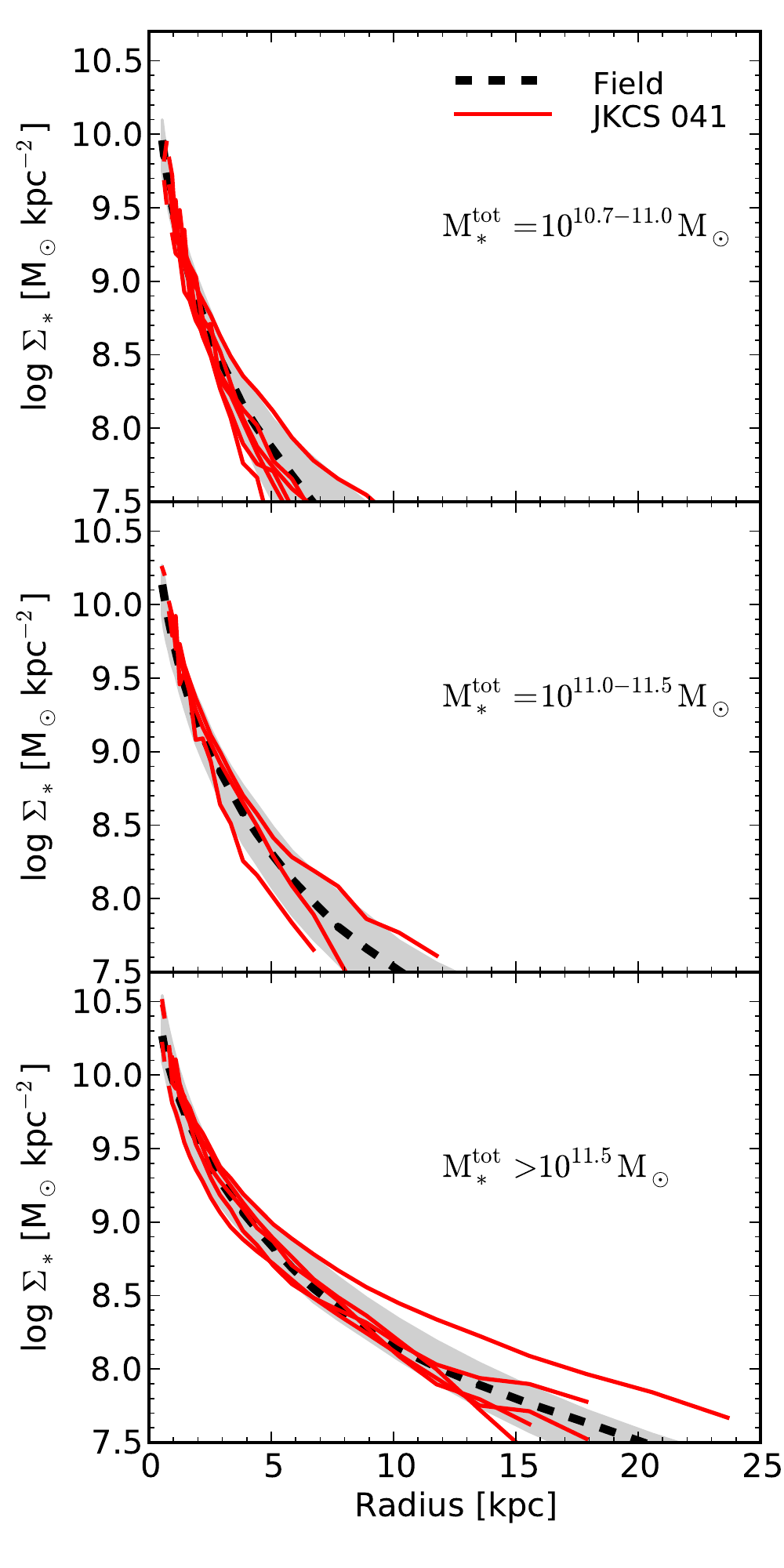}
\caption{Azimuthally averaged surface mass density $\Sigma_*$ profiles of \jkcs~members (red lines), plotted down to a limiting surface brightness of $H_{160} = 26$~mag~arcsec${}^{-2}$ and PSF-deconvolved as described in the text. In each of three stellar mass bins, we compare to the population of quiescent field galaxies at $z\sim1.8$ that have $q > 0.45$, excluding highly flattened galaxies that are absent in the cluster sample. The thick dashed line shows median surface density profile of the field sample derived from our S\'{e}rsic fits, and the gray region encloses 68\% of the field profiles at each radius.\label{fig:light_profiles}}
\end{figure}

In summary, there is no significant difference overall between the mass--radius relation defined by the quiescent \jkcs~members and that defined by our coeval field sample. There is a weak hint of a mass-dependent trend in which the most massive cluster members are offset to larger radii, if all color-selected quiescent galaxies are considered irrespective of morphology. However, a closer inspection reveals that this may arise because the cluster population is richer in spheroids, and spheroids are ``larger'' than quiescent disk-like galaxies. Figure~\ref{fig:light_profiles} supports this conclusion via a direct comparison the surface mass density profiles of the \jkcs~members to the field galaxies. Here we consider only field galaxies with $q > 0.45$ to better match the cluster sample. The \emph{HST} PSF was deconvolved from the observed F160W light profile using the technique proposed by \citet{Szomoru10}, and the resulting light profile was  converted to a stellar mass profile using a constant $M_*/L$ for each galaxy. There is no clear difference in the mass profiles in the field and \jkcs~samples. 

\begin{figure}
\includegraphics[width=\linewidth]{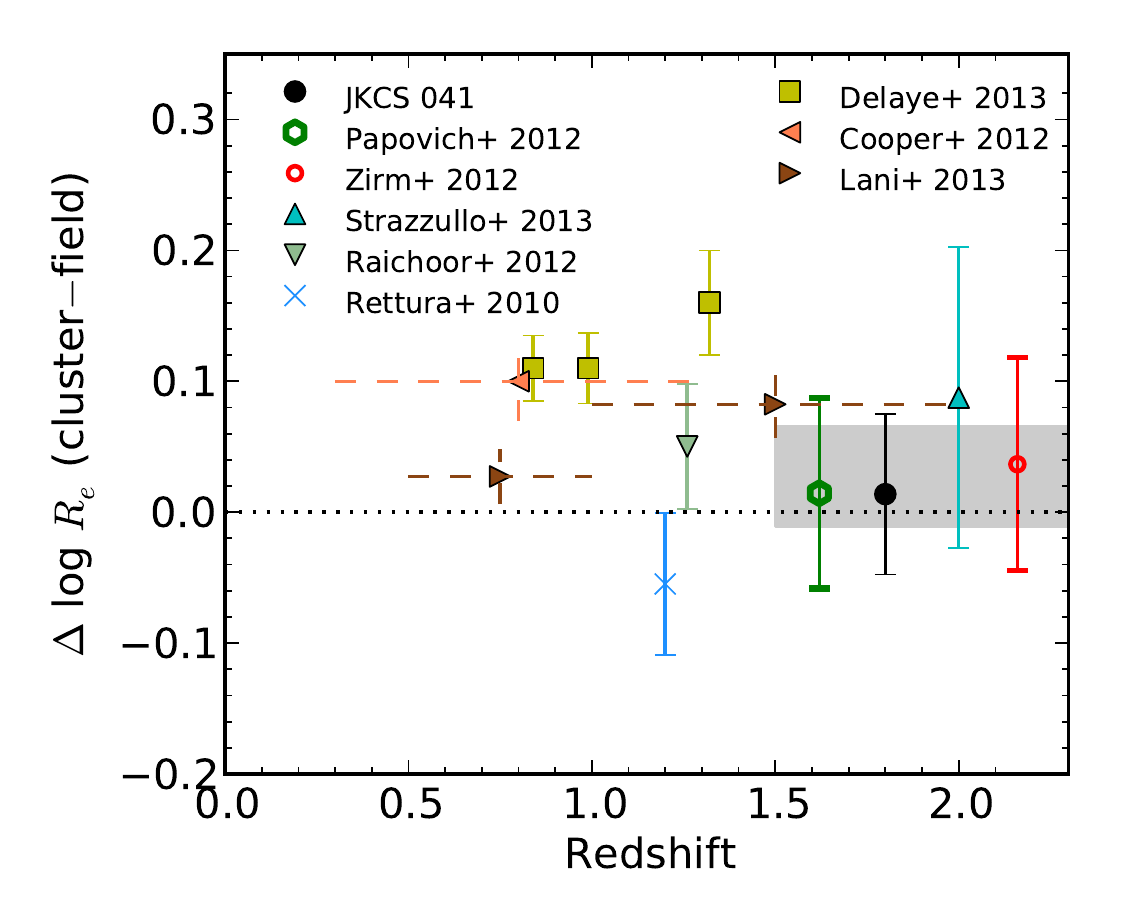}
\caption{Comparison of published results on the environmental dependence of the mass--radius relation of quiescent galaxies. Each point represents the mean offset $\Delta \log R_e$ from the field relation. For studies of individual clusters, listed in the upper-left legend, we compare to Equation~\ref{eqn:fit_full_field}. For ensembles of clusters \citep{Delaye14} and studies of group-scale overdensities \citep[][dashed error bars]{Cooper12,Lani13}, the published offsets from the authors' field relation are quoted directly. The shaded band denotes the weighted mean of the $z > 1.6$ clusters and its $1\sigma$ uncertainty. Appendix~C describes our method for compiling and harmonizing these diverse data sets and describes systematic uncertainties inherent in such a comparison.\label{fig:othersizestudies}}
\end{figure}

\subsection{Comparison to other studies of the environmental dependence of the mass--radius relation}
\label{sec:literatureMSR}

Several recent studies of the environmental dependence of galaxy sizes at high redshifts are compared in Figure~\ref{fig:othersizestudies}. The references in the upper-left legend refer to individual clusters, for which we have compiled published structural measurements of their quiescent or early type members. In Section~\ref{sec:discussion} we review the bulk physical properties of the $z > 1.6$ clusters themselves; our focus here on the mass--radius relation. To synthesize these published results into a quantity that can be compared as directly as possible, given the diversity of samples and methods (see Appendix~C for details), we compute the mean offset $\langle \Delta \log R_e^{\rm maj} \rangle$ between the quiescent members of each cluster and the field relation in Equation~\ref{eqn:fit_full_field}.  We regard Figure~\ref{fig:othersizestudies} as a first step toward synthesizing results from various high-$z$ studies, but caution that systematic differences in measurement techniques may affect a comparison of our field sample with other authors' cluster data; some of these are discussed in Appendix~C.

Considering the $z > 1.6$ clusters first, \citet[][see also \citealt{Bassett13}]{Papovich12}, \citet{Zirm12} and \citet{Strazzullo13} have all remarked on evidence for larger sizes among the quiescent members of the clusters they studied. (We note that many of these members are actually photometric candidates, whereas the members of \jkcs~are confirmed by grism redshifts.)  Based on Figure~\ref{fig:othersizestudies}, we regard the present evidence for a variation in the mass--size relation in the cores of these most distant clusters and proto-clusters as very marginal. On the other hand, present sample sizes are too small to rule out a modest size enhancement of $\sim 0.05$~dex. 
Moving to lower redshifts, \citet{Delaye14} studied 9 clusters at $z = 0.8-1.4$ along with a field sample selected and analyzed in a homogeneous way. They found significant evidence for an offset in the mass--radius relation by $\Delta \log R_e \simeq 0.1$~dex. In the two $z \sim 1.2$ clusters studied by \citet{Rettura10} and \citet{Raichoor12}, however, we find no significant offset.

\citet{Bassett13} noticed that the slight trend for the quiescent candidate members of the cluster they studied (IRC-0218A, $z=1.62$) to have larger $R_e$ and smaller $n$ was mostly driven by a population of disk galaxies located at large cluster-centric radii $R \approx 1-1.5$~Mpc.\footnote{As described in Appendix~C, we include only the members of this cluster within $R < 1$~Mpc in Figure~\ref{fig:othersizestudies} for a better comparison with other data sets.} Although their remark that differences in morphology can influence comparisons of the mass--radius relation is similar to our findings, we note that that the nearly pure disks they discuss ($n \sim 1$) have \emph{larger} $R_e$ than the mean quiescent galaxy --- consistent with faded spirals that have been starved of gas during infall --- whereas the disk-like quiescent field population discussed in Section~\ref{sec:sizes} is offset to \emph{smaller} $R_e$ and exhibits a wide range of $n$ indicating a significant build-up of bulges (see a similar trend in \citealt{HuertasCompany13}). Altogether, this points to a complex mixture of morphologies varying from the field to the cluster outskirts and core. 

In addition to these cluster studies, two recent studies have examined the dependence of the mass--radius relation on local density in blank field surveys, where the densest regions are typically groups or low-mass clusters. These results are distinguished with dashed error bars in Figure~\ref{fig:othersizestudies}. \citet{Cooper12} found a size enhancement of $\Delta \log R_e \simeq 0.1$~dex among early type galaxies in the densest regions in the DEEP3 survey fields. In the UDS field, \citet{Lani13} detected a similar enhancement that was dominated by the most massive and highest-redshift galaxies. This is the regime where we found that differences in the morphological mix could affect our interpretation of \jkcs. Lani et al.~considered such a possibility and tested it by cutting their sample in S\'{e}rsic index $n$. Although this is a reasonable first approach, we find the connection between the oblate, disk-like quiescent population and S\'{e}rsic index to be loose (Section~\ref{sec:sizes}). Additionally, while the $M_*$--$R_e^{\rm maj}$ relation likely varies with $q$ (Figure~\ref{fig:simpleMSR}), we find no such dependence on $n$ for quiescent galaxies. In future work, it would be useful to consider the $q$ distributions of samples whose mass--radius relations are being compared.\footnote{Interestingly, further testing by C.~Lani et al. (2013, private communication) following the submission of this paper has shown that their results are not affected by an axis ratio cut of $q > 0.4$.}

In contrast to these $z \gtrsim 1$ studies, there appears to be \emph{no} dependence at $z \sim 0$ of the size of early type galaxies on local density, halo mass, or position within the halo \citep{Weinmann09,Guo09,Nair10,HuertasCompany13local}. These $z \sim 0$ results, however, have been challenged by \citet{Valentinuzzi10}, who claim an excess of \emph{compact} massive galaxies in local clusters; interestingly, these compact galaxies show a tendency to have S0 morphologies. The only clear point of agreement is that the BCGs in very massive clusters are exceptionally large \citep[e.g.,][]{Bernardi07}. 

In summary, the evidence for environmental variation in the mass--radius relation in the most distant $z > 1.6$ clusters is still limited by small samples. At $z \sim 1$ there is good evidence for an offset to larger sizes in the cluster sample studied by \citet{Delaye14}, as well as in group-scale overdensities \citep{Cooper12,Lani13}. At $z \sim 0$, most evidence points toward a remarkable independence of early type galaxy structure on environment. There are contrary indications for many secondary trends that might shed light on an underlying physical picture: are galaxy sizes enhanced primarily in distant clusters' cores (Delaye et al.) or their outskirts \citep{Bassett13}? Is the enhancement stronger for higher (Lani et al.) or lower mass (Delaye et al.) galaxies? Furthermore, the evolutionary connection between $z \gtrsim 1$ results and the precise constraints available at $z \sim 0$ remains unclear.

\section{Discussion}
\label{sec:discussion}
 
\begin{deluxetable*}{lccccl}
\tablewidth{\linewidth}
\tablecaption{\jkcs~Compared to Other Spectroscopically Verified $z > 1.6$ Proto-clusters and Clusters} 
\tablehead{\colhead{Cluster} & \colhead{$z$} & \colhead{Mass $M_{200}$} & \colhead{Diffuse X-ray flux} & \colhead{$N_{\rm spec}$ /} & \colhead{References} \\
\colhead{} & \colhead{} & \colhead{$(\msol)$} & \colhead{(erg cm${}^{-2}$ s${}^{-1}$)} & \colhead{$N_{\rm spec~Q}$} & \colhead{}}
\startdata
JKCS~041${}^{\dagger}$ & 1.80 & $(2-3) \times 10^{14}$ & $2 \times 10^{-14}$ & 19 / 15 & This work, \citet{Andreon13} \\
IRC-0218A${}^*$ & 1.62 & $(2-7) \times 10^{13}$ & $\sim 3 \times 10^{-15}$ & 11 / 3 & P10, T10, P12, B13, L13, Pi12 \\
SpARCS J022427-032354 & 1.63 & \ldots & \ldots & 12 / 3 & \citet{Muzzin13} \\
IDCS J1426+3508${}^{\dagger}$ & 1.75 & $4 \times 10^{14}$ & $3 \times 10^{-14}$ & 7 / 2 & S12, B12 \\
IDCS J1433.2+3306${}^{\dagger}$ & 1.89 & $\sim 10^{14}$ & \ldots & 7 / 2 & B07, Z12 \\
Cl J1449+0856${}^{\dagger}$ & 2.00 & $5 \times 10^{13}$ & $9 \times 10^{-16}$ & 22 / 7 & G11, G13, S13 \\
MRC 0156-252 & 2.02 & \ldots & $\sim 2 \times 10^{-15\ddag}$ & 10 / 1 & O05, Ga13 \\
MRC 1138-262 & 2.16 & \ldots & \ldots & 11 / 4 & Zi12, T13, and references therein
\enddata
\tablecomments{$N_{\rm spec}$ is the number of spectroscopic members, of which $N_{\rm spec~Q}$ are quiescent. Masses and X-ray fluxes are only indicative, since various energy bands, apertures, and scaling relations are used. References: P10, P12: \citet{Papovich10,Papovich12}, Pi12: \citet{Pierre12}, S12: \citet{Stanford12}, B12: \citet{Brodwin12}, B07: \citet{Brodwin07}, Z12: \citet{Zeimann12}, G11, G13: \citet{Gobat11,Gobat13}, S13: \citet{Strazzullo13}, T10, T13: \citet{Tanaka10,Tanaka13}, B13: \citet{Bassett13}, L13: \citet{Lotz13}, Zi12: \citet{Zirm12}, O05: \citet{Overzier05}, Ga13: \citet{Galametz13}. ${}^{\dagger}$Based on WFC3 grism data.\label{tab:otherclusters} ${}^*$Also called XMM-LSS J02182-05102. ${}^{\ddag}$The X-ray emission is suspected to be associated with the radio galaxy rather than thermal ICM emission.}
\end{deluxetable*}
 
In addition to \jkcs, seven overdensities containing a red galaxy population have been identified at $z>1.6$ and confirmed spectroscopically.\footnote{In addition to these, we note that \citet{Spitler13} recently discovered a $z = 2.2$ cluster candidate containing a red galaxy population using medium-band photometric redshifts.} Although all have been labeled ``clusters'' or ``proto-clusters,'' these are in fact a diverse set of structures that span a wide range of masses and evolutionary states. The properties of these systems are summarized in Table~\ref{tab:otherclusters}. 

\jkcs~is remarkable in several ways. First, it appears to be a fairly massive cluster for its redshift. As \citet{Andreon13} describe, the X-ray luminosity, X-ray temperature, and galaxy richness give mass estimates of $\log M_{200} \simeq 14.2 - 14.5$ that are reasonably consistent given the uncertainties in the evolution of the relevant scaling relations. \citet{Culverhouse10} report the non-detection of an SZ signal in the direction of \jkcs, corresponding to an upper limit of $\log M_{200} \lesssim 14.5$. Therefore, given the depth of the observation, this non-detection is still consistent with the range of independent X-ray-- and richness--based estimates.\footnote{Here we use the \citet{Bonamente08} scaling relation between $Y_{2500}$ and $M_{2500}$ to estimate $\log M_{2500} < 13.7$, which corresponds to $\log M_{200} < 14.5$ assuming the  \citet{Duffy08} mass--concentration relation.} While deeper SZ observations of \jkcs~will be very valuable, we conclude that all present data are consistent with a mass in the range $M_{200} \simeq (2-3) \times 10^{14} \msol$. Compared to the other $z > 1.6$ clusters in Table~\ref{tab:otherclusters} with estimated masses, \jkcs~appears to be the most massive other than IDCS~J1426+3508, which is possibly more massive by a factor $\sim 1.5 - 2$.

Second, we have been able to confirm a large number of member galaxies via grism redshifts (see Table~\ref{tab:otherclusters}), especially those that are quiescent. This has allowed us to construct a spectroscopic sample that is fairly complete at radii $R < R_{500}$ and masses $M_* > 10^{10.6} \msol$, and is thus suitable for studying environmental effects on the member galaxies. We emphasize that comparing numbers of spectroscopic members is not the same as comparing the underlying galaxy populations, given the diversity of observations and analysis methods used in Table~\ref{tab:otherclusters}. However, the bright end of the red sequence is quite rich in \jkcs.  For a detailed comparison of its red sequence morphology with those of other high-redshift structures, we refer to \citet{Andreon13}.

Motivated by our unique data on the quiescent population in \jkcs, we have compared their structural and stellar population properties to coeval field samples. Considering first the structure and morphology of the cluster members, we found some evidence for a lack of quiescent disk-like galaxies relative to the field population. In the context of cluster studies at lower redshift, this is consistent with the idea that the cluster ellipticals are formed early ($z > 2$) in dissipative mergers, probably continuing to evolve via dry mergers, whereas many S0's are formed much later at $z \lesssim 0.5$ and decline in numbers toward higher redshifts \citep[e.g.,][]{Dressler97,Andreon97,Smith05,Postman05,Poggianti09}. 
An interesting related development is the observation that the fraction of quiescent galaxies in the field with disk-like components appears to \emph{increase} at $z > 1$, particular among massive ($M_* > 10^{11} \msol$) systems (see references in Section~\ref{sec:shapes}). The relative lack of these compact, disk-like quiescent galaxies in \jkcs~suggests that the cluster environment either inhibits their formation or else is effective in destroying the more loosely bound disk material, perhaps through tidal stripping or galaxy--galaxy encounters that build up the bulge. Larger samples of distant clusters in a range of evolutionary stages are needed to verify this trend.

Comparing the radial profiles of the cluster members to their field counterparts, we detect no statistically significant differences overall, but found a hint of a trend for larger effective radii among the most massive cluster members. One interpretation, which has been promoted in studies of other $z > 1.6$ clusters and proto-clusters \citep{Zirm12,Papovich12,Bassett13,Lotz13}, is that size growth proceeds at an accelerated rate in the cluster environment, perhaps due to a higher rate of mergers or a higher fraction that are dry. We cannot rule out this possibility, but we note that present constraints in these most distant clusters remain statistically weak (Section~\ref{sec:literatureMSR}). Furthermore, in the case of \jkcs, we found that a difference in the morphological mixture of color-selected quiescent galaxies relative to the field may account for our observations just as well. Although this explanation also points toward environment-dependent evolution, it suggests a more nuanced picture in which bulge growth and morphological transformation may play a role in shaping the mass--radius relation in clusters, and not only a pure acceleration of ``inside-out'' spheroid growth.

A weak environmental dependence of size among quiescent galaxies of the same mass and morphology would  indicate that either the galaxy merger rate does not vary substantially among the environments sampled, or that the rate of size growth is decoupled from the merger activity. This would be surprising given that mergers are thought to be the prime driver of spheroid growth (see \S1). Presently, however, it is not clear how to connect observations of the mass--radius relation in clusters at different redshifts into an evolutionary sequence. As discussed in Section~\ref{sec:literatureMSR}, results at $z \gtrsim 1.5$ are not conclusive, the $z \sim 1$ study with the most statistical power \citep{Delaye14} indicates that cluster members are enlarged by $\Delta \log R_e \approx 0.1$~dex, while at $z \sim 0$ there seems to be no relation between the structure of early type galaxies and their local environment or halo mass. One possibility is that cluster members experience an initially enhanced rate of galaxy--galaxy encounters and mergers during infall, as the cluster is forming, while the virialization of the cluster and the resulting high velocity dispersion then inhibits future merging (see, e.g., \citealt{Lotz13} and \citealt{Delaye14}). In this picture, the mass--radius relation of cluster members is offset to larger $R_e$ at high redshift, while at later times the field galaxies ``catch up'' and this offset declines. It will be interesting to test this hypothesis as larger samples of distant clusters and richer data sets become available.

While high-$z$ studies have used local density or cluster membership to quantify the environment, a galaxy's status as central galaxy in its dark matter halo may be more physically relevant. Central galaxies are expected grow more rapidly than satellites in some models, and they benefit from the accretion of stars that are tidally stripped from disrupted sinking satellites \citep[e.g.,][]{Shankar13}. This process of ``cannibalism'' becomes increasingly important in higher halo mass, with the giant BCGs being the most extreme examples. The BCG of \jkcs~indeed has the most extended light profile of all the cluster members, and it is the third nearest of the spectroscopic members to the cluster center. The BCG appears similar to that of the massive \citet{Stanford12} cluster at $z=1.75$, which is also exceptionally luminous and extended ($R_e = 18$~kpc).

A complementary approach is to quantify the rate of galaxy interactions and mergers more directly. \citet{Lotz13} indeed inferred a high ongoing merger rate --- exceeding that in the field by a factor of 3--10 --- in IRC-0218A at $z=1.62$, based on their estimation that $57^{+13}_{-14}\%$ of the massive proto-cluster members have double nuclei or a close satellite galaxy. By visual inspection of the 17 spectroscopic members of \jkcs~in our continuum-selected sample (Figure~\ref{fig:galfits}), we find that 3, i.e., $18^{+12}_{-6}\%$, have close companions within the same search radius used by Lotz et al.~(20 kpc comoving).\footnote{These are IDs 376 and 375, which are paired with one another and a faint, diffuse blue system (see Figure~\ref{fig:galfits}), and ID 286.} Although a full analysis would require accounting for projected pairs in the cluster, this suggests a lower rate of ongoing mergers in \jkcs, consistent with the latter being in a more dynamically evolved state.

Turning to the stellar populations of the galaxies in \jkcs, we found a high fraction of quenched systems compared to coeval field galaxies of the same mass (Figure~\ref{fig:fq}). Elevated quiescent fractions $f_Q$, indicating the early onset of a star-formation--density relation, have been reported in the cores of other $z > 1.6$ clusters  \citep{Quadri12,Strazzullo13}. When comparing our results with others, it is important to bear in mind several factors. First, some studies have emphasized the presence of galaxies in $z \gtrsim 1.4$ cluster cores that have unusually high levels of star formation compared with cluster galaxies at lower redshift \citep{Hilton10,Fassbender11}. While we also have located two massive galaxies with SFRs~$\sim 140 \msol$~yr${}^{-1}$ (Section~\ref{sec:galaxyprops}) in the core of \jkcs, we emphasize that they still represent a lower fraction of the galaxy population than in the field. Second, our grism-based study is confined to relatively massive galaxies in the cluster core ($M_* > 10^{10.6} \msol$, $R < R_{500} \approx 500$~kpc). Measurements of $f_Q$ that extend to lower stellar masses and larger cluster-centric radii are expected to be lower. Finally, there is likely a significant variation in $f_Q$ from cluster to cluster \citep[e.g.,][]{Brodwin13}, and the color-based selection method used to discover \jkcs~may prefer higher-$f_Q$ clusters relative to a cluster mass-limited sample. What we have clearly shown is that the cluster core environment does affect the fraction of massive galaxies that are quenched by $z=1.8$ in at least some clusters.

When considering the physical processes responsible for truncating star formation, it is common to distinguish internal quenching mechanisms (often referred to as mass- or self-quenching) from environmentally related processes that correlate with the local density or the position of a galaxy within its halo \citep[e.g.,][]{Peng10}. The clear signature of the environment on star formation activity in \jkcs~at $z = 1.8$ implies that truncation by cluster processes has been fairly rapid, since the galaxies must have fallen into the cluster fairly recently (see also \citealt{Quadri12}). Some semi-analytic models in fact predict the disappearance of environmental quenching beyond $z \gtrsim 1.5$ \citep{McGee09}, when the $\sim 2$~Gyr timescale for stripping of hot halo gas (``strangulation'') exceeds the time for which the necessary dense ICM has existed. Observations of a star formation--density relation at earlier epochs suggests that more rapid quenching mechanisms may be at work, such as ram-pressure stripping.

Although roughly half of the spectroscopic members of \jkcs~have been quenched by environmentally related processes (Section~\ref{sec:fq}), we nonetheless found that the mean ages of these galaxies does not differ greatly from similarly selected samples in the field. This indicates that the quenching mechanism had no large effect on \emph{when} truncation occurred. 
This finding is consistent with the idea that the environment modulates the \emph{fraction} of quiescent systems without much affecting their ages. Evidence at lower redshift for a null or weak ($\lesssim 0.4$~Gyr) environmental dependence of age among quiescent systems comes from studies of spectroscopic age diagnostics \citep{Thomas10,Moresco10,Muzzin12} and spectral energy distributions \citep{Andreon96,Raichoor11} at $z \simeq 0 - 1.2$, as well as from the evolution of the fundamental plane in clusters and the field at $z < 1.3$ \citep{vanDokkum07}. Our study extends earlier work by probing cluster galaxy ages through spectral diagnostics close to the epoch of their star formation and comparing these to similar observations of coeval field systems.

There are no AGN members with bright optical line emission in the core of \jkcs, as are present in several other $z > 1.6$ clusters \citep[e.g.,][]{Stanford12,Zeimann12,Gobat13}. Much fainter line emission can be reached in our composite spectra. Interestingly, there is no sign of the centrally concentrated, faint emission in H$\beta$ and [\ion{O}{3}] that was seen by \citet{Whitaker13} in their composite spectra of quiescent field galaxies. Equally strong line emission would have been detected in our stack of $M_* > 10^{11} \msol$ cluster members. If the field emission traces star formation, this finding would indicate that the dead cluster members lack the residual nuclear star formation present in field samples. Whitaker et al.~suggest that a LINER-type spectrum is more likely, given their estimate of the [\ion{O}{3}]/H$\beta$ line ratio and the line luminosity. At $z \sim 0$ the prevalence of faint [\ion{O}{3}] emission does not decrease in denser environments \citep{Kauffmann04}, so such a trend at $z \sim 2$ would be intriguing if verified in other clusters.

\section{Summary}
Based on our \emph{HST} WFC3 imaging and grism observations of \jkcs, along with associated multi-wavelength data, we conclude:
\begin{enumerate}
\item \jkcs~is a genuine rich, X-ray luminous cluster at $z = 1.80$, confirmed through the spectroscopic identification of 19 members that are spatially aligned with diffuse X-ray emission. The spectroscopic members include 15 quiescent galaxies, the largest number yet confirmed in any $z > 1.6$ cluster. Five of these are very massive galaxies having $M_*^{\rm tot} = 10^{11.6-12} \msol$.

\item High-quality composite grism spectra of the quiescent cluster members allow us to measure their stellar ages via the strengths of the H$\delta$, H$\gamma$, H$\beta$ and Mg~\emph{b} absorption lines. Less massive quiescent members with $M_* < 10^{11} \msol$ have mean luminosity-weighted ages of $0.9^{+0.2}_{-0.1}$ Gyr, while more massive galaxies are older ($1.4^{+0.3}_{-0.2}$ Gyr).

\item Comparing the spectra of the quiescent cluster members to those of similarly-selected field galaxies studied by \citet{Whitaker13}, we find that the field and cluster samples span a very similar range of ages. At the same time, the fraction of quenched galaxies at fixed stellar mass is much higher in \jkcs. This implies that the cluster environment is responsible for quenching of a substantial fraction of massive galaxies in \jkcs, but that the mode of quenching (environmental versus internal) does not have a large effect on \emph{when} star formation is truncated within the $\sim 0.3-0.5$~Gyr uncertainties in our comparison.

\item The centrally concentrated H$\beta$ and [\ion{O}{3}] emission seen by Whitaker et al.~in median spectra of quiescent field galaxies is absent in the \jkcs~members, at least among the more massive galaxies ($M_* > 10^{11} \msol$) where the high quality of the grism spectra permit a detailed comparison.

\item Comparing the quiescent members of \jkcs~to a large sample of coeval field galaxies, we find that the distribution of projected axis ratios suggests a lower fraction of disk-like systems among quiescent galaxies in the cluster.

\vspace{-0.2cm}
\item We find no statistically significant difference in the mass--radius relation or in the radial mass profiles of the quiescent cluster members compared to their field counterparts. While the most massive cluster members ($M_* > 10^{11.5} {\rm M}_{\odot}$) are marginally offset from the field mass--radius relation when considering all quiescent systems together, this apparent difference is weakened when the samples are better matched in morphology. Larger samples are still needed to clarify the structure of galaxies in distant, forming clusters, as well as to connect these results to studies at lower redshift.

\end{enumerate}

\acknowledgments
We thank the referee for a detailed report. It is a pleasure to acknowledge insightful conversations with Marc Huertas-Company, John Mulchaey, and Sirio Belli. We also thank Kate Whitaker, Rik Williams and the CSI team, Alessandro Rettura, Andrew Zirm, and Casey Papovich for sharing their data in an electronic format, as well as Nor Pirzkal and Beth Perriello for their assistance in planning and executing the \emph{HST} observations.  
Based on observations made with the NASA/ESA \emph{Hubble Space Telescope}, obtained at the Space Telescope Science Institute, which is operated by the Association of Universities for Research in Astronomy, Inc., under NASA contract NAS 5-26555. These observations are associated with program number GO-12927, which was supported under NASA contract NAS 5-26555. A.R.~acknowledges financial contribution from the agreement ASI-INAF I/009/10/0 and from Osservatorio Astronomico di Brera. Based on observations obtained with MegaPrime/MegaCam, a joint project of CFHT and CEA/IRFU, at the Canada--France--Hawaii Telescope (CFHT) which is operated by the National Research Council (NRC) of Canada, the Institut National des Science de l'Univers of the Centre National de la Recherche Scientifique (CNRS) of France, and the University of Hawaii. This work is based in part on data products produced at Terapix available at the Canadian Astronomy Data Centre as part of the Canada--France--Hawaii Telescope Legacy Survey, a collaborative project of NRC and CNRS.

\begin{appendix}

\section{Grism Redshift Catalog}

Table~4 lists the redshifts derived for the 98 galaxies described in Section~\ref{sec:z}. For emission line sources, we assign a quality flag `A' when more than one line is visible and `B' otherwise. For continuum sources, we qualitatively assign a quality flag based on the appearance of the spectrum and the posterior probability distribution $P(z)$. Spectra with a weak or absent continuum break, often with a multimodal $P(z)$, carry a `C' flag. The `B' flag corresponds to a more clearly detected continuum break; we expect the vast majority of these redshifts to be reliable. The `A' flag is reserved for the highest signal-to-noise objects with unambiguous continuum breaks and, in some cases, absorption lines.

\LongTables
\begin{deluxetable*}{lllccccc}
\tablewidth{0.6\linewidth}
\tablecaption{Grism Redshifts}
\tablehead{\colhead{ID} & \colhead{R.A.} & \colhead{Dec.} & \colhead{$H_{160}$} & \colhead{$z_{\rm grism}$} & \colhead{Type} & \colhead{Quality}}
\startdata
220 & 36.695309 & $-4.687007$ & 19.08 & $0.285 \pm 0.005$ & E & B \\
167 & 36.694981 & $-4.685004$ & 20.15 & $1.064 \pm 0.005$ & E & B \\
698 & 36.673634 & $-4.704338$ & 20.18 & $1.127 \pm 0.005$ & E & B \\
13 & 36.683534 & $-4.672097$ & 20.39 & $0.609 \pm 0.005$ & E & B \\
516 & 36.683296 & $-4.698129$ & 20.59 & $0.963 \pm 0.005$ & E & B \\
272 & 36.681727 & $-4.689340$ & 20.63 & $1.798 \pm 0.002$ & C & A \\
355 & 36.686442 & $-4.692394$ & 20.80 & $1.798 \pm 0.002$ & C & A \\
409 & 36.692244 & $-4.693913$ & 20.85 & $0.692 \pm 0.005$ & E & A \\
60 & 36.687740 & $-4.677383$ & 20.86 & $0.608 \pm 0.005$ & E & B \\
448 & 36.691822 & $-4.694914$ & 21.05 & $0.797 \pm 0.005$ & E & B \\
376 & 36.675006 & $-4.692865$ & 21.20 & $1.811 \pm 0.006$ & C & A \\
64 & 36.675602 & $-4.677701$ & 21.24 & $2.415 \pm 0.001$ & C & A \\
628 & 36.678489 & $-4.701768$ & 21.26 & $1.592 \pm 0.010$ & C & B \\
499 & 36.681694 & $-4.697093$ & 21.27 & $1.127 \pm 0.005$ & E & B \\
445 & 36.673416 & $-4.694926$ & 21.33 & $0.893 \pm 0.005$ & E & B \\
356 & 36.694233 & $-4.692351$ & 21.35 & $1.805 \pm 0.004$ & C & A \\
546 & 36.665075 & $-4.699060$ & 21.36 & $2.187 \pm 0.054$ & C & C \\
485 & 36.670279 & $-4.696597$ & 21.41 & $1.131 \pm 0.005$ & E & B \\
164 & 36.661774 & $-4.684718$ & 21.45 & $1.325 \pm 0.005$ & E & B \\
743 & 36.697722 & $-4.705844$ & 21.47 & $1.324 \pm 0.003$ & E & A \\
657 & 36.675567 & $-4.702566$ & 21.61 & $1.812 \pm 0.002$ & C & A \\
48 & 36.678011 & $-4.676309$ & 21.67 & $0.962 \pm 0.005$ & E & B \\
165 & 36.661849 & $-4.684869$ & 21.68 & $1.302 \pm 0.005$ & E & B \\
286 & 36.687899 & $-4.689939$ & 21.69 & $1.798 \pm 0.041$ & C & B \\
342 & 36.696650 & $-4.691744$ & 21.74 & $1.323 \pm 0.005$ & E & A \\
519 & 36.702752 & $-4.697865$ & 21.76 & $1.055 \pm 0.005$ & E & B \\
352 & 36.690511 & $-4.692148$ & 21.88 & $1.797 \pm 0.005$ & C & A \\
601 & 36.689218 & $-4.700765$ & 21.89 & $1.339 \pm 0.018$ & C & C \\
451 & 36.680181 & $-4.695045$ & 21.90 & $1.470 \pm 0.047$ & E & B \\
556 & 36.675557 & $-4.699295$ & 21.91 & $1.591 \pm 0.006$ & C & A \\
249 & 36.702231 & $-4.688053$ & 21.98 & $1.935 \pm 0.003$ & C & A \\
410 & 36.673327 & $-4.693843$ & 22.00 & $2.406 \pm 0.009$ & C & A \\
107 & 36.676193 & $-4.681298$ & 22.01 & $1.623 \pm 0.004$ & E & A \\
452 & 36.683320 & $-4.695092$ & 22.02 & $1.464 \pm 0.004$ & C & A \\
779 & 36.695368 & $-4.707747$ & 22.03 & $1.713 \pm 0.009$ & C & B \\
320 & 36.668857 & $-4.691090$ & 22.04 & $1.125 \pm 0.005$ & E & A \\
411 & 36.673819 & $-4.693840$ & 22.11 & $1.821 \pm 0.004$ & C & A \\
447 & 36.691213 & $-4.694868$ & 22.12 & $1.797 \pm 0.010$ & C & A \\
197 & 36.699141 & $-4.685847$ & 22.13 & $1.704 \pm 0.007$ & C & B \\
166 & 36.695278 & $-4.685600$ & 22.16 & $0.484 \pm 0.005$ & E & B \\
289 & 36.689652 & $-4.689939$ & 22.17 & $1.802 \pm 0.003$ & C & A \\
589 & 36.693715 & $-4.698247$ & 22.21 & $0.702 \pm 0.005$ & E & B \\
392 & 36.685294 & $-4.693101$ & 22.33 & $2.065 \pm 0.012$ & E & A \\
85 & 36.689254 & $-4.679838$ & 22.35 & $1.519 \pm 0.005$ & E & A \\
387 & 36.682313 & $-4.692964$ & 22.36 & $1.801 \pm 0.009$ & C & B \\
655 & 36.682254 & $-4.702452$ & 22.40 & $0.795 \pm 0.005$ & E & B \\
375 & 36.674884 & $-4.692784$ & 22.43 & $1.819 \pm 0.008$ & C & B \\
317 & 36.699109 & $-4.690911$ & 22.45 & $1.787 \pm 0.003$ & C & A \\
80 & 36.690513 & $-4.679514$ & 22.51 & $1.174 \pm 0.005$ & E & A \\
798 & 36.667559 & $-4.708978$ & 22.51 & $1.065 \pm 0.005$ & E & B \\
105 & 36.676666 & $-4.681000$ & 22.54 & $1.623 \pm 0.004$ & E & B \\
359 & 36.676956 & $-4.692278$ & 22.54 & $1.792 \pm 0.004$ & C & B \\
365 & 36.691019 & $-4.692373$ & 22.54 & $1.511 \pm 0.005$ & E & A \\
569 & 36.681467 & $-4.699630$ & 22.61 & $1.834 \pm 0.022$ & C & C \\
637 & 36.679943 & $-4.701682$ & 22.70 & $1.490 \pm 0.094$ & C & C \\
385 & 36.702109 & $-4.692868$ & 22.71 & $1.257 \pm 0.005$ & E & B \\
281 & 36.690609 & $-4.689444$ & 22.77 & $1.806 \pm 0.004$ & C & B \\
334 & 36.690954 & $-4.691279$ & 22.79 & $1.133 \pm 0.005$ & E & B \\
674 & 36.687376 & $-4.703028$ & 22.85 & $1.302 \pm 0.005$ & E & A \\
693 & 36.677710 & $-4.703786$ & 22.86 & $1.820 \pm 0.014$ & C & C \\
323 & 36.674250 & $-4.691128$ & 22.99 & $1.369 \pm 0.009$ & C & C \\
224 & 36.684922 & $-4.686954$ & 23.04 & $0.966 \pm 0.005$ & E & A \\
201 & 36.676671 & $-4.686139$ & 23.04 & $0.924 \pm 0.005$ & E & A \\
8 & 36.680094 & $-4.670625$ & 23.07 & $0.968 \pm 0.005$ & E & A \\
16 & 36.692232 & $-4.672568$ & 23.12 & $1.474 \pm 0.005$ & E & A \\
531 & 36.679186 & $-4.698393$ & 23.12 & $1.818 \pm 0.005$ & E & A \\
414 & 36.696719 & $-4.693920$ & 23.16 & $1.334 \pm 0.005$ & E & A \\
459 & 36.675068 & $-4.695578$ & 23.25 & $1.599 \pm 0.053$ & C & C \\
653 & 36.676695 & $-4.702391$ & 23.25 & $1.611 \pm 0.041$ & C & C \\
368 & 36.679813 & $-4.692496$ & 23.26 & $1.951 \pm 0.033$ & C & C \\
587 & 36.665178 & $-4.700129$ & 23.27 & $1.917 \pm 0.015$ & C & C \\
446 & 36.679765 & $-4.694762$ & 23.29 & $1.485 \pm 0.054$ & C & C \\
255 & 36.687932 & $-4.688383$ & 23.30 & $1.795 \pm 0.040$ & C & C \\
77 & 36.681823 & $-4.678888$ & 23.35 & $0.902 \pm 0.005$ & E & A \\
300 & 36.696786 & $-4.690403$ & 23.36 & $0.693 \pm 0.005$ & E & A \\
582 & 36.691929 & $-4.700078$ & 23.42 & $1.132 \pm 0.005$ & E & A \\
161 & 36.684522 & $-4.684455$ & 23.48 & $1.137 \pm 0.005$ & E & B \\
61 & 36.687598 & $-4.677597$ & 23.49 & $2.049 \pm 0.005$ & E & A \\
593 & 36.698937 & $-4.700375$ & 23.51 & $2.164 \pm 0.005$ & E & A \\
117 & 36.689081 & $-4.681802$ & 23.52 & $1.474 \pm 0.005$ & E & A \\
177 & 36.672820 & $-4.685007$ & 23.65 & $0.798 \pm 0.005$ & E & A \\
156 & 36.694895 & $-4.684186$ & 23.66 & $1.965 \pm 0.005$ & E & A \\
504 & 36.690690 & $-4.697156$ & 23.79 & $1.064 \pm 0.005$ & E & B \\
477 & 36.700287 & $-4.696110$ & 23.82 & $1.833 \pm 0.005$ & E & B \\
332 & 36.671646 & $-4.691251$ & 23.83 & $1.785 \pm 0.005$ & E & B \\
21 & 36.681790 & $-4.673701$ & 23.88 & $1.489 \pm 0.005$ & E & A \\
39 & 36.689767 & $-4.675155$ & 23.94 & $2.047 \pm 0.005$ & E & A \\
282 & 36.669934 & $-4.689446$ & 23.98 & $1.940 \pm 0.005$ & E & B \\
145 & 36.665669 & $-4.683570$ & 24.00 & $1.631 \pm 0.005$ & E & A \\
149 & 36.679709 & $-4.683808$ & 24.01 & $1.173 \pm 0.005$ & E & A \\
175 & 36.698451 & $-4.684924$ & 24.03 & $1.520 \pm 0.005$ & E & A \\
538 & 36.683767 & $-4.698538$ & 24.22 & $1.111 \pm 0.005$ & E & B \\
427 & 36.671179 & $-4.694407$ & 24.29 & $1.000 \pm 0.005$ & E & A \\
677 & 36.670007 & $-4.703092$ & 24.29 & $0.665 \pm 0.005$ & E & A \\
742 & 36.678980 & $-4.705792$ & 24.43 & $1.135 \pm 0.005$ & E & A \\
581 & 36.691821 & $-4.699903$ & 24.55 & $1.170 \pm 0.005$ & E & A \\
598 & 36.688125 & $-4.700496$ & 24.77 & $1.470 \pm 0.005$ & E & A \\
87 & 36.687851 & $-4.679866$ & 24.94 & $2.154 \pm 0.005$ & E & B 
\enddata
\tablecomments{Type `E' and `C' denote emission line and continuum-based grism redshifts, respectively. Uncertainties on emission line redshifts are listed as 0.005, based on our external comparison with higher-resolution data in Section~\ref{sec:emlines}; errors on the continuum-derived redshifts are based on the MCMC chains. Quality flags are explained in the text.\label{tab:z}}
\end{deluxetable*}

\section{Method for Fitting of Grism Spectra and Photometry}
\texttt{pyspecfit} is based on the MCMC sampler \texttt{MultiNest} \citep{Feroz09}. It accepts as input one or more spectra, with associated LSFs, along with broadband photometric data. For a given set of model parameters proposed by the sampler, the likelihood $\mathcal{L}$ is computed as follows. We begin with a grid of \citet{BC03} simple stellar population (SSP, or ``burst'') models with a \citet{Salpeter55} IMF. First, the SSPs are interpolated to the desired metallicity and integrated over the star formation history. We adopt an exponentially declining model with $\textrm{SFR} \propto e^{-t/\tau}$ for $t > -t_0$, where SFR is the star formation rate at time $t$, $\tau$ is the $e$-folding time, $t = 0$ at the epoch of observation, and  $t_0$ is the age. Gas lost during stellar evolution is not recycled. Next, dust attenuation is applied using the \citet{Calzetti00} law, parameterized by the attenuation $A_V$ at 5500~\AA. Finally, the spectrum is redshifted, and attenuation by the intergalactic medium blueward of Ly$\alpha$ is taken into account following \citet{Madau95}. 

The model is then binned to the wavelength grid of each observed spectrum and convolved by the LSF (i.e., the galaxy light profile) to produce model spectra, i.e., $M^{\rm G102}$ and $M^{\rm G141}$ for the fits described in Section~\ref{sec:contfit}. The model is also integrated over the filter transmission curves to obtain the model flux density $M^{\rm phot}_k$ through each observed filter. The likelihood is $\mathcal{L} = \exp(-\frac{1}{2} \chi^2)$, where
\begin{align}
\chi^2 = &\sum_i \left(\frac{D^{\rm G102}_i - P(\lambda_i) M^{\rm G102}_i}{\sigma^{\rm G102}_i}\right)^2 + 
\sum_j \left(\frac{D^{\rm G141}_j - P(\lambda_j) M^{\rm G141}_j}{\sigma^{\rm G141}_j}\right)^2 + 
\sum_k \left(\frac{D^{\rm phot}_k - M^{\rm phot}_k}{\sigma^{\rm phot}_k}\right)^2.
\end{align}
Here $D^{\rm G102}$ and $D^{\rm G141}$ are the observed spectra with associated uncertainties $\sigma^{\rm G102}$ and $\sigma^{\rm G141}$, $i$ and $j$ run over the pixels in each spectrum, and $D^{\rm phot}_k$ is flux density measured in filter $k$ with uncertainty $\sigma^{\rm phot}_k$. $P(\lambda)$ is a polynomial that scales and modulates the shape of the spectra. At minimum a constant is necessary to scale the spectra to the total flux, but it is also desirable to allow for some variation in the broadband spectral shape (see also \citealt{Brammer12}). We use a linear $P(\lambda)$, which is continuous across the entire wavelength range spanned by both grisms, and determine the coefficients that minimize $\chi^2$ for a given set of model parameters using a linear least-squares approach. Essentially, this procedure allows for a mild deformation of the spectral shape to match the photometric data, but the low polynomial order prevents the introduction of a spectral break. 

For our fits to the spectra and photometry of the individual galaxies in our continuum sample (Section~\ref{sec:contfit}), we chose uniform priors over $1 < z < 3$, $7 < \log \tau / {\rm yr} < 10$, $8 < \log t_0 / {\rm yr} < \log a(z) / {\rm yr}$, and $0 < A_V < 2$, and where $a(z)$ is the age of the universe at redshift $z$. The metallicity was fixed to solar. For our analysis of the continuum-normalized stacked spectra, we allow the metallicity to vary and fit simple stellar populations as described in Section~\ref{sec:stackspec}. \texttt{pyspecfit} produces samples from the posterior distribution for these parameters, as well as the stellar mass $M_*$ (including remnants) and SFR at the observation epoch. In this paper we primarily make use of the redshift and stellar mass estimates and report the median, with $1\sigma$ errors representing the 16th and 84th percentiles. We have compared our stellar mass estimates for the continuum sample of 40 galaxies to the estimates produced by \texttt{FAST}, which fits only the broadband photometry. The redshift was fixed to $z_{\rm grism}$ in \texttt{FAST}. We find that the median difference between the two mass estimates is consistent with zero, and there is no systematic trend with mass.

\section{Literature Compilation in Figure~16}

Here we describe our compilation of literature measurements of the variation of the stellar mass--size relation with environment used in Figure~\ref{fig:othersizestudies}. For the six individual clusters plotted, including \jkcs, we use the masses and radii of individual quiescent galaxies and compare these to the mean relation that we measured in the field (Equation~\ref{eqn:fit_full_field}). We take this field relation as a uniform basis of comparison for every cluster, since it is based on a much larger sample of field galaxies than those used in the following studies, but we note that this may introduce some systematic errors, which are estimated below. In each case, stellar masses are converted to a Salpeter IMF and a cut of $M_* > 10^{10.7} \msol$ is applied to ensure that similar mass ranges are probed. From \citet{Papovich12}, we take the 10 $UVJ$--quiescent galaxies above this limit with $P_z > 0.5$ and $R_{\rm proj} < 1$~Mpc. From \citet{Zirm12}, we take the eight photometrically-selected candidates in their Table 1; since their masses are derived using the \citet{Maraston05} models, we divide them by 0.69 to account for the typical offset from BC03-based masses found by \citet{Muzzin09}. From \citet{Strazzullo13} we take the four ``passive early type'' galaxies above our mass limit listed in their Figure~12. From \citet{Rettura10} we take the 18 galaxies in RDCS1252.9--2927 in their Table 1. From \citet{Raichoor11,Raichoor12} we take the sizes and BC03-based masses of 23 galaxies in the Lynx cluster E and W. For the Rettura et al.~and Raichoor et al.~data, we apply a mean offsets determined by \citet{Raichoor11} of $\Delta \log M_* = -0.05$, which includes an aperture correction ($+0.06$~dex) to total S\'{e}rsic magnitudes and the mean effect of including dust attenuation ($-0.11$~dex), which should better match our procedure.

For each cluster we compute the mean offset $\Delta \log R_e^{\rm maj}$ from Equation~\ref{eqn:fit_full_field} and estimate its uncertainty as $0.23 / \sqrt{N}$~dex, where $N$ is the number of cluster members, based on the scatter in the field relation. Several sources of systematic uncertainty may affect this comparison between our field relation and independently-measured masses and sizes of cluster galaxies. First, different authors use different photometric apertures. Using {\tt MAG\_AUTO}-scaled masses for our field galaxy sample, rather scaling to the total S\'{e}rsic  magnitude, produces a shift of only $\Delta \log R_e^{\rm maj} = -0.01$~dex in Equation~\ref{eqn:fit_full_field}, but larger offsets could apply to other data sets. Second, the inclusion with galaxies having questionable S\'{e}rsic fits can lead to shifts of $\sim0.02$~dex. Third, although we have tried to harmonize stellar mass to first order by applying offsets based on the IMF and the set of stellar population models used, other differences in the priors and fitting procedure remain. Since the Papovich et al.~sample overlaps our UDS data, we are able in this case to directly compare stellar mass estimates. For the 20 overlapping $UVJ$-quiescent galaxies with $M_* > 10^{10.7} \msol$, we find that our $M_*^{\rm tot}$ are offset from the Papovich et al.~measures by $-0.05$~dex, which corresponds to a shift in $\Delta \log R_e^{\rm maj}$ of $0.63 \times (-0.05) = -0.03$~dex. These uncertainties should be kept in mind pending future studies that homogeneously analyze data from an ensemble of high-$z$ clusters.

In addition to these studies of individual clusters, we also directly quote results from 3 studies of larger samples. From \citet{Delaye14}, we take the mean mass-normalized radii in their field and cluster samples in three redshift bins from their Table~9. From \citet{Cooper12} we take the difference in median sizes of matched galaxy samples in high- and low-density regions of the DEEP3 survey from their Figure~3. \citet{Lani13} publish relative sizes of red galaxies in high- and low-density regions in the UDS field, broken down by mass (their Figures 5 and 6). To better compare with the above works, we average these mass-dependent measurements in each of their redshift bins, weighting by the number of galaxies in each mass bin. Only mass bins with $M_* > 10^{10.7} \msol$ were used, after converting to a Salpeter IMF.

\end{appendix}

\bibliographystyle{apj}
\bibliography{jkcs041}

\begin{thebibliography}{145}
\expandafter\ifx\csname natexlab\endcsname\relax\def\natexlab#1{#1}\fi

\bibitem[{{Alberts} {et~al.}(2014){Alberts}, {Pope}, {Brodwin}, {Atlee}, {Lin},
  {Dey}, {Eisenhardt}, {Gettings}, {Gonzalez}, {Jannuzi}, {Mancone},
  {Moustakas}, {Snyder}, {Stanford}, {Stern}, {Weiner}, \&
  {Zeimann}}]{Alberts14}
{Alberts}, S., {Pope}, A., {Brodwin}, M., {et~al.} 2014, \mnras, 437, 437

\bibitem[{{Andreon}(1996)}]{Andreon96}
{Andreon}, S. 1996, \aap, 314, 763

\bibitem[{{Andreon}(2011)}]{Andreon11b}
---. 2011, \aap, 529, L5

\bibitem[{{Andreon} {et~al.}(1997){Andreon}, {Davoust}, \& {Heim}}]{Andreon97}
{Andreon}, S., {Davoust}, E., \& {Heim}, T. 1997, \aap, 323, 337

\bibitem[{{Andreon} {et~al.}(2008){Andreon}, {de Propris}, {Puddu}, {Giordano},
  \& {Quintana}}]{Andreon08}
{Andreon}, S., {de Propris}, R., {Puddu}, E., {Giordano}, L., \& {Quintana}, H.
  2008, \mnras, 383, 102

\bibitem[{{Andreon} \& {Huertas-Company}(2011)}]{Andreon11}
{Andreon}, S., \& {Huertas-Company}, M. 2011, \aap, 526, A11

\bibitem[{{Andreon} {et~al.}(2009){Andreon}, {Maughan}, {Trinchieri}, \&
  {Kurk}}]{Andreon09}
{Andreon}, S., {Maughan}, B., {Trinchieri}, G., \& {Kurk}, J. 2009, \aap, 507,
  147

\bibitem[{{Andreon} {et~al.}(2013){Andreon}, {Newman}, {Trinchieri},
  {Raichoor}, {Ellis}, \& {Treu}}]{Andreon13}
{Andreon}, S., {Newman}, A.~B., {Trinchieri}, G., {et~al.} 2013,
  arXiv:1311.4361

\bibitem[{{Atek} {et~al.}(2010){Atek}, {Malkan}, {McCarthy}, {Teplitz},
  {Scarlata}, {Siana}, {Henry}, {Colbert}, {Ross}, {Bridge}, {Bunker},
  {Dressler}, {Fosbury}, {Martin}, \& {Shim}}]{Atek10}
{Atek}, H., {Malkan}, M., {McCarthy}, P., {et~al.} 2010, \apj, 723, 104

\bibitem[{{Bassett} {et~al.}(2013){Bassett}, {Papovich}, {Lotz}, {Bell},
  {Finkelstein}, {Newman}, {Tran}, {Almaini}, {Lani}, {Cooper}, {Croton},
  {Dekel}, {Ferguson}, {Kocevski}, {Koekemoer}, {Koo}, {McGrath}, {McIntosh},
  \& {Wechsler}}]{Bassett13}
{Bassett}, R., {Papovich}, C., {Lotz}, J.~M., {et~al.} 2013, \apj, 770, 58

\bibitem[{{Bedregal} {et~al.}(2013){Bedregal}, {Scarlata}, {Henry}, {Atek},
  {Rafelski}, {Teplitz}, {Dominguez}, {Siana}, {Colbert}, {Malkan}, {Ross},
  {Martin}, {Dressler}, {Bridge}, {Hathi}, {Masters}, {McCarthy}, \&
  {Rutkowski}}]{Bedregal13}
{Bedregal}, A.~G., {Scarlata}, C., {Henry}, A.~L., {et~al.} 2013, \apj, 778,
  126

\bibitem[{{Bell} {et~al.}(2005){Bell}, {Papovich}, {Wolf}, {Le Floc'h},
  {Caldwell}, {Barden}, {Egami}, {McIntosh}, {Meisenheimer},
  {P{\'e}rez-Gonz{\'a}lez}, {Rieke}, {Rieke}, {Rigby}, \& {Rix}}]{Bell05}
{Bell}, E.~F., {Papovich}, C., {Wolf}, C., {et~al.} 2005, \apj, 625, 23

\bibitem[{{Bernardi} {et~al.}(2007){Bernardi}, {Hyde}, {Sheth}, {Miller}, \&
  {Nichol}}]{Bernardi07}
{Bernardi}, M., {Hyde}, J.~B., {Sheth}, R.~K., {Miller}, C.~J., \& {Nichol},
  R.~C. 2007, \aj, 133, 1741

\bibitem[{{Bertin} \& {Arnouts}(1996)}]{SExtractor}
{Bertin}, E., \& {Arnouts}, S. 1996, \aaps, 117, 393

\bibitem[{{Bessell}(1990)}]{Bessell90}
{Bessell}, M.~S. 1990, \pasp, 102, 1181

\bibitem[{{Bezanson} {et~al.}(2009){Bezanson}, {van Dokkum}, {Tal},
  {Marchesini}, {Kriek}, {Franx}, \& {Coppi}}]{Bezanson09}
{Bezanson}, R., {van Dokkum}, P.~G., {Tal}, T., {et~al.} 2009, \apj, 697, 1290

\bibitem[{{Bielby} {et~al.}(2012){Bielby}, {Hudelot}, {McCracken}, {Ilbert},
  {Daddi}, {Le F{\`e}vre}, {Gonzalez-Perez}, {Kneib}, {Marmo}, {Mellier},
  {Salvato}, {Sanders}, \& {Willott}}]{Bielby12}
{Bielby}, R., {Hudelot}, P., {McCracken}, H.~J., {et~al.} 2012, \aap, 545, A23

\bibitem[{{Bielby} {et~al.}(2010){Bielby}, {Finoguenov}, {Tanaka}, {McCracken},
  {Daddi}, {Hudelot}, {Ilbert}, {Kneib}, {Le F{\`e}vre}, {Mellier}, {Nandra},
  {Petitjean}, {Srianand}, {Stalin}, \& {Willott}}]{Bielby10}
{Bielby}, R.~M., {Finoguenov}, A., {Tanaka}, M., {et~al.} 2010, \aap, 523, A66

\bibitem[{{Bonamente} {et~al.}(2008){Bonamente}, {Joy}, {LaRoque}, {Carlstrom},
  {Nagai}, \& {Marrone}}]{Bonamente08}
{Bonamente}, M., {Joy}, M., {LaRoque}, S.~J., {et~al.} 2008, \apj, 675, 106

\bibitem[{{Brammer} {et~al.}(2008){Brammer}, {van Dokkum}, \&
  {Coppi}}]{Brammer08}
{Brammer}, G.~B., {van Dokkum}, P.~G., \& {Coppi}, P. 2008, \apj, 686, 1503

\bibitem[{{Brammer} {et~al.}(2012){Brammer}, {van Dokkum}, {Franx},
  {Fumagalli}, {Patel}, {Rix}, {Skelton}, {Kriek}, {Nelson}, {Schmidt},
  {Bezanson}, {da Cunha}, {Erb}, {Fan}, {F{\"o}rster Schreiber}, {Illingworth},
  {Labb{\'e}}, {Leja}, {Lundgren}, {Magee}, {Marchesini}, {McCarthy},
  {Momcheva}, {Muzzin}, {Quadri}, {Steidel}, {Tal}, {Wake}, {Whitaker}, \&
  {Williams}}]{Brammer12}
{Brammer}, G.~B., {van Dokkum}, P.~G., {Franx}, M., {et~al.} 2012, \apjs, 200,
  13

\bibitem[{{Brodwin} {et~al.}(2007){Brodwin}, {Gonzalez}, {Moustakas},
  {Eisenhardt}, {Stanford}, {Stern}, \& {Brown}}]{Brodwin07}
{Brodwin}, M., {Gonzalez}, A.~H., {Moustakas}, L.~A., {et~al.} 2007, \apjl,
  671, L93

\bibitem[{{Brodwin} {et~al.}(2012){Brodwin}, {Gonzalez}, {Stanford}, {Plagge},
  {Marrone}, {Carlstrom}, {Dey}, {Eisenhardt}, {Fedeli}, {Gettings}, {Jannuzi},
  {Joy}, {Leitch}, {Mancone}, {Snyder}, {Stern}, \& {Zeimann}}]{Brodwin12}
{Brodwin}, M., {Gonzalez}, A.~H., {Stanford}, S.~A., {et~al.} 2012, \apj, 753,
  162

\bibitem[{{Brodwin} {et~al.}(2013){Brodwin}, {Stanford}, {Gonzalez}, {Zeimann},
  {Snyder}, {Mancone}, {Pope}, {Eisenhardt}, {Stern}, {Alberts}, {Ashby},
  {Brown}, {Chary}, {Dey}, {Galametz}, {Gettings}, {Jannuzi}, {Miller},
  {Moustakas}, \& {Moustakas}}]{Brodwin13}
{Brodwin}, M., {Stanford}, S.~A., {Gonzalez}, A.~H., {et~al.} 2013, \apj, 779,
  138

\bibitem[{{Bruce} {et~al.}(2012){Bruce}, {Dunlop}, {Cirasuolo}, {McLure},
  {Targett}, {Bell}, {Croton}, {Dekel}, {Faber}, {Ferguson}, {Grogin},
  {Kocevski}, {Koekemoer}, {Koo}, {Lai}, {Lotz}, {McGrath}, {Newman}, \& {van
  der Wel}}]{Bruce12}
{Bruce}, V.~A., {Dunlop}, J.~S., {Cirasuolo}, M., {et~al.} 2012, \mnras, 427,
  1666

\bibitem[{{Bruzual} \& {Charlot}(2003)}]{BC03}
{Bruzual}, G., \& {Charlot}, S. 2003, \mnras, 344, 1000

\bibitem[{{Buitrago} {et~al.}(2008){Buitrago}, {Trujillo}, {Conselice},
  {Bouwens}, {Dickinson}, \& {Yan}}]{Buitrago08}
{Buitrago}, F., {Trujillo}, I., {Conselice}, C.~J., {et~al.} 2008, \apjl, 687,
  L61

\bibitem[{{Buitrago} {et~al.}(2013){Buitrago}, {Trujillo}, {Conselice}, \&
  {H{\"a}u{\ss}ler}}]{Buitrago13}
{Buitrago}, F., {Trujillo}, I., {Conselice}, C.~J., \& {H{\"a}u{\ss}ler}, B.
  2013, \mnras, 428, 1460

\bibitem[{{Calzetti} {et~al.}(2000){Calzetti}, {Armus}, {Bohlin}, {Kinney},
  {Koornneef}, \& {Storchi-Bergmann}}]{Calzetti00}
{Calzetti}, D., {Armus}, L., {Bohlin}, R.~C., {et~al.} 2000, \apj, 533, 682

\bibitem[{{Carollo} {et~al.}(2013){Carollo}, {Bschorr}, {Renzini}, {Lilly},
  {Capak}, {Cibinel}, {Ilbert}, {Onodera}, {Scoville}, {Cameron}, {Mobasher},
  {Sanders}, \& {Taniguchi}}]{Carollo13}
{Carollo}, C.~M., {Bschorr}, T.~J., {Renzini}, A., {et~al.} 2013,
  arXiv:1302.5115

\bibitem[{{Chabrier}(2003)}]{Chabrier03}
{Chabrier}, G. 2003, \pasp, 115, 763

\bibitem[{{Chang} {et~al.}(2013{\natexlab{a}}){Chang}, {van der Wel}, {Rix},
  {Wuyts}, {Zibetti}, {Ramkumar}, \& {Holden}}]{Chang13b}
{Chang}, Y.-Y., {van der Wel}, A., {Rix}, H.-W., {et~al.} 2013{\natexlab{a}},
  \apj, 762, 83

\bibitem[{{Chang} {et~al.}(2013{\natexlab{b}}){Chang}, {van der Wel}, {Rix},
  {Holden}, {Bell}, {McGrath}, {Wuyts}, {H{\"a}ussler}, {Barden}, {Faber},
  {Mozena}, {Ferguson}, {Guo}, {Galametz}, {Grogin}, {Kocevski}, {Koekemoer},
  {Dekel}, {Huang}, {Hathi}, \& {Donley}}]{Chang13}
---. 2013{\natexlab{b}}, \apj, 773, 149

\bibitem[{{Conroy} {et~al.}(2014){Conroy}, {Graves}, \& {van
  Dokkum}}]{Conroy14}
{Conroy}, C., {Graves}, G.~J., \& {van Dokkum}, P.~G. 2014, \apj, 780, 33

\bibitem[{{Cooper} {et~al.}(2012){Cooper}, {Griffith}, {Newman}, {Coil},
  {Davis}, {Dutton}, {Faber}, {Guhathakurta}, {Koo}, {Lotz}, {Weiner},
  {Willmer}, \& {Yan}}]{Cooper12}
{Cooper}, M.~C., {Griffith}, R.~L., {Newman}, J.~A., {et~al.} 2012, \mnras,
  419, 3018

\bibitem[{{Culverhouse} {et~al.}(2010){Culverhouse}, {Bonamente}, {Bulbul},
  {Carlstrom}, {Gralla}, {Greer}, {Hasler}, {Hawkins}, {Hennessy}, {Jetha},
  {Joy}, {Lamb}, {Leitch}, {Marrone}, {Miller}, {Mroczkowski}, {Muchovej},
  {Pryke}, {Sharp}, {Woody}, {Andreon}, {Maughan}, \&
  {Stanford}}]{Culverhouse10}
{Culverhouse}, T.~L., {Bonamente}, M., {Bulbul}, E., {et~al.} 2010, \apjl, 723,
  L78

\bibitem[{{Damjanov} {et~al.}(2009){Damjanov}, {McCarthy}, {Abraham},
  {Glazebrook}, {Yan}, {Mentuch}, {Le Borgne}, {Savaglio}, {Crampton},
  {Murowinski}, {Juneau}, {Carlberg}, {J{\o}rgensen}, {Roth}, {Chen}, \&
  {Marzke}}]{Damjanov09}
{Damjanov}, I., {McCarthy}, P.~J., {Abraham}, R.~G., {et~al.} 2009, \apj, 695,
  101

\bibitem[{{Damjanov} {et~al.}(2011){Damjanov}, {Abraham}, {Glazebrook},
  {McCarthy}, {Caris}, {Carlberg}, {Chen}, {Crampton}, {Green}, {J{\o}rgensen},
  {Juneau}, {Le Borgne}, {Marzke}, {Mentuch}, {Murowinski}, {Roth}, {Savaglio},
  \& {Yan}}]{Damjanov11}
{Damjanov}, I., {Abraham}, R.~G., {Glazebrook}, K., {et~al.} 2011, \apjl, 739,
  L44

\bibitem[{{Delaye} {et~al.}(2014){Delaye}, {Huertas-Company}, {Mei}, {Lidman},
  {Licitra}, {Newman}, {Raichoor}, {Shankar}, {Barrientos}, {Bernardi},
  {Cerulo}, {Couch}, {Demarco}, {Mu{\~n}oz}, {S{\'a}nchez-Janssen}, \&
  {Tanaka}}]{Delaye14}
{Delaye}, L., {Huertas-Company}, M., {Mei}, S., {et~al.} 2014, \mnras, 441, 203

\bibitem[{{Dressler} {et~al.}(2013){Dressler}, {Oemler}, {Poggianti},
  {Gladders}, {Abramson}, \& {Vulcani}}]{Dressler13}
{Dressler}, A., {Oemler}, Jr., A., {Poggianti}, B.~M., {et~al.} 2013, \apj,
  770, 62

\bibitem[{{Dressler} {et~al.}(2004){Dressler}, {Oemler}, {Poggianti}, {Smail},
  {Trager}, {Shectman}, {Couch}, \& {Ellis}}]{Dressler04}
---. 2004, \apj, 617, 867

\bibitem[{{Dressler} {et~al.}(1997){Dressler}, {Oemler}, {Couch}, {Smail},
  {Ellis}, {Barger}, {Butcher}, {Poggianti}, \& {Sharples}}]{Dressler97}
{Dressler}, A., {Oemler}, Jr., A., {Couch}, W.~J., {et~al.} 1997, \apj, 490,
  577

\bibitem[{{Duffy} {et~al.}(2008){Duffy}, {Schaye}, {Kay}, \& {Dalla
  Vecchia}}]{Duffy08}
{Duffy}, A.~R., {Schaye}, J., {Kay}, S.~T., \& {Dalla Vecchia}, C. 2008,
  \mnras, 390, L64

\bibitem[{{Eisenhardt} {et~al.}(2008){Eisenhardt}, {Brodwin}, {Gonzalez},
  {Stanford}, {Stern}, {Barmby}, {Brown}, {Dawson}, {Dey}, {Doi}, {Galametz},
  {Jannuzi}, {Kochanek}, {Meyers}, {Morokuma}, \& {Moustakas}}]{Eisenhardt08}
{Eisenhardt}, P.~R.~M., {Brodwin}, M., {Gonzalez}, A.~H., {et~al.} 2008, \apj,
  684, 905

\bibitem[{{Fakhouri} {et~al.}(2010){Fakhouri}, {Ma}, \&
  {Boylan-Kolchin}}]{Fakhouri10}
{Fakhouri}, O., {Ma}, C.-P., \& {Boylan-Kolchin}, M. 2010, \mnras, 406, 2267

\bibitem[{{Fan} {et~al.}(2010){Fan}, {Lapi}, {Bressan}, {Bernardi}, {De Zotti},
  \& {Danese}}]{Fan10}
{Fan}, L., {Lapi}, A., {Bressan}, A., {et~al.} 2010, \apj, 718, 1460

\bibitem[{{Fan} {et~al.}(2008){Fan}, {Lapi}, {De Zotti}, \& {Danese}}]{Fan08}
{Fan}, L., {Lapi}, A., {De Zotti}, G., \& {Danese}, L. 2008, \apjl, 689, L101

\bibitem[{{Fassbender} {et~al.}(2011){Fassbender}, {Nastasi}, {B{\"o}hringer},
  {{\v S}uhada}, {Santos}, {Rosati}, {Pierini}, {M{\"u}hlegger}, {Quintana},
  {Schwope}, {Lamer}, {de Hoon}, {Kohnert}, {Pratt}, \& {Mohr}}]{Fassbender11}
{Fassbender}, R., {Nastasi}, A., {B{\"o}hringer}, H., {et~al.} 2011, \aap, 527,
  L10

\bibitem[{{Feroz} {et~al.}(2009){Feroz}, {Hobson}, \& {Bridges}}]{Feroz09}
{Feroz}, F., {Hobson}, M.~P., \& {Bridges}, M. 2009, \mnras, 398, 1601

\bibitem[{{Finn} {et~al.}(2010){Finn}, {Desai}, {Rudnick}, {Poggianti}, {Bell},
  {Hinz}, {Jablonka}, {Milvang-Jensen}, {Moustakas}, {Rines}, \&
  {Zaritsky}}]{Finn10}
{Finn}, R.~A., {Desai}, V., {Rudnick}, G., {et~al.} 2010, \apj, 720, 87

\bibitem[{{Fumagalli} {et~al.}(2013){Fumagalli}, {Labbe}, {Patel}, {Franx},
  {van Dokkum}, {Brammer}, {da Cunha}, {Forster Schreiber}, {Kriek}, {Quadri},
  {Rix}, {Wake}, {Whitaker}, {Lundgren}, {Marchesini}, {Maseda}, {Momcheva},
  {Nelson}, {Pacifici}, \& {Skelton}}]{Fumagalli13}
{Fumagalli}, M., {Labbe}, I., {Patel}, S.~G., {et~al.} 2013, arXiv:1308.4132

\bibitem[{{Galametz} {et~al.}(2013){Galametz}, {Stern}, {Pentericci}, {De
  Breuck}, {Vernet}, {Wylezalek}, {Fassbender}, {Hatch}, {Kurk}, {Overzier},
  {Rettura}, \& {Seymour}}]{Galametz13}
{Galametz}, A., {Stern}, D., {Pentericci}, L., {et~al.} 2013, arXiv:1309.6645

\bibitem[{{Gobat} {et~al.}(2008){Gobat}, {Rosati}, {Strazzullo}, {Rettura},
  {Demarco}, \& {Nonino}}]{Gobat08}
{Gobat}, R., {Rosati}, P., {Strazzullo}, V., {et~al.} 2008, \aap, 488, 853

\bibitem[{{Gobat} {et~al.}(2011){Gobat}, {Daddi}, {Onodera}, {Finoguenov},
  {Renzini}, {Arimoto}, {Bouwens}, {Brusa}, {Chary}, {Cimatti}, {Dickinson},
  {Kong}, \& {Mignoli}}]{Gobat11}
{Gobat}, R., {Daddi}, E., {Onodera}, M., {et~al.} 2011, \aap, 526, A133

\bibitem[{{Gobat} {et~al.}(2013){Gobat}, {Strazzullo}, {Daddi}, {Onodera},
  {Carollo}, {Renzini}, {Finoguenov}, {Cimatti}, {Scarlata}, \&
  {Arimoto}}]{Gobat13}
{Gobat}, R., {Strazzullo}, V., {Daddi}, E., {et~al.} 2013, arXiv:1305.3576

\bibitem[{{Grogin} {et~al.}(2011){Grogin}, {Kocevski}, {Faber}, {Ferguson},
  {Koekemoer}, {Riess}, {Acquaviva}, {Alexander}, {Almaini}, {Ashby}, {Barden},
  {Bell}, {Bournaud}, {Brown}, {Caputi}, {Casertano}, {Cassata}, {Castellano},
  {Challis}, {Chary}, {Cheung}, {Cirasuolo}, {Conselice}, {Roshan Cooray},
  {Croton}, {Daddi}, {Dahlen}, {Dav{\'e}}, {de Mello}, {Dekel}, {Dickinson},
  {Dolch}, {Donley}, {Dunlop}, {Dutton}, {Elbaz}, {Fazio}, {Filippenko},
  {Finkelstein}, {Fontana}, {Gardner}, {Garnavich}, {Gawiser}, {Giavalisco},
  {Grazian}, {Guo}, {Hathi}, {H{\"a}ussler}, {Hopkins}, {Huang}, {Huang},
  {Jha}, {Kartaltepe}, {Kirshner}, {Koo}, {Lai}, {Lee}, {Li}, {Lotz}, {Lucas},
  {Madau}, {McCarthy}, {McGrath}, {McIntosh}, {McLure}, {Mobasher},
  {Moustakas}, {Mozena}, {Nandra}, {Newman}, {Niemi}, {Noeske}, {Papovich},
  {Pentericci}, {Pope}, {Primack}, {Rajan}, {Ravindranath}, {Reddy}, {Renzini},
  {Rix}, {Robaina}, {Rodney}, {Rosario}, {Rosati}, {Salimbeni}, {Scarlata},
  {Siana}, {Simard}, {Smidt}, {Somerville}, {Spinrad}, {Straughn}, {Strolger},
  {Telford}, {Teplitz}, {Trump}, {van der Wel}, {Villforth}, {Wechsler},
  {Weiner}, {Wiklind}, {Wild}, {Wilson}, {Wuyts}, {Yan}, \& {Yun}}]{Grogin11}
{Grogin}, N.~A., {Kocevski}, D.~D., {Faber}, S.~M., {et~al.} 2011, \apjs, 197,
  35

\bibitem[{{Guo} {et~al.}(2009){Guo}, {McIntosh}, {Mo}, {Katz}, {van den Bosch},
  {Weinberg}, {Weinmann}, {Pasquali}, \& {Yang}}]{Guo09}
{Guo}, Y., {McIntosh}, D.~H., {Mo}, H.~J., {et~al.} 2009, \mnras, 398, 1129

\bibitem[{{Hilton} {et~al.}(2010){Hilton}, {Lloyd-Davies}, {Stanford}, {Stott},
  {Collins}, {Romer}, {Hosmer}, {Hoyle}, {Kay}, {Liddle}, {Mehrtens}, {Miller},
  {Sahl{\'e}n}, \& {Viana}}]{Hilton10}
{Hilton}, M., {Lloyd-Davies}, E., {Stanford}, S.~A., {et~al.} 2010, \apj, 718,
  133

\bibitem[{{Hopkins} {et~al.}(2009){Hopkins}, {Bundy}, {Murray}, {Quataert},
  {Lauer}, \& {Ma}}]{Hopkins09}
{Hopkins}, P.~F., {Bundy}, K., {Murray}, N., {et~al.} 2009, \mnras, 398, 898

\bibitem[{{Huertas-Company} {et~al.}(2013{\natexlab{a}}){Huertas-Company},
  {Shankar}, {Mei}, {Bernardi}, {Aguerri}, {Meert}, \&
  {Vikram}}]{HuertasCompany13local}
{Huertas-Company}, M., {Shankar}, F., {Mei}, S., {et~al.} 2013{\natexlab{a}},
  \apj, 779, 29

\bibitem[{{Huertas-Company} {et~al.}(2013{\natexlab{b}}){Huertas-Company},
  {Mei}, {Shankar}, {Delaye}, {Raichoor}, {Covone}, {Finoguenov}, {Kneib},
  {Le}, \& {Povic}}]{HuertasCompany13}
{Huertas-Company}, M., {Mei}, S., {Shankar}, F., {et~al.} 2013{\natexlab{b}},
  \mnras, 428, 1715

\bibitem[{{Jian} {et~al.}(2012){Jian}, {Lin}, \& {Chiueh}}]{Jian12}
{Jian}, H.-Y., {Lin}, L., \& {Chiueh}, T. 2012, \apj, 754, 26

\bibitem[{{Kajisawa} {et~al.}(2011){Kajisawa}, {Ichikawa}, {Tanaka}, {Yamada},
  {Akiyama}, {Suzuki}, {Tokoku}, {Katsuno Uchimoto}, {Konishi}, {Yoshikawa},
  {Nishimura}, {Omata}, {Ouchi}, {Iwata}, {Hamana}, \& {Onodera}}]{Kajisawa11}
{Kajisawa}, M., {Ichikawa}, T., {Tanaka}, I., {et~al.} 2011, \pasj, 63, 379

\bibitem[{{Kampczyk} {et~al.}(2013){Kampczyk}, {Lilly}, {de Ravel}, {Le
  F{\`e}vre}, {Bolzonella}, {Carollo}, {Diener}, {Knobel}, {Kova{\v c}},
  {Maier}, {Renzini}, {Sargent}, {Vergani}, {Abbas}, {Bardelli}, {Bongiorno},
  {Bordoloi}, {Caputi}, {Contini}, {Coppa}, {Cucciati}, {de la Torre},
  {Franzetti}, {Garilli}, {Iovino}, {Kneib}, {Koekemoer}, {Lamareille}, {Le
  Borgne}, {Le Brun}, {Leauthaud}, {Mainieri}, {Mignoli}, {Pello}, {Peng},
  {Perez Montero}, {Ricciardelli}, {Scodeggio}, {Silverman}, {Tanaka}, {Tasca},
  {Tresse}, {Zamorani}, {Zucca}, {Bottini}, {Cappi}, {Cassata}, {Cimatti},
  {Fumana}, {Guzzo}, {Kartaltepe}, {Marinoni}, {McCracken}, {Memeo}, {Meneux},
  {Oesch}, {Porciani}, {Pozzetti}, \& {Scaramella}}]{Kampczyk13}
{Kampczyk}, P., {Lilly}, S.~J., {de Ravel}, L., {et~al.} 2013, \apj, 762, 43

\bibitem[{{Kauffmann} {et~al.}(2004){Kauffmann}, {White}, {Heckman},
  {M{\'e}nard}, {Brinchmann}, {Charlot}, {Tremonti}, \&
  {Brinkmann}}]{Kauffmann04}
{Kauffmann}, G., {White}, S.~D.~M., {Heckman}, T.~M., {et~al.} 2004, \mnras,
  353, 713

\bibitem[{{Kelson} {et~al.}(2014){Kelson}, {Williams}, {Dressler}, {McCarthy},
  {Shectman}, {Mulchaey}, {Villanueva}, {Crane}, \& {Quadri}}]{Kelson14}
{Kelson}, D.~D., {Williams}, R.~J., {Dressler}, A., {et~al.} 2014, \apj, 783,
  110

\bibitem[{{Kewley} {et~al.}(2004){Kewley}, {Geller}, \& {Jansen}}]{Kewley04}
{Kewley}, L.~J., {Geller}, M.~J., \& {Jansen}, R.~A. 2004, \aj, 127, 2002

\bibitem[{{Koekemoer} {et~al.}(2011){Koekemoer}, {Faber}, {Ferguson}, {Grogin},
  {Kocevski}, {Koo}, {Lai}, {Lotz}, {Lucas}, {McGrath}, {Ogaz}, {Rajan},
  {Riess}, {Rodney}, {Strolger}, {Casertano}, {Dahlen}, {Dickinson}, {Dolch},
  {Fontana}, {Giavalisco}, {Grazian}, {Guo}, {Hathi}, {Huang}, {van der Wel},
  {Yan}, {Acquaviva}, {Almaini}, {Ashby}, {Barden}, {Bell}, {Bournaud},
  {Brown}, {Caputi}, {Cassata}, {Challis}, {Chary}, {Cheung}, {Cirasuolo},
  {Conselice}, {Roshan Cooray}, {Croton}, {Daddi}, {Dav{\'e}}, {de Mello}, {de
  Ravel}, {Dekel}, {Donley}, {Dunlop}, {Dutton}, {Elbaz}, {Fazio},
  {Filippenko}, {Finkelstein}, {Frazer}, {Gardner}, {Garnavich}, {Gawiser},
  {Gruetzbauch}, {Hartley}, {H{\"a}ussler}, {Herrington}, {Hopkins}, {Huang},
  {Jha}, {Johnson}, {Kartaltepe}, {Khostovan}, {Kirshner}, {Lani}, {Lee}, {Li},
  {Madau}, {McCarthy}, {McIntosh}, {McLure}, {McPartland}, {Mobasher},
  {Moreira}, {Mortlock}, {Moustakas}, {Mozena}, {Nandra}, {Newman}, {Nielsen},
  {Niemi}, {Noeske}, {Papovich}, {Pentericci}, {Pope}, {Primack},
  {Ravindranath}, {Reddy}, {Renzini}, {Rix}, {Robaina}, {Rosario}, {Rosati},
  {Salimbeni}, {Scarlata}, {Siana}, {Simard}, {Smidt}, {Snyder}, {Somerville},
  {Spinrad}, {Straughn}, {Telford}, {Teplitz}, {Trump}, {Vargas}, {Villforth},
  {Wagner}, {Wandro}, {Wechsler}, {Weiner}, {Wiklind}, {Wild}, {Wilson},
  {Wuyts}, \& {Yun}}]{Koekemoer11}
{Koekemoer}, A.~M., {Faber}, S.~M., {Ferguson}, H.~C., {et~al.} 2011,
  arXiv:1105.3754

\bibitem[{{Kormendy} \& {Kennicutt}(2004)}]{Kormendy04}
{Kormendy}, J., \& {Kennicutt}, Jr., R.~C. 2004, \araa, 42, 603

\bibitem[{{Kriek} {et~al.}(2009){Kriek}, {van Dokkum}, {Franx}, {Illingworth},
  \& {Magee}}]{Kriek09b}
{Kriek}, M., {van Dokkum}, P.~G., {Franx}, M., {Illingworth}, G.~D., \&
  {Magee}, D.~K. 2009, \apjl, 705, L71

\bibitem[{{Kriek} {et~al.}(2010){Kriek}, {Labb{\'e}}, {Conroy}, {Whitaker},
  {van Dokkum}, {Brammer}, {Franx}, {Illingworth}, {Marchesini}, {Muzzin},
  {Quadri}, \& {Rudnick}}]{Kriek10}
{Kriek}, M., {Labb{\'e}}, I., {Conroy}, C., {et~al.} 2010, \apjl, 722, L64

\bibitem[{{K{\"u}mmel} {et~al.}(2009){K{\"u}mmel}, {Walsh}, {Pirzkal},
  {Kuntschner}, \& {Pasquali}}]{Kummel09}
{K{\"u}mmel}, M., {Walsh}, J.~R., {Pirzkal}, N., {Kuntschner}, H., \&
  {Pasquali}, A. 2009, \pasp, 121, 59

\bibitem[{{Lani} {et~al.}(2013){Lani}, {Almaini}, {Hartley}, {Mortlock},
  {Haeussler}, {Chuter}, {Simpson}, {van der Wel}, {Grutzbauch}, {Conselice},
  {Bradshaw}, {Cooper}, {Faber}, {Grogin}, {Kocevski}, {Koekemoer}, \&
  {Lai}}]{Lani13}
{Lani}, C., {Almaini}, O., {Hartley}, W.~G., {et~al.} 2013, arXiv:1307.3247

\bibitem[{{Lawrence} {et~al.}(2007){Lawrence}, {Warren}, {Almaini}, {Edge},
  {Hambly}, {Jameson}, {Lucas}, {Casali}, {Adamson}, {Dye}, {Emerson},
  {Foucaud}, {Hewett}, {Hirst}, {Hodgkin}, {Irwin}, {Lodieu}, {McMahon},
  {Simpson}, {Smail}, {Mortlock}, \& {Folger}}]{Lawrence07}
{Lawrence}, A., {Warren}, S.~J., {Almaini}, O., {et~al.} 2007, \mnras, 379,
  1599

\bibitem[{{Le F{\`e}vre} {et~al.}(2013){Le F{\`e}vre}, {Cassata}, {Cucciati},
  {Garilli}, {Ilbert}, {Le Brun}, {Maccagni}, {Moreau}, {Scodeggio}, {Tresse},
  {Zamorani}, {Adami}, {Arnouts}, {Bardelli}, {Bolzonella}, {Bondi},
  {Bongiorno}, {Bottini}, {Cappi}, {Charlot}, {Ciliegi}, {Contini}, {de la
  Torre}, {Foucaud}, {Franzetti}, {Gavignaud}, {Guzzo}, {Iovino}, {Lemaux},
  {L{\'o}pez-Sanjuan}, {McCracken}, {Marano}, {Marinoni}, {Mazure}, {Mellier},
  {Merighi}, {Merluzzi}, {Paltani}, {Pell{\`o}}, {Pollo}, {Pozzetti},
  {Scaramella}, {Tasca}, {Vergani}, {Vettolani}, {Zanichelli}, \&
  {Zucca}}]{LeFevre13}
{Le F{\`e}vre}, O., {Cassata}, P., {Cucciati}, O., {et~al.} 2013, \aap, 559,
  A14

\bibitem[{{Lin} {et~al.}(2010){Lin}, {Cooper}, {Jian}, {Koo}, {Patton}, {Yan},
  {Willmer}, {Coil}, {Chiueh}, {Croton}, {Gerke}, {Lotz}, {Guhathakurta}, \&
  {Newman}}]{Lin10}
{Lin}, L., {Cooper}, M.~C., {Jian}, H.-Y., {et~al.} 2010, \apj, 718, 1158

\bibitem[{{Lotz} {et~al.}(2013){Lotz}, {Papovich}, {Faber}, {Ferguson},
  {Grogin}, {Guo}, {Kocevski}, {Koekemoer}, {Lee}, {McIntosh}, {Momcheva},
  {Rudnick}, {Saintonge}, {Tran}, {van der Wel}, \& {Willmer}}]{Lotz13}
{Lotz}, J.~M., {Papovich}, C., {Faber}, S.~M., {et~al.} 2013, \apj, 773, 154

\bibitem[{{Madau}(1995)}]{Madau95}
{Madau}, P. 1995, \apj, 441, 18

\bibitem[{{Mantz} {et~al.}(2014){Mantz}, {Abdulla}, {Carlstrom}, {Greer},
  {Leitch}, {Marrone}, {Muchovej}, {Adami}, {Birkinshaw}, {Bremer}, {Clerc},
  {Giles}, {Horellou}, {Maughan}, {Pacaud}, {Pierre}, \& {Willis}}]{Mantz14}
{Mantz}, A.~B., {Abdulla}, Z., {Carlstrom}, J.~E., {et~al.} 2014,
  arXiv:1401.2087

\bibitem[{{Maraston}(2005)}]{Maraston05}
{Maraston}, C. 2005, \mnras, 362, 799

\bibitem[{{McGee} {et~al.}(2009){McGee}, {Balogh}, {Bower}, {Font}, \&
  {McCarthy}}]{McGee09}
{McGee}, S.~L., {Balogh}, M.~L., {Bower}, R.~G., {Font}, A.~S., \& {McCarthy},
  I.~G. 2009, \mnras, 400, 937

\bibitem[{{McGrath} {et~al.}(2008){McGrath}, {Stockton}, {Canalizo}, {Iye}, \&
  {Maihara}}]{McGrath08}
{McGrath}, E.~J., {Stockton}, A., {Canalizo}, G., {Iye}, M., \& {Maihara}, T.
  2008, \apj, 682, 303

\bibitem[{{McIntosh} {et~al.}(2008){McIntosh}, {Guo}, {Hertzberg}, {Katz},
  {Mo}, {van den Bosch}, \& {Yang}}]{McIntosh08}
{McIntosh}, D.~H., {Guo}, Y., {Hertzberg}, J., {et~al.} 2008, \mnras, 388, 1537

\bibitem[{{Moran} {et~al.}(2007){Moran}, {Ellis}, {Treu}, {Smith}, {Rich}, \&
  {Smail}}]{Moran07}
{Moran}, S.~M., {Ellis}, R.~S., {Treu}, T., {et~al.} 2007, \apj, 671, 1503

\bibitem[{{Moresco} {et~al.}(2010){Moresco}, {Pozzetti}, {Cimatti}, {Zamorani},
  {Mignoli}, {di Cesare}, {Bolzonella}, {Zucca}, {Lilly}, {Kova{\v c}},
  {Scodeggio}, {Cassata}, {Tasca}, {Vergani}, {Halliday}, {Carollo}, {Contini},
  {Kneib}, {Le F{\'e}vre}, {Mainieri}, {Renzini}, {Bardelli}, {Bongiorno},
  {Caputi}, {Coppa}, {Cucciati}, {de la Torre}, {de Ravel}, {Franzetti},
  {Garilli}, {Iovino}, {Kampczyk}, {Knobel}, {Lamareille}, {Le Borgne}, {Le
  Brun}, {Maier}, {Pell{\`o}}, {Peng}, {Perez Montero}, {Ricciardelli},
  {Silverman}, {Tanaka}, {Tresse}, {Abbas}, {Bottini}, {Cappi}, {Guzzo},
  {Koekemoer}, {Leauthaud}, {Maccagni}, {Marinoni}, {McCracken}, {Memeo},
  {Meneux}, {Nair}, {Oesch}, {Porciani}, {Scaramella}, {Scarlata}, \&
  {Scoville}}]{Moresco10}
{Moresco}, M., {Pozzetti}, L., {Cimatti}, A., {et~al.} 2010, \aap, 524, A67

\bibitem[{{Muzzin} {et~al.}(2009){Muzzin}, {Marchesini}, {van Dokkum},
  {Labb{\'e}}, {Kriek}, \& {Franx}}]{Muzzin09}
{Muzzin}, A., {Marchesini}, D., {van Dokkum}, P.~G., {et~al.} 2009, \apj, 701,
  1839

\bibitem[{{Muzzin} {et~al.}(2013){Muzzin}, {Wilson}, {Demarco}, {Lidman},
  {Nantais}, {Hoekstra}, {Yee}, \& {Rettura}}]{Muzzin13}
{Muzzin}, A., {Wilson}, G., {Demarco}, R., {et~al.} 2013, \apj, 767, 39

\bibitem[{{Muzzin} {et~al.}(2012){Muzzin}, {Wilson}, {Yee}, {Gilbank},
  {Hoekstra}, {Demarco}, {Balogh}, {van Dokkum}, {Franx}, {Ellingson}, {Hicks},
  {Nantais}, {Noble}, {Lacy}, {Lidman}, {Rettura}, {Surace}, \&
  {Webb}}]{Muzzin12}
{Muzzin}, A., {Wilson}, G., {Yee}, H.~K.~C., {et~al.} 2012, \apj, 746, 188

\bibitem[{{Naab} {et~al.}(2009){Naab}, {Johansson}, \& {Ostriker}}]{Naab09}
{Naab}, T., {Johansson}, P.~H., \& {Ostriker}, J.~P. 2009, \apjl, 699, L178

\bibitem[{{Nair} {et~al.}(2010){Nair}, {van den Bergh}, \& {Abraham}}]{Nair10}
{Nair}, P.~B., {van den Bergh}, S., \& {Abraham}, R.~G. 2010, \apj, 715, 606

\bibitem[{{Newman} {et~al.}(2012){Newman}, {Ellis}, {Bundy}, \&
  {Treu}}]{Newman12}
{Newman}, A.~B., {Ellis}, R.~S., {Bundy}, K., \& {Treu}, T. 2012, \apj, 746,
  162

\bibitem[{{Nipoti} {et~al.}(2009){Nipoti}, {Treu}, {Auger}, \&
  {Bolton}}]{Nipoti09}
{Nipoti}, C., {Treu}, T., {Auger}, M.~W., \& {Bolton}, A.~S. 2009, \apjl, 706,
  L86

\bibitem[{{Nipoti} {et~al.}(2012){Nipoti}, {Treu}, {Leauthaud}, {Bundy},
  {Newman}, \& {Auger}}]{Nipoti12}
{Nipoti}, C., {Treu}, T., {Leauthaud}, A., {et~al.} 2012, \mnras, 422, 1714

\bibitem[{{Overzier} {et~al.}(2005){Overzier}, {Harris}, {Carilli},
  {Pentericci}, {R{\"o}ttgering}, \& {Miley}}]{Overzier05}
{Overzier}, R.~A., {Harris}, D.~E., {Carilli}, C.~L., {et~al.} 2005, \aap, 433,
  87

\bibitem[{{Padilla} \& {Strauss}(2008)}]{Padilla08}
{Padilla}, N.~D., \& {Strauss}, M.~A. 2008, \mnras, 388, 1321

\bibitem[{{Papovich} {et~al.}(2010){Papovich}, {Momcheva}, {Willmer},
  {Finkelstein}, {Finkelstein}, {Tran}, {Brodwin}, {Dunlop}, {Farrah}, {Khan},
  {Lotz}, {McCarthy}, {McLure}, {Rieke}, {Rudnick}, {Sivanandam}, {Pacaud}, \&
  {Pierre}}]{Papovich10}
{Papovich}, C., {Momcheva}, I., {Willmer}, C.~N.~A., {et~al.} 2010, \apj, 716,
  1503

\bibitem[{{Papovich} {et~al.}(2012){Papovich}, {Bassett}, {Lotz}, {van der
  Wel}, {Tran}, {Finkelstein}, {Bell}, {Conselice}, {Dekel}, {Dunlop}, {Guo},
  {Faber}, {Farrah}, {Ferguson}, {Finkelstein}, {H{\"a}ussler}, {Kocevski},
  {Koekemoer}, {Koo}, {McGrath}, {McLure}, {McIntosh}, {Momcheva}, {Newman},
  {Rudnick}, {Weiner}, {Willmer}, \& {Wuyts}}]{Papovich12}
{Papovich}, C., {Bassett}, R., {Lotz}, J.~M., {et~al.} 2012, \apj, 750, 93

\bibitem[{{Peng} {et~al.}(2002){Peng}, {Ho}, {Impey}, \& {Rix}}]{Peng02}
{Peng}, C.~Y., {Ho}, L.~C., {Impey}, C.~D., \& {Rix}, H.-W. 2002, \aj, 124, 266

\bibitem[{{Peng} {et~al.}(2010){Peng}, {Lilly}, {Kova{\v c}}, {Bolzonella},
  {Pozzetti}, {Renzini}, {Zamorani}, {Ilbert}, {Knobel}, {Iovino}, {Maier},
  {Cucciati}, {Tasca}, {Carollo}, {Silverman}, {Kampczyk}, {de Ravel},
  {Sanders}, {Scoville}, {Contini}, {Mainieri}, {Scodeggio}, {Kneib}, {Le
  F{\`e}vre}, {Bardelli}, {Bongiorno}, {Caputi}, {Coppa}, {de la Torre},
  {Franzetti}, {Garilli}, {Lamareille}, {Le Borgne}, {Le Brun}, {Mignoli},
  {Perez Montero}, {Pello}, {Ricciardelli}, {Tanaka}, {Tresse}, {Vergani},
  {Welikala}, {Zucca}, {Oesch}, {Abbas}, {Barnes}, {Bordoloi}, {Bottini},
  {Cappi}, {Cassata}, {Cimatti}, {Fumana}, {Hasinger}, {Koekemoer},
  {Leauthaud}, {Maccagni}, {Marinoni}, {McCracken}, {Memeo}, {Meneux}, {Nair},
  {Porciani}, {Presotto}, \& {Scaramella}}]{Peng10}
{Peng}, Y.-j., {Lilly}, S.~J., {Kova{\v c}}, K., {et~al.} 2010, \apj, 721, 193

\bibitem[{{Pierre} {et~al.}(2012){Pierre}, {Clerc}, {Maughan}, {Pacaud},
  {Papovich}, \& {Willmer}}]{Pierre12}
{Pierre}, M., {Clerc}, N., {Maughan}, B., {et~al.} 2012, \aap, 540, A4

\bibitem[{{Poggianti} {et~al.}(2009){Poggianti}, {Arag{\'o}n-Salamanca},
  {Zaritsky}, {De Lucia}, {Milvang-Jensen}, {Desai}, {Jablonka}, {Halliday},
  {Rudnick}, {Varela}, {Bamford}, {Best}, {Clowe}, {Noll}, {Saglia},
  {Pell{\'o}}, {Simard}, {von der Linden}, \& {White}}]{Poggianti09}
{Poggianti}, B.~M., {Arag{\'o}n-Salamanca}, A., {Zaritsky}, D., {et~al.} 2009,
  \apj, 693, 112

\bibitem[{{Poggianti} {et~al.}(2013){Poggianti}, {Calvi}, {Bindoni},
  {D'Onofrio}, {Moretti}, {Valentinuzzi}, {Fasano}, {Fritz}, {De Lucia},
  {Vulcani}, {Bettoni}, {Gullieuszik}, \& {Omizzolo}}]{Poggianti13}
{Poggianti}, B.~M., {Calvi}, R., {Bindoni}, D., {et~al.} 2013, \apj, 762, 77

\bibitem[{{Postman} {et~al.}(2005){Postman}, {Franx}, {Cross}, {Holden},
  {Ford}, {Illingworth}, {Goto}, {Demarco}, {Rosati}, {Blakeslee}, {Tran},
  {Ben{\'{\i}}tez}, {Clampin}, {Hartig}, {Homeier}, {Ardila}, {Bartko},
  {Bouwens}, {Bradley}, {Broadhurst}, {Brown}, {Burrows}, {Cheng}, {Feldman},
  {Golimowski}, {Gronwall}, {Infante}, {Kimble}, {Krist}, {Lesser}, {Martel},
  {Mei}, {Menanteau}, {Meurer}, {Miley}, {Motta}, {Sirianni}, {Sparks}, {Tran},
  {Tsvetanov}, {White}, \& {Zheng}}]{Postman05}
{Postman}, M., {Franx}, M., {Cross}, N.~J.~G., {et~al.} 2005, \apj, 623, 721

\bibitem[{{Quadri} {et~al.}(2012){Quadri}, {Williams}, {Franx}, \&
  {Hildebrandt}}]{Quadri12}
{Quadri}, R.~F., {Williams}, R.~J., {Franx}, M., \& {Hildebrandt}, H. 2012,
  \apj, 744, 88

\bibitem[{{Raichoor} \& {Andreon}(2012)}]{Raichoor12b}
{Raichoor}, A., \& {Andreon}, S. 2012, \aap, 543, A19

\bibitem[{{Raichoor} {et~al.}(2011){Raichoor}, {Mei}, {Nakata}, {Stanford},
  {Holden}, {Rettura}, {Huertas-Company}, {Postman}, {Rosati}, {Blakeslee},
  {Demarco}, {Eisenhardt}, {Illingworth}, {Jee}, {Kodama}, {Tanaka}, \&
  {White}}]{Raichoor11}
{Raichoor}, A., {Mei}, S., {Nakata}, F., {et~al.} 2011, \apj, 732, 12

\bibitem[{{Raichoor} {et~al.}(2012){Raichoor}, {Mei}, {Stanford}, {Holden},
  {Nakata}, {Rosati}, {Shankar}, {Tanaka}, {Ford}, {Huertas-Company},
  {Illingworth}, {Kodama}, {Postman}, {Rettura}, {Blakeslee}, {Demarco}, {Jee},
  \& {White}}]{Raichoor12}
{Raichoor}, A., {Mei}, S., {Stanford}, S.~A., {et~al.} 2012, \apj, 745, 130

\bibitem[{{Reddy} {et~al.}(2006){Reddy}, {Steidel}, {Fadda}, {Yan}, {Pettini},
  {Shapley}, {Erb}, \& {Adelberger}}]{Reddy06}
{Reddy}, N.~A., {Steidel}, C.~C., {Fadda}, D., {et~al.} 2006, \apj, 644, 792

\bibitem[{{Rettura} {et~al.}(2010){Rettura}, {Rosati}, {Nonino}, {Fosbury},
  {Gobat}, {Menci}, {Strazzullo}, {Mei}, {Demarco}, \& {Ford}}]{Rettura10}
{Rettura}, A., {Rosati}, P., {Nonino}, M., {et~al.} 2010, \apj, 709, 512

\bibitem[{{Salpeter}(1955)}]{Salpeter55}
{Salpeter}, E.~E. 1955, \apj, 121, 161

\bibitem[{{Shankar} {et~al.}(2013){Shankar}, {Marulli}, {Bernardi}, {Mei},
  {Meert}, \& {Vikram}}]{Shankar13}
{Shankar}, F., {Marulli}, F., {Bernardi}, M., {et~al.} 2013, \mnras, 428, 109

\bibitem[{{Shen} {et~al.}(2003){Shen}, {Mo}, {White}, {Blanton}, {Kauffmann},
  {Voges}, {Brinkmann}, \& {Csabai}}]{Shen03}
{Shen}, S., {Mo}, H.~J., {White}, S.~D.~M., {et~al.} 2003, \mnras, 343, 978

\bibitem[{{Smith} {et~al.}(2005){Smith}, {Treu}, {Ellis}, {Moran}, \&
  {Dressler}}]{Smith05}
{Smith}, G.~P., {Treu}, T., {Ellis}, R.~S., {Moran}, S.~M., \& {Dressler}, A.
  2005, \apj, 620, 78

\bibitem[{{Spitler} {et~al.}(2012){Spitler}, {Labb{\'e}}, {Glazebrook},
  {Persson}, {Monson}, {Papovich}, {Tran}, {Poole}, {Quadri}, {van Dokkum},
  {Kelson}, {Kacprzak}, {McCarthy}, {Murphy}, {Straatman}, \&
  {Tilvi}}]{Spitler13}
{Spitler}, L.~R., {Labb{\'e}}, I., {Glazebrook}, K., {et~al.} 2012, \apjl, 748,
  L21

\bibitem[{{Stanford} {et~al.}(2012){Stanford}, {Brodwin}, {Gonzalez},
  {Zeimann}, {Stern}, {Dey}, {Eisenhardt}, {Snyder}, \& {Mancone}}]{Stanford12}
{Stanford}, S.~A., {Brodwin}, M., {Gonzalez}, A.~H., {et~al.} 2012, \apj, 753,
  164

\bibitem[{{Stockton} {et~al.}(2008){Stockton}, {McGrath}, {Canalizo}, {Iye}, \&
  {Maihara}}]{Stockton08}
{Stockton}, A., {McGrath}, E., {Canalizo}, G., {Iye}, M., \& {Maihara}, T.
  2008, \apj, 672, 146

\bibitem[{{Strazzullo} {et~al.}(2013){Strazzullo}, {Gobat}, {Daddi}, {Onodera},
  {Carollo}, {Dickinson}, {Renzini}, {Arimoto}, {Cimatti}, {Finoguenov}, \&
  {Chary}}]{Strazzullo13}
{Strazzullo}, V., {Gobat}, R., {Daddi}, E., {et~al.} 2013, arXiv:1305.3577

\bibitem[{{Szomoru} {et~al.}(2010){Szomoru}, {Franx}, {van Dokkum}, {Trenti},
  {Illingworth}, {Labb{\'e}}, {Bouwens}, {Oesch}, \& {Carollo}}]{Szomoru10}
{Szomoru}, D., {Franx}, M., {van Dokkum}, P.~G., {et~al.} 2010, \apjl, 714,
  L244

\bibitem[{{Tanaka} {et~al.}(2010){Tanaka}, {Finoguenov}, \& {Ueda}}]{Tanaka10}
{Tanaka}, M., {Finoguenov}, A., \& {Ueda}, Y. 2010, \apjl, 716, L152

\bibitem[{{Tanaka} {et~al.}(2013{\natexlab{a}}){Tanaka}, {Finoguenov},
  {Mirkazemi}, {Wilman}, {Mulchaey}, {Ueda}, {Xue}, {Brandt}, \&
  {Cappelluti}}]{Tanaka13b}
{Tanaka}, M., {Finoguenov}, A., {Mirkazemi}, M., {et~al.} 2013{\natexlab{a}},
  \pasj, 65, 17

\bibitem[{{Tanaka} {et~al.}(2013{\natexlab{b}}){Tanaka}, {Toft}, {Marchesini},
  {Zirm}, {De Breuck}, {Kodama}, {Koyama}, {Kurk}, \& {Tanaka}}]{Tanaka13}
{Tanaka}, M., {Toft}, S., {Marchesini}, D., {et~al.} 2013{\natexlab{b}}, \apj,
  772, 113

\bibitem[{{Taylor} {et~al.}(2009){Taylor}, {Franx}, {van Dokkum}, {Quadri},
  {Gawiser}, {Bell}, {Barrientos}, {Blanc}, {Castander}, {Damen},
  {Gonzalez-Perez}, {Hall}, {Herrera}, {Hildebrandt}, {Kriek}, {Labb{\'e}},
  {Lira}, {Maza}, {Rudnick}, {Treister}, {Urry}, {Willis}, \&
  {Wuyts}}]{Taylor09}
{Taylor}, E.~N., {Franx}, M., {van Dokkum}, P.~G., {et~al.} 2009, \apjs, 183,
  295

\bibitem[{{Thomas} {et~al.}(2010){Thomas}, {Maraston}, {Schawinski}, {Sarzi},
  \& {Silk}}]{Thomas10}
{Thomas}, D., {Maraston}, C., {Schawinski}, K., {Sarzi}, M., \& {Silk}, J.
  2010, \mnras, 404, 1775

\bibitem[{{Toft} {et~al.}(2009){Toft}, {Franx}, {van Dokkum}, {F{\"o}rster
  Schreiber}, {Labbe}, {Wuyts}, \& {Marchesini}}]{Toft09}
{Toft}, S., {Franx}, M., {van Dokkum}, P., {et~al.} 2009, \apj, 705, 255

\bibitem[{{Tran} {et~al.}(2010){Tran}, {Papovich}, {Saintonge}, {Brodwin},
  {Dunlop}, {Farrah}, {Finkelstein}, {Finkelstein}, {Lotz}, {McLure},
  {Momcheva}, \& {Willmer}}]{Tran10}
{Tran}, K.-V.~H., {Papovich}, C., {Saintonge}, A., {et~al.} 2010, \apjl, 719,
  L126

\bibitem[{{Treu} {et~al.}(2003){Treu}, {Ellis}, {Kneib}, {Dressler}, {Smail},
  {Czoske}, {Oemler}, \& {Natarajan}}]{Treu03}
{Treu}, T., {Ellis}, R.~S., {Kneib}, J.-P., {et~al.} 2003, \apj, 591, 53

\bibitem[{{Treu} {et~al.}(2005){Treu}, {Ellis}, {Liao}, \& {van
  Dokkum}}]{Treu05L}
{Treu}, T., {Ellis}, R.~S., {Liao}, T.~X., \& {van Dokkum}, P.~G. 2005, \apjl,
  622, L5

\bibitem[{{Valentinuzzi} {et~al.}(2010){Valentinuzzi}, {Fritz}, {Poggianti},
  {Cava}, {Bettoni}, {Fasano}, {D'Onofrio}, {Couch}, {Dressler}, {Moles},
  {Moretti}, {Omizzolo}, {Kj{\ae}rgaard}, {Vanzella}, \&
  {Varela}}]{Valentinuzzi10}
{Valentinuzzi}, T., {Fritz}, J., {Poggianti}, B.~M., {et~al.} 2010, \apj, 712,
  226

\bibitem[{{van der Wel} {et~al.}(2011){van der Wel}, {Rix}, {Wuyts}, {McGrath},
  {Koekemoer}, {Bell}, {Holden}, {Robaina}, \& {McIntosh}}]{vanderWel11}
{van der Wel}, A., {Rix}, H.-W., {Wuyts}, S., {et~al.} 2011, \apj, 730, 38

\bibitem[{{van Dokkum} \& {van der Marel}(2007)}]{vanDokkum07}
{van Dokkum}, P.~G., \& {van der Marel}, R.~P. 2007, \apj, 655, 30

\bibitem[{{van Dokkum} {et~al.}(2008){van Dokkum}, {Franx}, {Kriek}, {Holden},
  {Illingworth}, {Magee}, {Bouwens}, {Marchesini}, {Quadri}, {Rudnick},
  {Taylor}, \& {Toft}}]{vanDokkum08}
{van Dokkum}, P.~G., {Franx}, M., {Kriek}, M., {et~al.} 2008, \apjl, 677, L5

\bibitem[{{van Dokkum} {et~al.}(2010){van Dokkum}, {Whitaker}, {Brammer},
  {Franx}, {Kriek}, {Labb{\'e}}, {Marchesini}, {Quadri}, {Bezanson},
  {Illingworth}, {Muzzin}, {Rudnick}, {Tal}, \& {Wake}}]{vanDokkum10}
{van Dokkum}, P.~G., {Whitaker}, K.~E., {Brammer}, G., {et~al.} 2010, \apj,
  709, 1018

\bibitem[{{Weinmann} {et~al.}(2009){Weinmann}, {Kauffmann}, {van den Bosch},
  {Pasquali}, {McIntosh}, {Mo}, {Yang}, \& {Guo}}]{Weinmann09}
{Weinmann}, S.~M., {Kauffmann}, G., {van den Bosch}, F.~C., {et~al.} 2009,
  \mnras, 394, 1213

\bibitem[{{Weinzirl} {et~al.}(2011){Weinzirl}, {Jogee}, {Conselice},
  {Papovich}, {Chary}, {Bluck}, {Gr{\"u}tzbauch}, {Buitrago}, {Mobasher},
  {Lucas}, {Dickinson}, \& {Bauer}}]{Weinzirl11}
{Weinzirl}, T., {Jogee}, S., {Conselice}, C.~J., {et~al.} 2011, \apj, 743, 87

\bibitem[{{Whitaker} {et~al.}(2012){Whitaker}, {Kriek}, {van Dokkum},
  {Bezanson}, {Brammer}, {Franx}, \& {Labb{\'e}}}]{Whitaker12b}
{Whitaker}, K.~E., {Kriek}, M., {van Dokkum}, P.~G., {et~al.} 2012, \apj, 745,
  179

\bibitem[{{Whitaker} {et~al.}(2011){Whitaker}, {Labb{\'e}}, {van Dokkum},
  {Brammer}, {Kriek}, {Marchesini}, {Quadri}, {Franx}, {Muzzin}, {Williams},
  {Bezanson}, {Illingworth}, {Lee}, {Lundgren}, {Nelson}, {Rudnick}, {Tal}, \&
  {Wake}}]{Whitaker11}
{Whitaker}, K.~E., {Labb{\'e}}, I., {van Dokkum}, P.~G., {et~al.} 2011, \apj,
  735, 86

\bibitem[{{Whitaker} {et~al.}(2013){Whitaker}, {van Dokkum}, {Brammer},
  {Momcheva}, {Skelton}, {Franx}, {Kriek}, {Labbe}, {Fumagalli}, {Lundgren},
  {Nelson}, {Patel}, \& {Rix}}]{Whitaker13}
{Whitaker}, K.~E., {van Dokkum}, P.~G., {Brammer}, G., {et~al.} 2013,
  arXiv:1305.1943

\bibitem[{{Williams} {et~al.}(2011){Williams}, {Quadri}, \&
  {Franx}}]{Williams11}
{Williams}, R.~J., {Quadri}, R.~F., \& {Franx}, M. 2011, \apjl, 738, L25

\bibitem[{{Williams} {et~al.}(2009){Williams}, {Quadri}, {Franx}, {van Dokkum},
  \& {Labb{\'e}}}]{Williams09}
{Williams}, R.~J., {Quadri}, R.~F., {Franx}, M., {van Dokkum}, P., \&
  {Labb{\'e}}, I. 2009, \apj, 691, 1879

\bibitem[{{Williams} {et~al.}(2010){Williams}, {Quadri}, {Franx}, {van Dokkum},
  {Toft}, {Kriek}, \& {Labb{\'e}}}]{Williams10}
{Williams}, R.~J., {Quadri}, R.~F., {Franx}, M., {et~al.} 2010, \apj, 713, 738

\bibitem[{{Wuyts} {et~al.}(2008){Wuyts}, {Labb{\'e}}, {Schreiber}, {Franx},
  {Rudnick}, {Brammer}, \& {van Dokkum}}]{Wuyts08}
{Wuyts}, S., {Labb{\'e}}, I., {Schreiber}, N.~M.~F., {et~al.} 2008, \apj, 682,
  985

\bibitem[{{Zeimann} {et~al.}(2012){Zeimann}, {Stanford}, {Brodwin}, {Gonzalez},
  {Snyder}, {Stern}, {Eisenhardt}, {Mancone}, \& {Dey}}]{Zeimann12}
{Zeimann}, G.~R., {Stanford}, S.~A., {Brodwin}, M., {et~al.} 2012, \apj, 756,
  115

\bibitem[{{Zhang} {et~al.}(2011){Zhang}, {Andernach}, {Caretta}, {Reiprich},
  {B{\"o}hringer}, {Puchwein}, {Sijacki}, \& {Girardi}}]{Zhang10}
{Zhang}, Y.-Y., {Andernach}, H., {Caretta}, C.~A., {et~al.} 2011, \aap, 526,
  A105

\bibitem[{{Zibetti} {et~al.}(2013){Zibetti}, {Gallazzi}, {Charlot}, {Pierini},
  \& {Pasquali}}]{Zibetti13}
{Zibetti}, S., {Gallazzi}, A., {Charlot}, S., {Pierini}, D., \& {Pasquali}, A.
  2013, \mnras, 428, 1479

\bibitem[{{Zirm} {et~al.}(2012){Zirm}, {Toft}, \& {Tanaka}}]{Zirm12}
{Zirm}, A.~W., {Toft}, S., \& {Tanaka}, M. 2012, \apj, 744, 181

\end{thebibliography}
\clearpage

\end{document}